\documentclass[reprint,amsmath,amssymb,aps,prx,superscriptaddress,10pt,nofootinbib]{revtex4-2}

\usepackage{amsmath,amssymb,booktabs,graphicx,subcaption,upgreek}
\usepackage[normalem]{ulem}

\usepackage{caption}
\usepackage{ragged2e}

\usepackage[dvipsnames,table]{xcolor}
\usepackage[colorlinks=true,linkcolor=RoyalBlue,citecolor=Brown,urlcolor=Brown,filecolor=Brown]{hyperref}

\usepackage{float,fancyhdr,todonotes,enumerate}
\usepackage[shortcuts]{extdash}

\usepackage[version=4]{mhchem}
\usepackage{siunitx,multirow,braket}
\DeclareSIUnit\bohr{\text{\ensuremath{a}}_{0}}
\DeclareSIUnit\angstrom{\text{Å}}
\DeclareSIUnit\hartree{\text{\ensuremath{E}}_{\mathrm{h}}}
\usepackage{threeparttable}
\usepackage{comment}
\usepackage{soul, xcolor}

\usepackage[most]{tcolorbox}

\begin{document}

\title{\texorpdfstring{Formation of gaseous, doubly charged cerium monofluoride CeF$^{2+}$ and its sensitivity to new physics}{Formation of gaseous, doubly charged cerium monofluoride CeF2+ and its sensitivity to new physics}}

\author{R. Simpson}
\email{ranes@triumf.ca}
\affiliation{TRIUMF, Vancouver, British Columbia, V6T 2A3, Canada}
\affiliation{Department of Physics \& Astronomy, University of British Columbia, Vancouver, British Columbia V6T 1Z1, Canada}

\author{C. Z\"ulch}
\email{zuelchc4@staff.uni-marburg.de}
\affiliation{Fachbereich Chemie, Philipps-Universit\"at Marburg, Hans-Meerwein-Str. 4, 35032 Marburg, Germany}

\author{K. B. Ng}
\email{kbng@triumf.ca}
\affiliation{TRIUMF, Vancouver, British Columbia, V6T 2A3, Canada}

\author{I. Belosevic}\thanks{Present Address: IRFU, CEA, Université Paris-Saclay}
\affiliation{TRIUMF, Vancouver, British Columbia, V6T 2A3, Canada}

\author{C. Charles}
\affiliation{TRIUMF, Vancouver, British Columbia, V6T 2A3, Canada}
\affiliation{Department of Chemistry, Simon Fraser University. 8888 University Drive, Burnaby, British Columbia, V5A 1S6, Canada}

\author{P. Justus}
\affiliation{Ruprecht-Karls Universit\"at Heidelberg, Grabengasse 1, 69117 Heidelberg, Germany}
\affiliation{Max-Planck-Institut für Kernphysik, Saupfercheckweg 1, 69117 Heidelberg, Germany}

\author{R.~Berger}
\email{robert.berger@uni-marburg.de}
\affiliation{Fachbereich Chemie, Philipps-Universit\"at Marburg, Hans-Meerwein-Str. 4, 35032 Marburg, Germany}

\author{S.~Malbrunot-Ettenauer}
\email{sette@triumf.ca}
\affiliation{TRIUMF, Vancouver, British Columbia, V6T 2A3, Canada}
\affiliation{Department of Physics, University of Toronto, Toronto, Ontario, M5S 1A7, Canada}

\author{A. A. Kwiatkowski}
\affiliation{TRIUMF, Vancouver, British Columbia, V6T 2A3, Canada}
\affiliation{Department of Physics and Astronomy, University of Victoria, 3800 Finnerty Road, Victoria, British Columbia, V8O 5C2, Canada}

\author{M. P. Reiter}
\affiliation{School of Physics and Astronomy, The University of Edinburgh, Peter Guthrie Tait
Road, Edinburgh, EH9 3FD, Scotland, UK}

\author{J. Ash}
\affiliation{TRIUMF, Vancouver, British Columbia, V6T 2A3, Canada}

\author{C. Babcock}
\affiliation{TRIUMF, Vancouver, British Columbia, V6T 2A3, Canada}

\author{J. Bergmann}
\affiliation{II. Physikalisches Institut, Justus-Liebig-Universit¨at, 35392 Gießen, Germany}

\author{E. Brisley}
\affiliation{TRIUMF, Vancouver, British Columbia, V6T 2A3, Canada}
\affiliation{Department of Physics \& Astronomy, University of British Columbia, Vancouver, British Columbia V6T 1Z1, Canada}

\author{J. D. Cardona}
\affiliation{TRIUMF, Vancouver, British Columbia, V6T 2A3, Canada}
\affiliation{Department of Physics and Astronomy, University of Manitoba, 30A Sifton Road, Winnipeg, Manitoba, R3T 2N2 Canada}

\author{C. Chambers}
\affiliation{TRIUMF, Vancouver, British Columbia, V6T 2A3, Canada}

\author{A. Czihaly}
\affiliation{TRIUMF, Vancouver, British Columbia, V6T 2A3, Canada}
\affiliation{Department of Physics and Astronomy, University of Victoria, 3800 Finnerty Road, Victoria, British Columbia, V8O 5C2, Canada}

\author{A. Gottberg}
\affiliation{TRIUMF, Vancouver, British Columbia, V6T 2A3, Canada}

\author{S. Kakkar}
\affiliation{TRIUMF, Vancouver, British Columbia, V6T 2A3, Canada}
\affiliation{Department of Physics and Astronomy, University of Manitoba, 30A Sifton Road, Winnipeg, Manitoba, R3T 2N2 Canada}

\author{J. Lassen}
\affiliation{TRIUMF, Vancouver, British Columbia, V6T 2A3, Canada}
\affiliation{Department of Physics, Simon Fraser University. 8888 University Drive, Burnaby, British Columbia, V5A 1S6, Canada}
\affiliation{Department of Physics and Astronomy, University of Manitoba, 30A Sifton Road, Winnipeg, Manitoba, R3T 2N2 Canada}

\author{F. Maldonado Mil\'{a}n}
\affiliation{TRIUMF, Vancouver, British Columbia, V6T 2A3, Canada}

\author{A. Mollaebrahimi}
\affiliation{GSI Helmholtz Center for Heavy Ion Research, Campus Gie\ss en, 35392, Gie\ss en, Germany}

\author{V.~Radchenko}
\affiliation{TRIUMF, Vancouver, British Columbia, V6T 2A3, Canada}
\affiliation{Department of Chemistry, University of British Columbia, Vancouver, British Columbia V6T 1Z1, Canada}

\author{E. Taylor}
\affiliation{TRIUMF, Vancouver, British Columbia, V6T 2A3, Canada}
\affiliation{Department of Physics and Astronomy, University of Western Ontario, 1151 Richmond Street, London, Ontario, N6A 3K7 Canada}

\author{A. Teigelh\"ofer}
\affiliation{TRIUMF, Vancouver, British Columbia, V6T 2A3, Canada}

\author{C. Walls}
\affiliation{TRIUMF, Vancouver, British Columbia, V6T 2A3, Canada}
\affiliation{Department of Physics and Astronomy, University of Manitoba, 30A Sifton Road, Winnipeg, Manitoba, R3T 2N2 Canada}

\author{A. Weaver}
\affiliation{TRIUMF, Vancouver, British Columbia, V6T 2A3, Canada}

\author{P. Weligampola}
\affiliation{TRIUMF, Vancouver, British Columbia, V6T 2A3, Canada}
\affiliation{Department of Physics and Astronomy, University of Manitoba, 30A Sifton Road, Winnipeg, Manitoba, R3T 2N2 Canada}

\begin{abstract}
Tricationic protactinium monofluoride ($^{229}$PaF$^{3+}$) has been proposed as a candidate for probing physics beyond the Standard Model of particle physics. Since studies with $^{229}$PaF$^{3+}$ require significant experimental advances, we exploit the stable, valence-isoelectronic dicationic cerium monofluoride (CeF$^{2+}$) as a surrogate. Gas-phase fluorinated-cerium molecular ions are formed and identified using the Off-Line Ion Source and TITAN mass measurement facilities at TRIUMF. Quantum chemical calculations are performed on the electronic structure of CeF$^{2+}$, revealing a parallel to that of $^{229}$PaF$^{3+}$. Moreover, these calculations provide estimates on the sensitivity of CeF$^{2+}$ itself to various $\mathcal{P,T}$-odd properties. A brief discourse on the specifics of the quantum control of CeF$^{2+}$ is presented which anticipates future searches for symmetry violations.
\end{abstract}

\maketitle

\section{Introduction}

The Standard Model of particle physics provides presently the most successful description of the universe at a microscopic level \cite{Patrignani_2016}.
Nevertheless, it is considered to be incomplete and many efforts aim to detect new physics beyond the Standard Model. One class of such experiments is the search for electric dipole moments (EDMs) \cite{POSPELOV2005119,ENGEL201321,Chupp2019electric}, in which molecular systems have proven to be powerful tools \cite{Cho1991search,hudson2011improved,acme2018improved,Roussy2023an}.
This is made possible by the exquisite quantum control of molecules to, for example, exploit their rich internal structures for enhancements in sensitivities to new physics \cite{demille2001search}. Upgrades for these experiments are underway \cite{Norrgard2017hyperfine,Fitch:2021:2058-9565/abc931,ng2022spectroscopy,wu2022electrostatic}, and there is building momentum to use novel or more exotic molecules as probes for other sources of physics beyond the Standard Model and to provide access to additional control techniques like laser cooling \cite{DiRosa:04,isaev:2010,isaev:2016,verma2020electron,klos2022prospects,zulch2022cool,marc2023candidate,anderegg2023quantum,arrowsmithkron:2024}. Among these, radioactive molecules have recently been introduced as powerful precision probes for new physics \cite{ronaldNature, arrowsmithkron:2024}. Despite the short half-lives and minute sample sizes of the radionuclides involved, remarkable progress has been achieved, both in forming and controlling such molecular systems \cite{fan2021, AU2023375, MiaAC} and in elucidating their structure, \cite{Udrescu2024, MAK2024, Wilkins2025, MAK2025, MAKArxiv} all reinforcing their strong EDM sensitivities.

An example of an EDM measurement with new physics implications is the search for a nuclear Schiff moment in atomic or molecular systems \cite{Flambaum2020electric}. The protactinium-229 nucleus promises extremely high enhancement of the Schiff moment due to octupole deformation \cite{Spevak1997,Flambaum2020electric}. Tricationic protactinium-229 monofluoride ($^{229}$PaF$^{3+}$) has been proposed as a protactinium containing molecular system that possesses high sensitivity to a nuclear Schiff moment and is also predicted to have favorable properties for quantum control \cite{zulch2022cool}. Hence, $^{229}$PaF$^{3+}$ is an attractive candidate for probing physics beyond the Standard Model through the measurement of its exotic nuclear moment that arises from violation of parity-inversion ($\mathcal{P}$) and time-reversal ($\mathcal{T}$) symmetry. The particularly favorable opportunities of a $^{229}$PaF$^{3+}$ experiment in this context have been highlighted recently in the framework of a global analysis \cite{Gaul2024} that accounts for previous as well as currently planned measurements of $\mathcal{P},\mathcal{T}$-odd signatures in atoms and molecules.

However, the formation of $^{229}$PaF$^{3+}$ has not yet been demonstrated for three main reasons. First, $^{229}$Pa (half-life of 1.5~days) is a radioactive isotope, so experiments have to be performed at rare isotope laboratories. Second, protactinium is a heavy, refractory element which cannot be easily delivered to experimenters with conventional techniques available at radioactive ion beam facilities. Third, highly charged molecules are prone to Coulomb explosions, so creation and control of the molecule are non-trivial. The former two are being tackled by various rare isotope facilities in the world, see for example, in Ref. \cite{jost2013, griswold2018, BumrahPa, ShigekawaPa}. This paper aims to address the latter.

Since all protactinium isotopes are radioactive, lanthanide analogues constitute interesting alternatives to develop experimental schemes for subsequent studies of $^{229}$Pa. In particular, dicationic cerium (Ce$^{2+}$) is valence-isoelectronic to tricationic protactinium (Pa$^{3+}$) by virtue of their ground state electronic configurations [Xe]4f$^2$ and [Rn]5f$^2$, respectively. Hence, demonstrating the formation of dicationic cerium monofluoride (CeF$^{2+}$) at a rare isotope facility using stable cerium represents a rewarding steppingstone to the formation and quantum control of $^{229}$PaF$^{3+}$. Moreover, CeF$^{2+}$ is an interesting molecule in its own right within the context of precision measurements and searches for physics beyond the Standard Model. For instance, there is evidence suggesting nuclear octupole deformation in various cerium isotopes \cite{Egido1992,Cheal2003,sharipov2007collective,brewer2018octupole,Beckers2020,Alexa2022}, one of the key properties for enhancements in the experimental sensitivity to nuclear Schiff moments. Furthermore, as a topic of the present work, precision measurements performed on CeF$^{2+}$ can help to constrain new-physics parameters of effective field theory through a global analysis of measurement results from other atomic and molecular species \cite{Gaul2024}. While CeF$^{2+}$ was not included in the analysis of Ref.~\cite{Gaul2024}, the work has reinforced the general role of precision studies with medium-heavy systems. Despite sensitivities typically weaker than in heavier homologues, these experiments serve to tighten the coverage volume in the relevant $\mathcal{P}$,$\mathcal{T}$-odd parameter space, because the various molecular enhancement factors depend differently on the nuclear charge $Z$. With $Z=58$ cerium lies just slightly above the window referred to specifically in Ref.~\cite{Gaul2024}. Additionally, it has been suggested recently \cite{zuelch2025} that precision spectroscopy of CeF$^{2+}$ may also provide constraints on a possible variation of the fine-structure constant $\alpha$ that is discussed to emerge in the context of quantum gravity or as a signature of scalar bosonic dark matter \cite{uzan2003}.

Earlier experimental work has been performed on the formation of CeF$^{2+}$ in solutions \cite{sarma:1967,lyle:1967,walker:1967,arisaka:1999,migdisov:2009} as well as of gaseous CeF$^{+}$ \cite{koyanagi:2005,cheng:2006}. 
The observation of gaseous CeF$^{2+}$ was briefly mentioned in earlier work involving argon-ion bombardment of a CeF$_3$ crystal \cite{Lorincik}. However, this method required macroscopic amounts of CeF$_3$ and is thus not transferable to  the formation of radioactive $^{229}$PaF$^{3+}$.
 Several theoretical studies on CeF$^{2+}$ and closely related molecules have also been reported. These include thermodynamic properties of CeF$^{2+}$ \cite{haas:1995,arisaka:1999}, stability of various fluorination states of cerium \cite{gibson:1996}, ionization of neutral CeF \cite{gotkis:1991}, amongst many others. The present paper is dedicated to the experimental creation of CeF$^{2+}$ in the gas phase using instrumentation available at radioactive ion beam facilities and, thus, potentially transferable to $^{229}$PaF$^{3+}$, the theoretical discourse of the molecular level structure in CeF$^{2+}$, as well as its possible applications in precision measurements to probe for new physics.

We begin with Section \ref{sec:2}, where we demonstrate formation of CeF$^{2+}$ at TRIUMF and discuss its relevance to the formation of PaF$^{3+}$. We then present calculations of the electronic structure of CeF$^{2+}$ in Section \ref{sec:electronic structure} and review the results in the context of experimental observations and possibility of state manipulation via optical cycling transitions. Subsequently, we investigate sensitivities of various electronic states in CeF$^{2+}$ to $\mathcal{P},\mathcal{T}$-odd physics in Section \ref{sec:4}. Finally, we discuss prospects of performing quantum control of CeF$^{2+}$ in pursuit of a precision measurement of physics beyond present models in Section \ref{sec:5} before concluding with Section \ref{sec:6}.

\section{Experimental Formation and Identification of C\MakeLowercase{e}F Molecules}\label{sec:2}

\begin{figure*}[t]
    \centering
    \includegraphics[width=2\columnwidth]{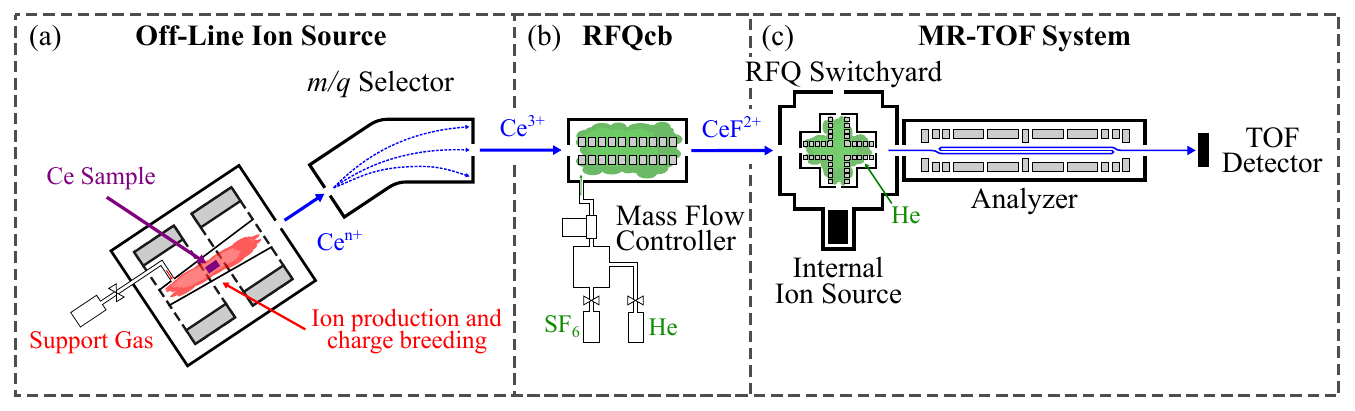}
    \caption{\justifying Schematic of the experimental apparatus. (a) Cerium ion production within the Multiple-Charge Ion Source at the Off-Line Ion Source, and separation of the desired ion species according to its mass-over-charge ratio $m/q$. (b)  Formation of diatomic cerium fluoride ions and simultaneous cooling and bunching inside the RFQcb. (c) MR-TOF system for mass determination and subsequent ion species identification. See text for details.}
    \label{fig:Ce_outline}
\end{figure*}

The experimental work is conducted in the low-energy hall of the Isotope Separator and ACcelerator (ISAC) complex \cite{ISAC,Ball2020} at TRIUMF utilizing its Off-Line Ion Source (OLIS) facility \cite{yayamanna:2010} 
%for the creation and delivery of stable isotope beams,
and TRIUMF’s Ion Trap for Atomic and Nuclear science (TITAN) \cite{Kwiatkowski:2024}. While details of the respective steps are discussed in subsequent sections, the general procedures of these studies are outlined in Figure~\ref{fig:Ce_outline}. 

Beams of stable $^{140}$Ce cations in different charge states are created by electron cyclotron resonance (ECR) ionization at the OLIS facility, see Section \ref{ssec:Ce Production}. After mass-over-charge ($m/q$) selection via a dipole magnet, a continuous beam of $^{140}$Ce$^{+}$ or $^{140}$Ce$^{3+}$ ions is sent to TITAN where the ions are stopped and confined by a buffer-gas-filled radio-frequency quadrupole ion beam cooler and buncher (RFQcb) \cite{Brunner:2012}, see Figure~\ref{fig:Ce_outline}(b). For the formation of $^{19}$F containing molecules, a small amount of sulfur hexafluoride (SF$_6$) is added to the helium (He) buffer-gas admixture so that various molecules of interest are synthesized in the RFQcb. After a well-defined formation period, the trapped ions are transferred into TITAN's multiple-reflection time-of-flight (MR-TOF)\cite{Reiter:2021} system for mass identification, see Figure~\ref{fig:Ce_outline}(c). 

The MR-TOF instrument consists of a helium-filled low energy RFQ switchyard for additional ion-bunch preparation and a differentially pumped time-of-flight (TOF) mass analyzer. In the latter, the ions are revolving in-between two electrostatic mirrors such that an extended flight path of up to a few kilometers is folded into a device of about $\sim0.5$~m in length. Ions of different mass $m$ and charge $q$ are separated in space which is measured on a TOF detector upon the ions' extraction from the MR-TOF device. When operated in high resolution mode accomplished by several hundred revolutions in the analyzer, mass resolving powers in excess of $R=m/\Delta m>10^5$ are achieved. The internal RFQ switchyard \cite{Plass2015} further accepts a beam of stable potassium (K) and rubidium (Rb) ions delivered by a separate surface ion source. These ions of well-known masses serve as calibrants for the recorded TOF spectra in which the molecules of interest are unambiguously identified. 

First molecular formation experiments are conducted with an incoming beam of $^{140}$Ce$^{+}$ ions to demonstrate the experimental procedures for the creation of molecules in  the RFQcb, see Section~\ref{ssec:Single Charge Ce}. To this end, we have upgraded our gas handling system to add  SF$_6$ into the buffer gas. In Section~\ref{ssec:multicharge Ce}, we describe the newly developed methods when exploiting incoming cerium beams of higher charge states which allow us to form and identify $^{140}$CeF$^{2+}$ in gaseous phase. 
As TITAN is directly coupled to TRIUMF's rare isotope production facilities, all employed techniques can be readily transferred to the formation of protactinium-containing molecules, once a Pa beam becomes available at ISAC. The prospect for the creation of PaF$^{3+}$ in general and specifically at TRIUMF is given in Section~\ref{ssec:PaFprospects}. 

\subsection{Formation of Cerium Cation Beams} \label{ssec:Ce Production}

Naturally occurring cerium is composed of four stable isotopes with atomic mass numbers ${A} =$ 136, 138, 140, and 142, of which $^{140}\text{Ce}$ has the highest abundance at 88.45\%. In the present work, $^{140}\text{Ce}$ cations are formed at OLIS for the first time, accomplished via ECR ionization  utilizing a `Supernanogan' Multi-Charge Ion Source (MCIS) \cite{yayamanna:2010}. Details of the extensive beam development campaign for the ion beam production using multiple source and sputtering parameters as well as support gases are summarized in \cite{JustusCeriumOlis}.

\begin{figure}[htbp]%
    \centering    
    \includegraphics[width=\columnwidth]{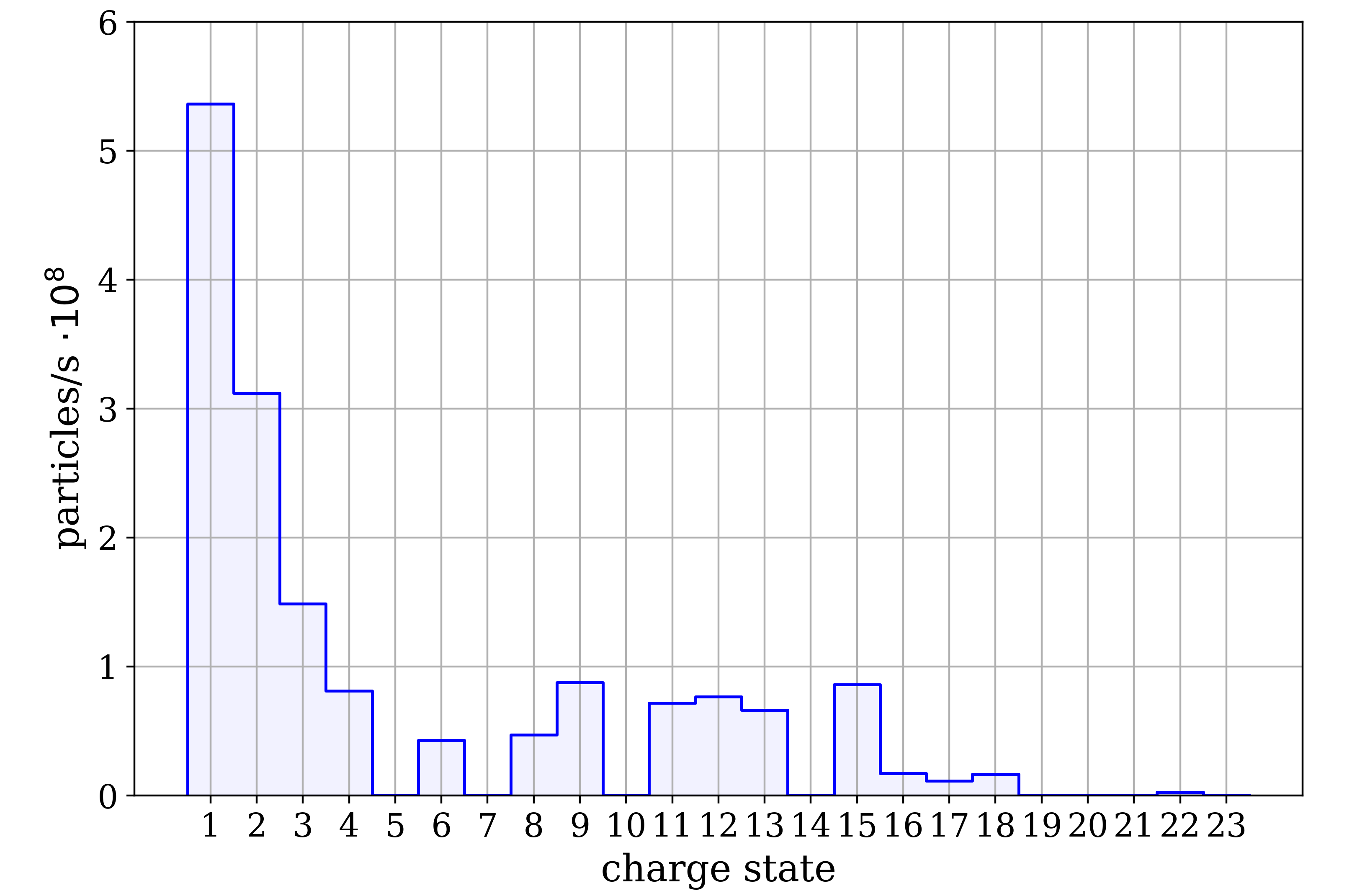}%
    \caption{\justifying Typical cerium cation charge state distribution from the Multi-charge ion source with helium} support gas. The distribution is shown as the yield of ions per second at $8 \times 10^{-9}~\text{mbar}$ gas pressure.%
    \label{fig: Cerium ion yield}
\end{figure}

In short, a $10 \times 3 \times 2$ mm$^3$ strip of 99\% pure cerium metal is held suspended inside the plasma chamber of the MCIS ion source along its longitudinal axis. The source is floated to $14.5 ~\text{kV}$ and a permanent $1.2~\text{T}$ semi-Helmholtz magnetic field ($B_{z}$) surrounds the plasma chamber. The cerium metal strip is sputtered at $-400~\text{V}$ relative to the plasma chamber with helium support gas using dual-frequency radiofrequency (RF) heating \cite{Jayamanna_2024}. This dual-frequency ECR sputtering of the cerium strip produces multiple positive charge-states of $^{140}\text{Ce}$ up to and beyond $q=+22$. The MCIS is normally operated to obtain higher charge states, typically $q>+10$. As it will be explained in Section \ref{ssec:Single Charge Ce}, the present program requires Ce ions with $q=+1$ and $q=+3$. Thus, operational parameters of the MCIS including higher sputter gas pressures, sputter gas chemical composition, distance of the cerium metal strip inside the plasma chamber, and sputter voltage are optimized in order to maximize the abundances of these lower charge states, see Figure~\ref{fig: Cerium ion yield}.
The cerium ions are accelerated to an ion beam energy of $E=14.5\cdot q~\text{keV}$ and are subsequently selected by a bending magnet ($m/\Delta m \sim 540$) downstream after the ion source, see Figure 1(a). Thus, a clean ion beam of either $^{140}\text{Ce}^{+}$ or $^{140}\text{Ce}^{3+}$ is transported via electrostatic beamlines to TITAN. There,  over $5 \times 10^{8}$ particles per second (pps) of $^{140}\text{Ce}^{+}$ and $1.5 \times 10^{8}$ pps of $^{140}\text{Ce}^{3+}$ are typically received for molecule formation experiments. For future campaigns, higher yields of cerium cations are anticipated by further optimizing source and sputter conditions.

\subsection{\texorpdfstring{Formation of Fluorine}-Containing Molecules with Incoming Ce$^+$ Beams}{} \label{ssec:Single Charge Ce}

The molecular formation and subsequent mass identification occur within the ion-trap system TITAN. For the purposes of this experiment, we are not using TITAN in its usual operating regime. Although TITAN draws on its capabilities to work with highly charged ions for Penning-trap mass measurements \cite{PhysRevLett.107.272501}, the RFQcb usually only accepts ion beams with a charge state of $q=+1$, and occasionally $q=+2$. 
Moreover, the RFQcb's gas mixing system has never been used in previous experimental campaigns. Thus, our first objective is to establish methods to produce fluorine-containing molecules in the RFQcb using the OLIS beam of Ce$^+$. 

The creation of molecules in gas-filled RFQcbs is not uncommon at radioactive ion beam facilities and it usually relates to the presence of residual-gas components in the ion trap. While undesirable in many applications, this effect can be harnessed to intentionally form molecules. Recently, this approach has been successfully applied in Ref.~\cite{AU2023375}. To increase the control of the in-trap formation of molecular ions and to expand it to fluorine-containing species, a suitable gas mixing system is of advantage. This method is often followed in analytical sciences when reaction or collision cells are employed \cite{TANNER20021361}, for example, in the context of mass spectrometry.

In our implementation at a rare isotope facility, trace amounts of sulfur hexafluoride (SF$_6$) are introduced into the trapping region of the RFQcb in addition to the normally used helium buffer gas, see Figure~\ref{fig:Ce_outline}(b). To achieve this, a reservoir of approximately 16~L is pumped to a vacuum pressure of $\sim10^{-3}$~mbar and then isolated from the vacuum pump. SF$_6$ is subsequently admitted into the reservoir to a pressure of $\sim20$~mbar. Finally, the reservoir is pressurized to 1500~mbar with helium gas to create a high pressure environment from which a slow leak is supplied to the RFQcb, providing stable operation for up to 4 hours before needing to be reset. In the present work, the RFQcb is held at a pressure of $\approx10^{-1}$~mbar balanced between a differential pumping system, which isolates the trapping region from the rest of the beamline, and a steady flow of 5 sccm of the gas mixture. The partial pressure of SF$_6$ in the RFQcb is thus $\sim10^{-3}$~mbar.

In order to form fluorine-containing cerium molecular ions,  the incoming continuous beam of singly charged cerium cations enters the RFQcb which is biased to an electrostatic potential just below the ion-beam energy. A precisely timed, variable gating window is activated by applying deflecting potentials to steering electrodes in front of the RFQcb. This defines the ion loading period and, thus, controls the number of injected ions. Once captured in the trap, the ions are thermalized via collisions with helium buffer-gas atoms that dissipate the ions' kinetic energy. Additionally, they may react with SF$_6$ molecules to form cerium fluoride. By allowing the formation period of typically 5-25~ms to persist significantly longer than the loading time (usually $\sim 500~\upmu$s), it can be ensured that all ions spend a similar time in the trap for cooling and synthesis of molecules. Upon the completion of the formation time, a well-defined ion bunch is released from the RFQcb. A pulsed drift tube is used to set the ion energy to a value suitable for recapture in the MR-TOF's internal RFQ switchyard.

Once the bunch is released into the MR-TOF mass analyzer, see Figure~\ref{fig:Ce_outline}(c), the ion species are physically separated in space based on their TOF. Since all ions pass the same acceleration potential before entering the analyzer, as the bunch revolves, individual ion species travel at a velocity inversely proportional to the square-root of their $m/q$ ratio. This separation can be detected as a series of temporally distinct peaks on the TOF detector. The quality in this temporal isolation of different ion species and, therefore, the mass resolving power is a function of the ions' path length, which depends on the number of turns (total revolutions completed in the analyzer) performed by the ion of interest. 

\begin{figure}[t]
    \centering
    \includegraphics[width=1\columnwidth]{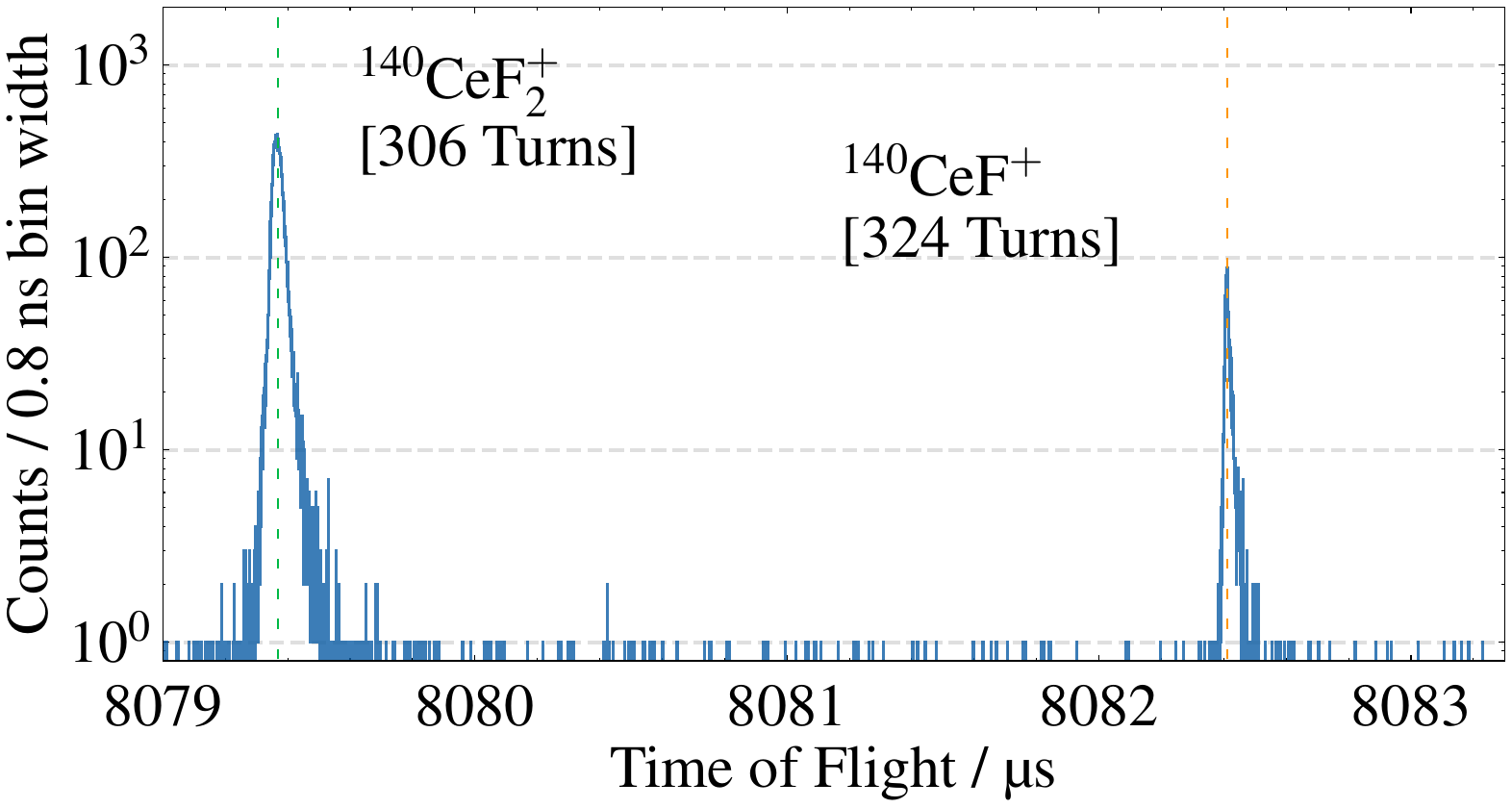}
    \caption{\justifying A high turn TOF spectrum of $^{140}$CeF$^+$ and $^{140}$CeF$_2^+$ ions. Here, CeF$_2^+$ ions have traversed over 306 turns, providing a mass resolving power of 106,000. The TOF is recorded as the elapsed time since the ion bunch has been injected into the MR-TOF analyzer section [see Figure \ref{fig:Ce_outline}(c)]. The second peak in the spectrum, $^{140}$CeF$^+$, although being the lighter of the pair, is seen at a later TOF. This is due to the higher number of turns completed, as indicated in the peak label. The number of turns is carefully chosen such that the release point of the ions allows both species to be seen clearly in the same TOF spectrum. This measurement represents the first high resolution confirmation of the formation of cerium monofluoride monocations in the RFQcb supported by the AME database values for the corresponding masses translated into corresponding TOF values, shown as vertical coloured lines.}
    \label{fig:CeF1+}
\end{figure}

Utilizing the MR-TOF control and data aquisition (DAQ) system described in Ref.~\cite{JBergmann2019}, initial TOF spectra are recorded at 1-5 turns, leading to a mass resolving power on the order of $R\approx5 \times 10^3$, which provide a broad overview of the ion-beam composition after the molecular formation in the RFQcb. Subsequently, high resolution measurements are performed at 250-500 turns resulting in a mass resolving power of up to $R=5 \times 10^5$. Once the peaks in the TOF spectrum are observed, comparisons in TOF are made to known values provided by the Atomic Mass Evaluation (AME) \cite{AME2020} in order to identify ionic species present in the MR-TOF device. An example of such a TOF spectrum is shown in Figure~\ref{fig:CeF1+} which constitutes our first high-resolution confirmation of cerium monofluoride $^{140}$CeF$^+$ and cerium difluoride $^{140}$CeF$_2^+$ ions formed at TITAN.

\subsection{Formation of Multiply Charged Cerium Fluoride Molecules} \label{ssec:multicharge Ce}
Following the demonstration of the formation and identification of singly charged cerium fluoride molecules within the TITAN ion-trap system, subsequent studies focus on multiply charged molecules along two main objectives. First, we aim to create doubly charged cerium monofluoride CeF$^{2+}$ as a molecule that is valence isoelectronic to triply charged protactinium monofluoride, PaF$^{3+}$. Second, we attempt to form triply charged cerium monofluoride CeF$^{3+}$ to explore the suitability of our experimental approach to produce molecules in higher charge states as a preparatory step towards the synthesis of PaF$^{3+}$. 

While the formation is energetically allowed, ionic molecular formation is heavily affected by the charge state of the various constituents, see Ref.~\cite{Bohme2022}. When forming doubly or triply charged molecules in an RFQcb, one challenge is accepting the incoming cerium atomic ions in higher charge states without any loss of charge to He buffer gas atoms. Due to the ionization energies\footnote{Here, we use the notation IE(X$^{n+}$) to refer to the energy required to ionize X$^{n+}$ into X$^{(n+1)+}$.}: IE(Ce$^{2+}$) = 20.198~eV, IE(Ce$^{3+}$) = 36.758~eV \cite{Martin1978},  IE(He) = 24.587~eV \cite{Kandule2010}, it is energetically possible that $^{140}$Ce$^{4+}$ undergoes charge exchange with He buffer-gas atoms. This may reduce the availability of $^{140}$Ce$^{4+}$ ions for reactions with SF$_6$. Since IE(Ce$^{2+})<$~IE(He), however, we do not expect charge exchange between He atoms and low-energy $^{140}$Ce$^{3+}$ ions. For these reasons, $^{140}$Ce$^{3+}$ is selected in an attempt to form both $^{140}$CeF$^{2+}$ and $^{140}$CeF$^{3+}$ ions. In addition to He atoms, other gas impurities may be present in the RFQcb's ion-confinement region, either through outgassing from material in the ion trap or as contamination in the He gas. Thus, $^{140}$Ce$^{3+}$ ions might be lost in charge exchange reactions with other residual gas components. To confirm that the RFQcb is capable of handling $^{140}$Ce$^{3+}$, its successful transmission is demonstrated, without leaking SF$_6$ into the RFQcb.

\begin{figure}[t]
    \centering
    \includegraphics[width=1\columnwidth]{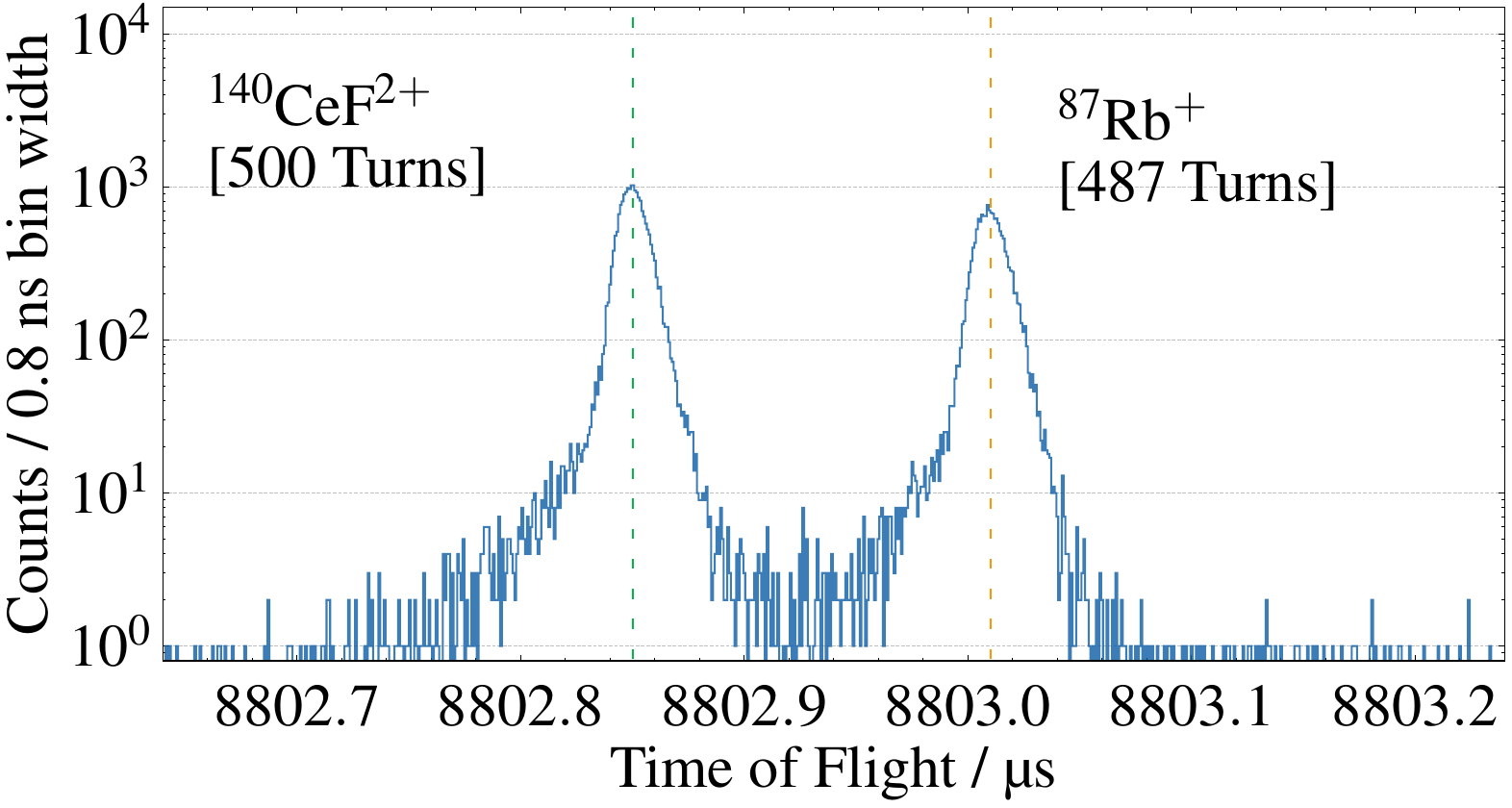}
    \caption{\justifying A high resolution TOF spectrum ($R$=296,000) identifying $^{140}$CeF$^{2+}$ after molecular formation with incoming $^{140}$Ce$^{3+}$ ions. In this 2000 second long measurement, the mass range selector \cite{Reiter:2021} is engaged which provided a timed window allowing only the CeF$^{2+}$ and the $^{87}$Rb$^+$ calibrant to pass through while trapped within the MR-TOF analyzer, deflecting all other species.}
    \label{fig:CeF_2+}
\end{figure}

Once SF$_6$ is added to the He buffer gas, fluorine-containing molecules can be formed.  Figure~\ref{fig:CeF_2+} illustrates the successful formation and detection of $^{140}$CeF$^{2+}$ ions after injecting $^{140}$Ce$^{3+}$ ions through the RFQcb. In addition to the TOF peak associated with $^{140}$CeF$^{2+}$, the TOF spectrum contains $^{85}$Rb$^+$ ions produced in the internal source which serve as calibrant and thus facilitate the unambiguous identification of $^{140}$CeF$^{2+}$. Once the gas reservoir is pumped out again and refilled with He but without the addition of SF$_6$, the peak associated with $^{140}$CeF$^{2+}$ disappears from the TOF spectrum.  

\begin{figure}[t]
    \centering
    \includegraphics[width=1\columnwidth]{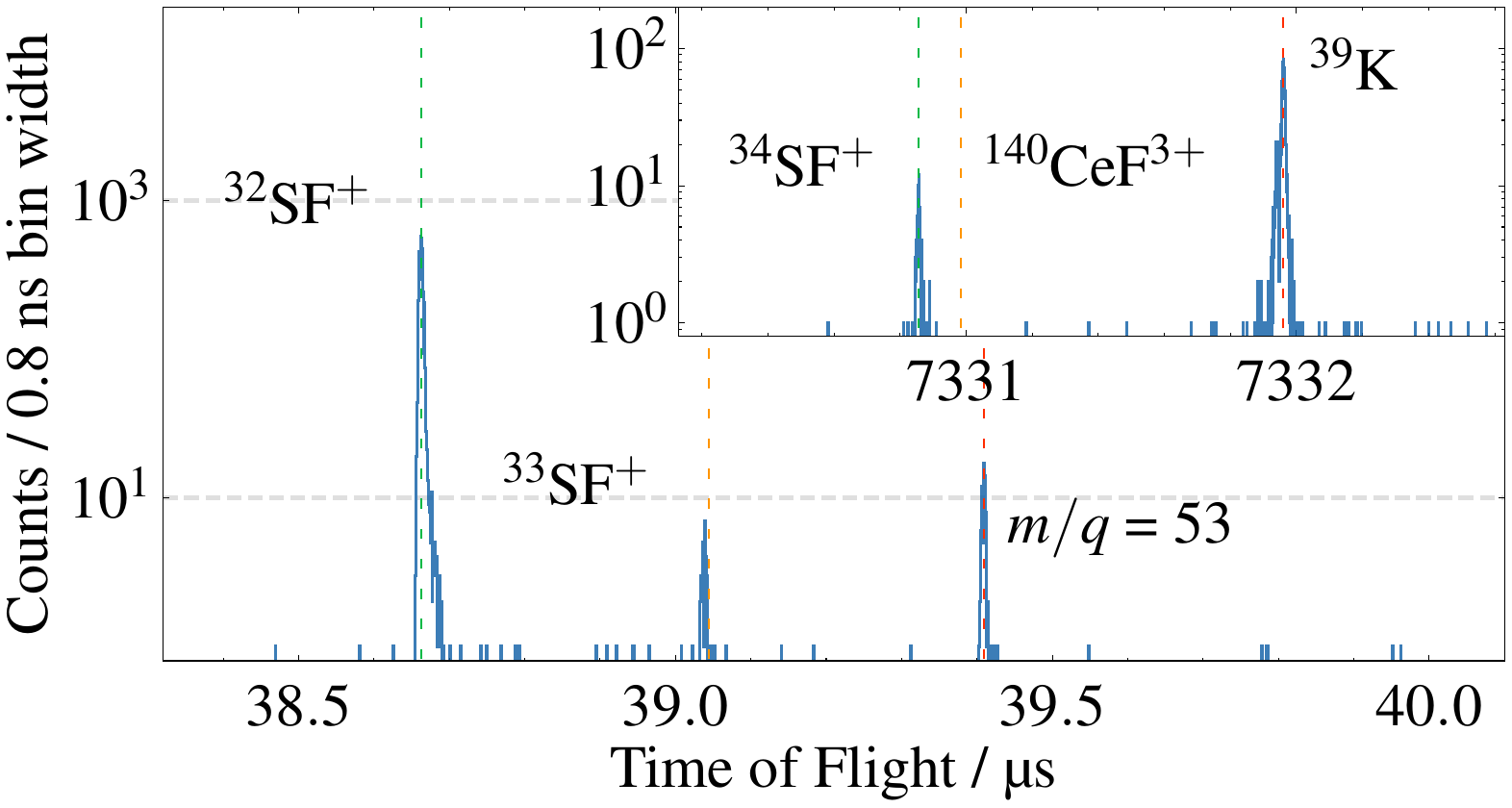}
    \caption{\justifying Identifying the $m/q = 53$ peak. A TOF spectrum recorded after a single turn in the MR-TOF analyzer shows a peak corresponding to the $m/q$ of the $^{140}$CeF$^{3+}$ molecule. Inset: A high-resolution TOF spectrum reveals that the peak is actually a fragment of the SF$_6$ gas that has been ionized in the process of forming the cerium fluoride.}
    \label{fig:CeF3+}
\end{figure}

The secondary goal of producing $^{140}$CeF$^{3+}$ ions proves to be more challenging, see Figure~\ref{fig:CeF3+}, which also underscores the importance of a high-performance mass spectrometer for these types of studies. In the low-resolution spectrum with $R\approx$~4,000, a TOF peak would correspond to the approximate $m/q$ ratio of $^{140}$CeF$^{3+}$. However, once a MR-TOF measurement with 510 turns is performed, inset of Figure~\ref{fig:CeF3+}, the corresponding ion species is in fact identified as $^{34}$SF$^+$, resulting from a reaction between the incoming $^{140}$Ce$^{3+}$ ions and the neutral SF$_6$ in which the latter is fragmented. Although we cannot strictly exclude its feasibility, our present studies do not provide evidence of a $^{140}$CeF$^{3+}$ formation via our present method and an incoming $^{140}$Ce$^{3+}$, despite various changes in the experimental conditions. 

In light of the successful formation of various fluorine-containing molecules, the method shows the relative ease at which the TITAN infrastructure can produce the $^{140}$CeF$^{2+}$ ions by taking advantage of the gas injection setup connected to the RFQcb. The high resolution as well as the various beam-cleaning capabilities of the MR-TOF system result in highly confident ion identification. This result is a proof-of-principle for further studies in molecular formation in cooler-bunchers including PaF$^{3+}$.

The objective of the present work has been to demonstrate the suitability of instrumentation available at radioactive ion beam facilities, designed to handle the small sample sizes inherent to accelerator-produced radionuclides, for the formation of (doubly charged) molecules. At this stage, we thus do not consider the efficiency of transforming the incoming ion beam into molecules of interest to be maximal.  To illustrate the potential of our method, we nevertheless quantify the molecular-ion production rates and efficiencies achieved in these first results and outline upgrades that will enhance them, providing guidance on what can be expected in future work to be focused on optimizing overall performance. To this end, the rate of single-ion CeF$^{2+}$ hits on the TOF detector after MR-TOF mass identification is compared with the intensity of the continuous Ce$^{3+}$ beam delivered from OLIS, as measured on a Faraday cup upstream of the RFQcb. In measurements in which both data is reliably recorded, the typical rate of MR-TOF mass separated CeF$^{2+}$ is $\approx350$~ions per second.

High resolution measurements further reveal that the ion beam extracted from the RFQcb at $m/q = 159/2$ is entirely due to CeF$^{2+}$, and is free from other contaminating ions. Hence, while the MR-TOF system is essential for unambiguous mass identification in the present work, it could readily be replaced by lower-resolution methods such as a magnetic dipole separator that operate with nearly unit efficiency. For atomic ions, the efficiency of the MR-TOF system is reported to be approximately 30\% \cite{Reiter:2021}. For (multiply charged) molecules, we observe collision-induced dissociation \cite{Jacobs2022} as well as intensified charge exchange in the MR-TOF's RFQ, further reducing the efficiency. Therefore, at least 1,000 CeF$^{2+}$ ions per second are formed and delivered by the RFQcb and, thus, already available for subsequent precision studies. This corresponds to an efficiency of at least $3\cdot10^{-6}$, or $\approx10^{-6}$ after the MR-TOF mass identification, when compared to the number of Ce$^{3+}$ ions injected into the RFQcb.

Ongoing upgrades to our method focus on the gas-handling system of the RFQcb for better control of the gas flow into the cooler buncher. This will allow reproducible and precise adjustment of the mixing ratio between the SF$_6$ reactant gas and the He buffer gas, facilitating systematic optimization. It will also enable an environment for molecular formation which remains stable over an entire experimental campaign, in contrast to the repeated refilling of the mixing reservoir required in the present configuration. Other experimental parameters to be optimized are the energy of the stored Ce$^{3+}$ ions, the reaction and ion storage time in the RFQcb, or minimizing gas impurities in the RFQcb system which presently lead to additional undesired chemical reactions with the incoming Ce$^{3+}$ ion beam, all reducing the CeF$^{2+}$ yield. Once implemented, all of these upgrades as well as the full optimization of all experimental parameters promise to significantly enhance the efficiency in creating (multiply charged) molecules of interest and thus provide a realistic prospect to expand the developed method for the formation of PaF$^{3+}$, even when aiming for the short-lived isotope $^{229}$Pa.

\subsection{Prospects of Formation of Highly Charged PaF at TRIUMF}\label{ssec:PaFprospects}

\begin{figure*}[htb]
    \centering
    \includegraphics[width=0.7\linewidth]{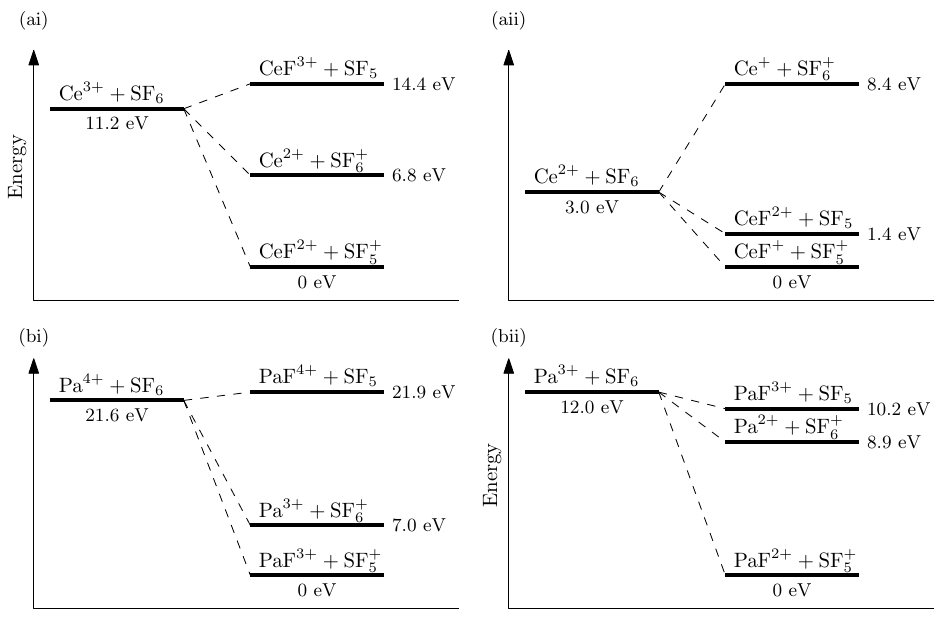} \\
    \includegraphics[width=0.6\linewidth]{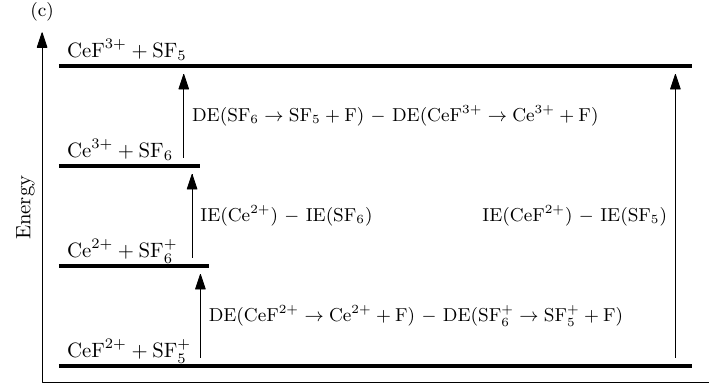}
    \caption{\justifying Energy level diagrams (not to scale) for reactions of various charges states of cerium [row (a)] and protactinium [row (b)] with neutral sulfur hexafluoride. We only include a selected set of products of interest in the diagrams. Activation barriers and transition states are not included in the diagrams. Energy assignments are performed via minimization of the least squares residuals (residuals correspond to 0.3~eV strain on the energy levels) of a collation of ionization (IE) and dissociation energies (DE) from literature and this work (see Table~\ref{tab:dissociations} and references therein as well as Figure}~\ref{fig13}) where applicable. As an example, panel (c) shows the Born-Haber diagram used to evaluate the energy levels in panel (ai). We note that SF$_6^+$ is unstable and dissociates with $-0.99~\mathrm{eV}$ into SF$_5^+ +$ F. See main text for implications on the formation of PaF$^{3+}$.
    \label{fig:reactionenergy}
\end{figure*}

The production of multiply charged molecules in the RFQcb through the use of a He-SF$_6$ mixture not only allows for the identification of the previously discussed CeF$^{2+}$, but also provides insight into the viability of transferring this method to highly sought-after PaF$^{3+}$ ions. Combining our experimental observations with a thermochemistry picture, the results of which are shown in Figure~\ref{fig:reactionenergy}, allows for a heuristic prediction of the reaction products when injecting protactinium ions into the same molecular-formation environment. 

Our observation of CeF$^{2+}$ with the injection of Ce$^{3+}$ (Figure~\ref{fig:CeF_2+}) is consistent with the overall exothermic reaction as shown in panel (ai) in Figure~\ref{fig:reactionenergy}. The lack of CeF$^{3+}$ formation with the same reactants is also supported by the fact that the reaction is endothermic [panel (ai) in Figure~\ref{fig:reactionenergy}]. Moreover, we note that CeF$^{3+}$ is unstable with respect to Coulomb explosion into Ce$^{2+}$ and F$^+$, which could further reduce the chance of detection of CeF$^{3+}$. An in-depth discussion of the meta-stability of CeF$^{3+}$ follows in Section~\ref{sec:electronic structure}.

Charge exchange of the ionic reactants with the neutral reactants and buffer gas in the system could limit the range of usable charge states of the ions. Neglecting meta-stability of the highly charged molecular products for now, row (a) in Figure~\ref{fig:reactionenergy} suggests a trend where one could obtain CeF$^{n+}$ through a reaction with SF$_6$ and Ce$^{(n+1)+}$. One might be tempted to use Ce$^{4+}$ as a reactant to form CeF$^{3+}$. However, the idea of starting with Ce$^{4+}$ as a reactant is dismissed as the charge exchange between Ce$^{4+}$ (IE(Ce$^{3+}) \approx 37~\mathrm{eV}$ \cite{cao2003theoretical}) and the helium buffer gas is highly exothermic; thus, in practice, the Ce$^{4+}$ ions would likely turn into Ce$^{3+}$ before they have a chance to react with SF$_6$ to form CeF$^{3+}$.

In a next step, we extend our heuristic thermochemistry model to the case of protactinium and investigate prospects of forming PaF$^{3+}$ with methods similar to those used for cerium. First, panel (bii) in Figure~\ref{fig:reactionenergy} indicates that the formation of PaF$^{3+}$ from incoming Pa$^{3+}$ is expected to be exothermic. However, the channel that yields PaF$^{2+}$ and SF$_5^+$ is even more strongly favored by exothermicity. Additional studies are thus required to determine the dominant reaction pathways from $\mathrm{Pa}^{3+} + \mathrm{SF}_6$ and their respective cross sections to determine if one obtained sufficient yields of $\mathrm{PaF}^{3+}$. This set of reactants would then represent a possible means to form PaF$^{3+}$, purely from the consideration of the relative energies of the reactants and products.

Second, panel (bi) of Figure~\ref{fig:reactionenergy} suggests that one can obtain PaF$^{3+}$ molecules with incoming Pa$^{4+}$ ions. However, similar to the case of Ce$^{4+}$, the charge exchange of Pa$^{4+}$ (IE(Pa$^{3+}) \approx 31~\mathrm{eV}$ \cite{cao2003theoretical}) with helium is highly exothermic. One could consider an alternative experimental scheme in which Pa$^{4+}$ ions survive long enough such that the chemical reaction to form PaF$^{3+}$ can occur. For these reasons, ongoing work at TRIUMF focuses on the development of a dedicated gas exchange cell. Trapping Pa$^{4+}$ in a chamber containing only neutral SF$_6$ will facilitate the production of the desired PaF$^{3+}$ before sending the resulting molecular ions downstream towards the RFQcb for cooling and bunching. One notes that PaF$^{3+}$ can perform charge exchange with SF$_6$ (IE(SF$_6) \approx 15~\mathrm{eV}$ \cite{yencha:1997}). Hence, the mean free paths of Pa$^{4+}$ and the subsequent PaF$^{3+}$ through the gas exchange cell should be designed to be enough for only one collision with SF$_6$. In the RFQcb, the PaF$^{3+}$ ions are not expected to engage in charge exchange with the inert helium buffer gas, and can be effectively cooled and bunched. 

An alternative method to obtain PaF$^{3+}$ could involve electron impact ionization of the desired species after the molecular formation in a lower charge state, e.g., PaF$^+$. In this way the convoluted chemistry in the cooler-buncher is avoided at the expense of possible dissociation of the molecule through the electron impact. Interestingly, the formation of tricationic uranium fluoride (UF$^{3+}$) has been demonstrated through this approach \cite{Schwarz1999, Schroder2004}. We are presently extending the technique to form PaF$^{3+}$ at TRIUMF, for instance, by exploiting TITAN's Electron-Beam Ion Trap (EBIT) \cite{LAPIERRE201054}, which is designed to access higher charge states in ions through electron-impact ionization. 

Performing any of these molecular formation experiments requires access to Pa beams which are notoriously difficult to deliver, especially at ISOL facilities such as ISAC given the refractory nature of Pa. Sources of the long-lived isotope $^{231}$Pa ($T_{1/2}\approx3\cdot10^{4}$ years) are available at selected sites.
An efficient ion source utilizing a $^{231}$Pa sample could be instrumental in demonstrating the proposed experimental techniques for forming PaF$^{3+}$ molecules. Efforts continue to extract shorter lived Pa isotopes from ISAC targets,
although Pa's vanishing mobility in the target material poses a serious challenge. Molecular sidebands are often used to increase mobility. However, the vapor pressure of Pa is lower than that of the ThC$_x$ or UC$_x$ target material, making this approach challenging to implement without volatilizing an abundance of the target material itself. A more promising strategy towards Pa beams at TRIUMF may thus be found in the laboratory's multi-user isotope production facilities using high-energy proton beams, allowing for additional production of rare isotopes. In particular, the irradiation of high $Z$ materials, e.g. thorium, at TRIUMF's BL1A beamline provides via spallation reactions and subsequent chemical separation access to valuable medical radionuclides such as $\alpha$ emitters for targeted $\alpha$ therapy (TAT). This method is exploited, e.g., to attain the actinium isotope $^{225}$Ac \cite{Robertson2018,Radchenko1495}. A similar approach has previously been developed at Los Alamos National Laboratory to obtain and isolate $^{230}$Pa ($T_{1/2}\approx17$ days)\cite{Radchenko2016}. Effort is currently under way to maximize the production of $^{230}$Pa at TRIUMF which could ultimately open a path to $^{229}$Pa. Once a chemically separated Pa sample is available, an ion beam could be formed by utilizing laser-resonance-ionization  inside a hot cavity ion source \cite{PhysRevA.98.022505}, followed by charge state breeding in TITAN's EBIT, ISAC’s charge state booster \cite{Ames2014,Adegun_2024}, or CANREB’s EBIS  \cite{Schultz_2022}. $^{229}$Pa's short half-life implies that all preparations for such a beam formation have to proceed at a timescale of at most a few days after isotope production and that the initial sample size is sufficiently large to accommodate radioactive decay losses, as well as efficiencies in the charge breeding and in the formation of the desired PaF$^{3+}$ molecules. Radiological safety impacts from $\alpha$ emitters in the decay chains  must also be taken into account, as these impact on facility operation, operator safety, and waste management. Although challenging, the requirements are believed to be feasible in principle.

\section{\texorpdfstring{Electronic Structure of C\MakeLowercase{e}F$^{2+}$}{}} \label{sec:electronic structure}

In order to assess the prospects of quantum control of $\ce{CeF^2+}$, we investigate the nine energetically lowest electronic states by $ab \hspace{1mm} initio$ calculations. To this end we predict excitation energies and transition moments. We provide in addition estimates of various molecular parameters relevant for assessing quantum control experiments that aim to reveal signatures from violation of fundamental $\mathcal{P,T}$-symmetry. In section \ref{sec:4}, we also predict the sensitivity of the molecular dication to various sources of $\mathcal{P,T}$-violation to identify possible `science' states, i.e. electronic
states promising for such experiments. Whereas one typically uses only a single science state for a given molecule, one could also envision to take advantage of different sensitivities to new physics that electronic states of different nature can provide in \ce{CeF^2+}. Especially in relation to global analysis, the particular prospects of using molecules with varying sensitivities to  $\mathcal{P,T}$-odd effects has been highlighted recently \cite{Gaul2024}.

Beyond such interest in \ce{CeF^2+} in itself, the ion also serves as a prototype system for subsequent studies in the valence isoelectronic trication \ce{PaF^3+}. Similarities between the two systems are naively expected when considering the electronic ground state configuration of \ce{Ce^2+} and \ce{Pa^3+}, which are [Xe]4f$^2$ and [Rn]5f$^2$, respectively. An increasing molecular charge causes different energy lowerings for s, p, d and f orbitals, respectively. For f orbitals, this energy drop is larger than for orbitals with lower angular momentum. A clear separation of electronic states into a manifold of levels corresponding to the respective orbital angular momentum is observed. In addition, a dense level structure is expected within these blocks for molecular ions. This feature has been studied in some detail for \ce{PaF^3+} \cite{zulch2022cool} and is, thus, also expected in \ce{CeF^2+}. Related concepts were also applied successfully to \ce{UF^3+} \cite{zulch:2023}. 

These general expectations are confirmed by our explicit \textit{ab initio} calculations performed in the present work for \ce{CeF^2+} on the level of unrestricted coupled-cluster calculations including single and double excitations iteratively and triple excitations perturbatively [UCCSD(T)], as well as on the level of two-component complex generalized Hartree--Fock (2c-ZORA-cGHF). Details on the methodology used for these calculations are provided in the appendix~\ref{sec:comps}.

\begin{figure}[!htb]
    \centering
    \includegraphics[width=1\columnwidth]{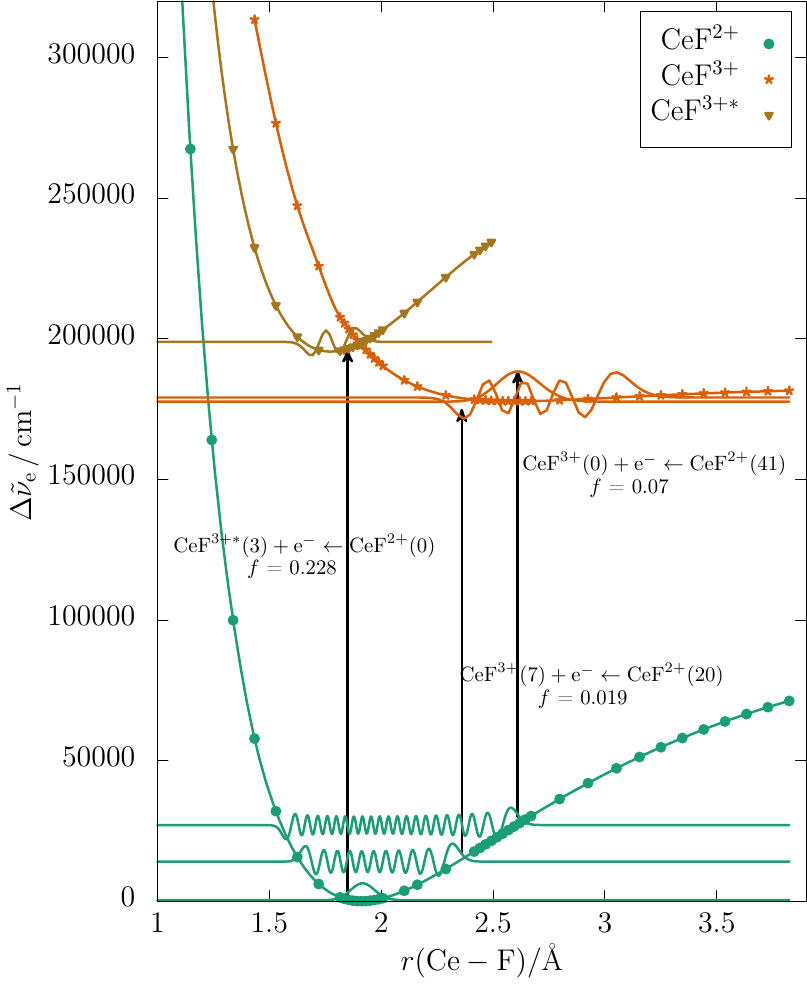}
    \caption{\justifying Potential energy surface of the electronic ground states of
    CeF$^{2+}$ and CeF$^{3+}$ together with an electronically excited state of
    the trication (CeF$^{3+*}$) computed on the level of 2c-ZORA-cGHF. Lines
    are shown to guide the eye. The wavefunctions of several vibrational states
    are shown with scaled intensity for visualization indicating possible
    pathways for quantum control. Exemplary vibronic transitions are shown with
    vibrational levels in brackets and the Franck--Condon factor $f$.}
    \label{fig:cef3+2+_pots}
\end{figure}

The electronic ground state potential of \ce{CeF^2+} forms a deep well (see Figure~\ref{fig:cef3+2+_pots}), suggesting a strong bond in the gas phase and protection against Coulomb explosion. This stability with respect to Coulomb explosion is ordinarily expected for doubly charged diatomic fluorides \ce{MF^{2+}} when the ionization energy of F exceeds the one of its cationic bonding partner \cite{bates1955,Schwarz1999}. This is the case for \ce{Ce^+} and, thus, the ground state potential of \ce{CeF^2+} connects adiabatically with the energetically lowest dissociation channel \ce{Ce^2+ + F}, with associated dissociation energy of \SI{6.5}{eV} $\approx$ \SI{50000}{cm^{-1}} as predicted from UCCSD(T) computations, suggesting a slightly stronger covalent bond when compared to \ce{PaF^3+}.

The equilibrium bond length, with $\SI{3.60}{\bohr}=\SI{1.91}{\angstrom}$, is predicted from UCCSD(T) to be shorter than the bond length in isoelectronic BaF (\SI{2.159}{\angstrom}, \cite{ryzlewicz1980}) and its lighter homologues SrF (\SI{2.076}{\angstrom}, \cite{barrow1967}) and CaF (\SI{1.967}{\angstrom}, \cite{field1975}), but slightly longer than predicted for \ce{PaF^3+} \cite{zulch2022cool}. Bond lengths from 2c-ZORA-cGHF calculations for \ce{CeF^2+} deviate by less than $1\%$ from those predicted on the coupled cluster level.

The corresponding trication \ce{CeF^3+} exhibits a rather shallow potential diabatically connected to the Ce$^{3+}+$F dissociation channel with a dissociation energy of $\SI{1.7}{eV}$, whilst the diabatic potential related to the asymptotic Ce$^{2+}+$F$^+$ dissociation channel crosses and lies with $\SI{-0.3}{eV}$ slightly below the energy of \ce{CeF^3+} in its equilibrium arrangement. For the valence isoelectronic \ce{PaF^4+}, the situation is qualitatively similar with dissociation to neutral F being endothermic by $\SI{3.7}{eV}$ \cite{zulch2022cool}, whereas the Coulomb explosion pathway to afford \ce{Pa^3+} and \ce{F^+} is found to be stronger exothermic with $\SI{-9.7}{eV}$. This implies that both \ce{CeF^3+} and \ce{PaF^4+} are metastable, with lower stability being expected for the latter. The bond length in \ce{CeF^3+} optimized on the level of 2c-ZORA-cGHF is predicted at $\SI{4.90}{\bohr}=\SI{2.59}{\angstrom}$ and, thus, far from the equilibrium bond length in \ce{CeF^2+}. An excited state CeF$^{3+*}$ at approximately $\SI{24.3}{eV}\approx\SI{196000}{cm^{-1}}$ ($\SI{18000}{cm^{-1}}$ above \ce{CeF^3+} ground state) was found on the level of 2c-ZORA-cGHF with a deep potential well resembling more the ground state of \ce{PaF^4+}, as it has a similar equilibrium bond length as the lower charge state (equilibrium bond length of $\SI{3.43}{\bohr}=\SI{1.82}{\angstrom}$). 

The adiabatic ionization energy for \ce{CeF^2+} with UCCSD(T) is estimated at $\SI{24.1}{eV} \approx \SI{195000}{cm^{-1}}$ and with cGHF $\SI{22.0}{eV} \approx \SI{177000}{cm^{-1}}$, which is roughly $\SI{2}{eV}$ to $\SI{4}{eV}$ higher than the ionization energy for \ce{Ce^2+} that lies at \SI{20.2}{eV} \cite{sugar1973}.

Efficiency aspects of a possible ionization process from \ce{CeF^2+} to \ce{CeF^3+} can be estimated by the highest overlap of the vibrational wavefunctions in \ce{CeF^2+} with a vibrational wavefunction in \ce{CeF^3+}, with the norm squared of these overlap integrals being the Franck--Condon (FC) factors. The highest FC factor was found for the transition $\nu_\mathrm{\ce{CeF^3+}}(0) \leftarrow \nu_\mathrm{\ce{CeF^2+}}(41)$ with $0.07$, whereas starting from $\nu_\mathrm{\ce{CeF^2+}}(0)$ all FC factors to bound vibrational levels of the trication are practically zero. This would  correspond to a low probability of an ionization process without subsequent dissociation. Ionization to the energetically higher CeF$^{3+*}$ state on the other hand has large FC factors due to the similar bond lengths. For the transition from the ground state in \ce{CeF^2+} to the third vibrational state in CeF$^{3+*}$ a FC factor of $0.228$ is predicted. Thus, in this simple picture a lower propensity to direct dissociation in the ionization process would be expected.

\begin{figure}[!htb]
  \centering
  \includegraphics[width=1\columnwidth]{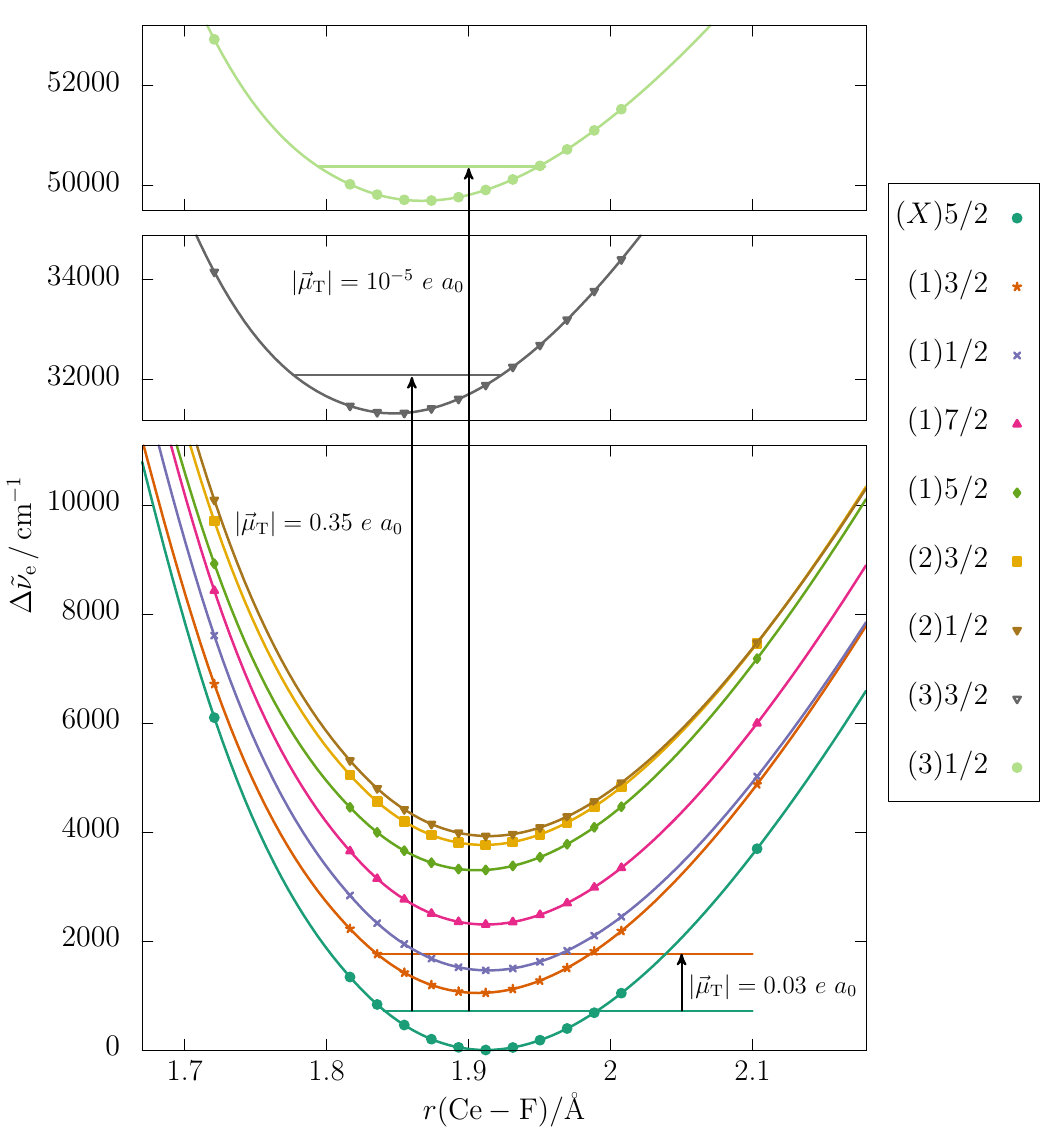}
  \caption{\justifying Born--Oppenheimer potential energy curves of the nine energetically
  lowest electronic states of CeF$^{2+}$ computed on the level of
  2c-ZORA-cGHF. The lines are there to guide the eye. The transition dipole
  moments for the vibronic excitations $\vec{\mu}_\mathrm{T}$ are
  calculated from the electric transition dipole moments $\vec{\mu}$ at the ground state equilibrium structure as
  ${f^{(i)}}^{1/2}\times \vec{\mu}$ with the Franck--Condon factor $f^{(i)}$ of the
  specific vibronic transition which here for simplicity is always set to
  $f^{(0)}$.}
  \label{fig:pes_cef}
\end{figure}

Focusing on \ce{CeF^2+}, the Born--Oppenheimer potential energy curves of the
nine energetically lowest electronic states are visually compared in
Figure~\ref{fig:pes_cef}.

The seven energetically lowest electronic states obtained in these calculations form a set of quasi-parallel potentials exhibiting a well-ordered compressed structure all within $\approx\SI{4000}{cm^{-1}}$. The energetically closest states to these seven states lie at $\SI{32000}{cm^{-1}}$ and \SI{50000}{cm^{-1}} with varying equilibrium bond lengths. The manifold of quasi-parallel states is connected with low electric transition dipole moments, as e.g. the transition from the ground state to the electronically first excited state remains at $0.03~{e}~\si{\bohr}$. In contrast, the transition dipole moment from the ground state to the electronically excited state at around $\SI{32000}{cm^{-1}}$ is predicted at $0.16~{e}~\si{\bohr}$, whereas the transition to the highest here computed electronically excited state is electric dipole forbidden in the Franck--Condon approximation and, thus, has a diminished transition dipole moment. When compared to \ce{PaF^3+} \cite{zulch2022cool}, the qualitative picture found for \ce{CeF^2+} herein is in good accordance and the sequence of electronic states identical, implying the before indicated similar properties. A detailed description of the electronic states is given in Table~\ref{tab:elec_states} reflecting upon this resemblance.

\begin{table*}[!htb]
  \caption{\justifying Characterization of the nine energetically lowest electronic states
  of CeF$^{2+}$. Equilibrium bond length $r_\mathrm{e}$, rotational constant $B_\mathrm{e}$, harmonic
  vibrational wavenumber $\tilde{\omega}_\mathrm{e}$ and vertical excitation
  wavenumber $\Delta\tilde{\nu}_\mathrm{e}$ are shown at the level of
  2c-ZORA-cGHF together with the projection of the electronic orbital angular
  momentum on the molecular axis $\Lambda$ and the electric
  dipole moment $\left| \vec{\mu}_\mathrm{e} \right|$ in the intrinsic frame of the molecule. $M$ states the results
  of a Mulliken analysis on the singly occupied molecular orbital indicated as
  percentages. The individual states are further characterized by the complex two-component spinors of their singly occupied molecular orbital ($\Psi_{1\alpha}$ and $\Psi_{1\beta}$) with the respective norm shown by the isosurfaces and the corresponding complex phase indicated by the color code (Ce is located at the bottom (yellow) and F at the top (green)). The isotopes $^{140}$Ce and $^{19}$F are considered.}
  \label{tab:elec_states}
  \resizebox{\textwidth}{!}{%
    \begin{tabular}{
      c
      c
      S[table-format=2.2,round-mode=places,round-precision=1  ]
      S[table-format=2.3,round-mode=figures,round-precision=3 ]
      S[table-format=1.3,round-mode=figures,round-precision=3 ]
      S[table-format=4.0,round-mode=figures,round-precision=3 ]
      S[table-format=6.0,round-mode=figures,round-precision=3 ]
      S[table-format=-1.2,round-mode=figures,round-precision=3]
      c
      c
      c
      }
      \toprule
      {State} &
      {Term Symbol} &
      {$\Lambda$} &
      {$r_\mathrm{e} /\si{\angstrom}$} &
      {$B_\mathrm{e} /\si{cm^{-1}}$} &
      {$\tilde{\omega}_\mathrm{e} /\si{\per\centi\metre}$} &
      {$\tilde{T}_\mathrm{e} /\si{\per\centi\metre}$} &
      {$\left| \vec{\mu}_\mathrm{e}\right| / (e~\si{\bohr})$} &
      {$M \times 100$} &
      {$\Psi_{1\alpha}$} &
      {$\Psi_{1\beta }$}
      \\
      \midrule
      \multirow{1}{*}{$(X)5/2$}  & $^2\Phi  _{5/2}$ & 2.939 & 1.9120 & 0.27564 & 722 & {\textemdash} & 1.49386 & 0.99f          & \raisebox{-.5\height}{\includegraphics[trim={4.0cm 2.3cm 2.2cm 3.3cm},clip,height=1.4cm,angle=90]{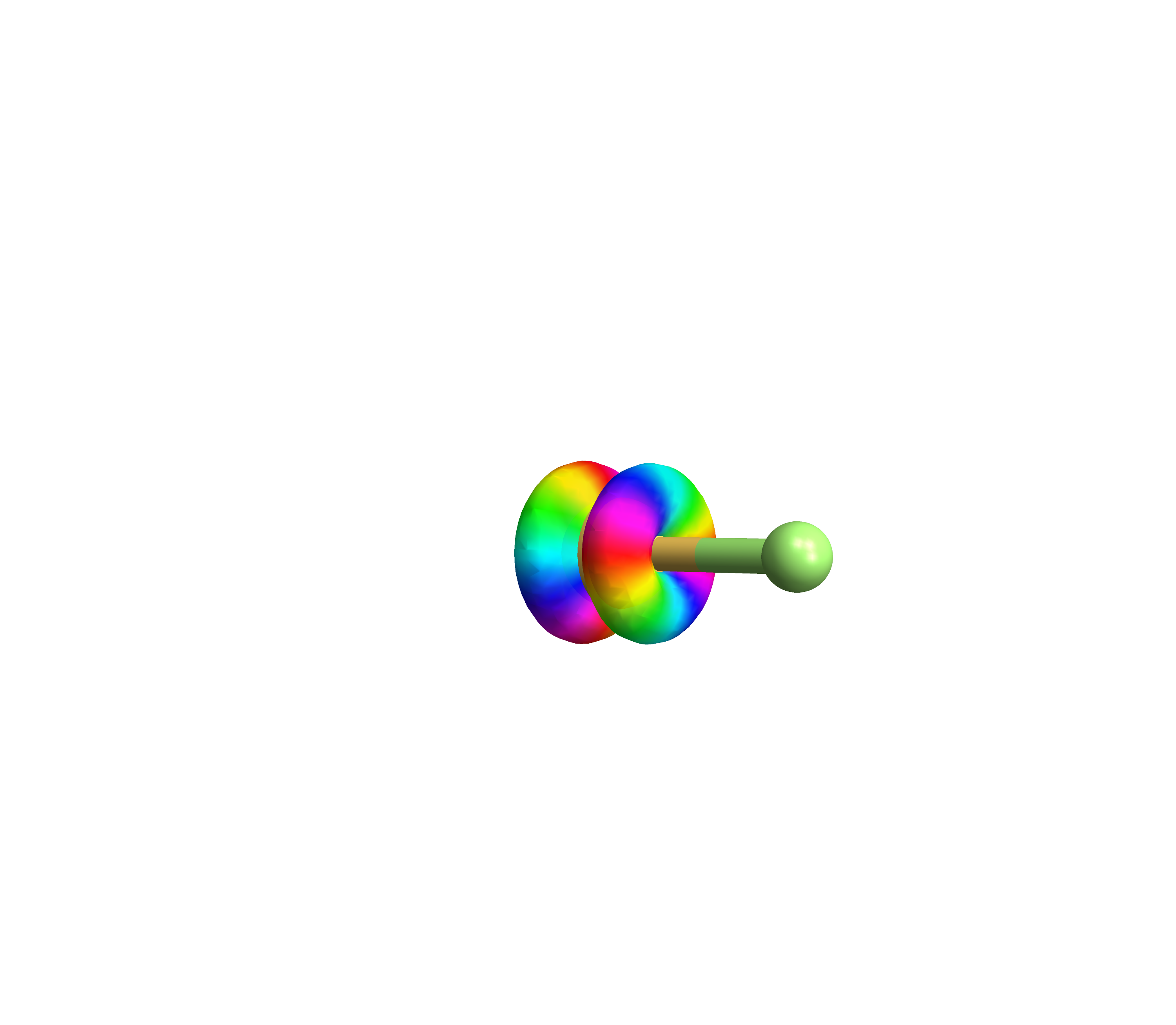}} & \raisebox{-.5\height}{\includegraphics[trim={4.0cm 2.3cm 2.2cm 3.3cm},clip,height=1.4cm,angle=90]{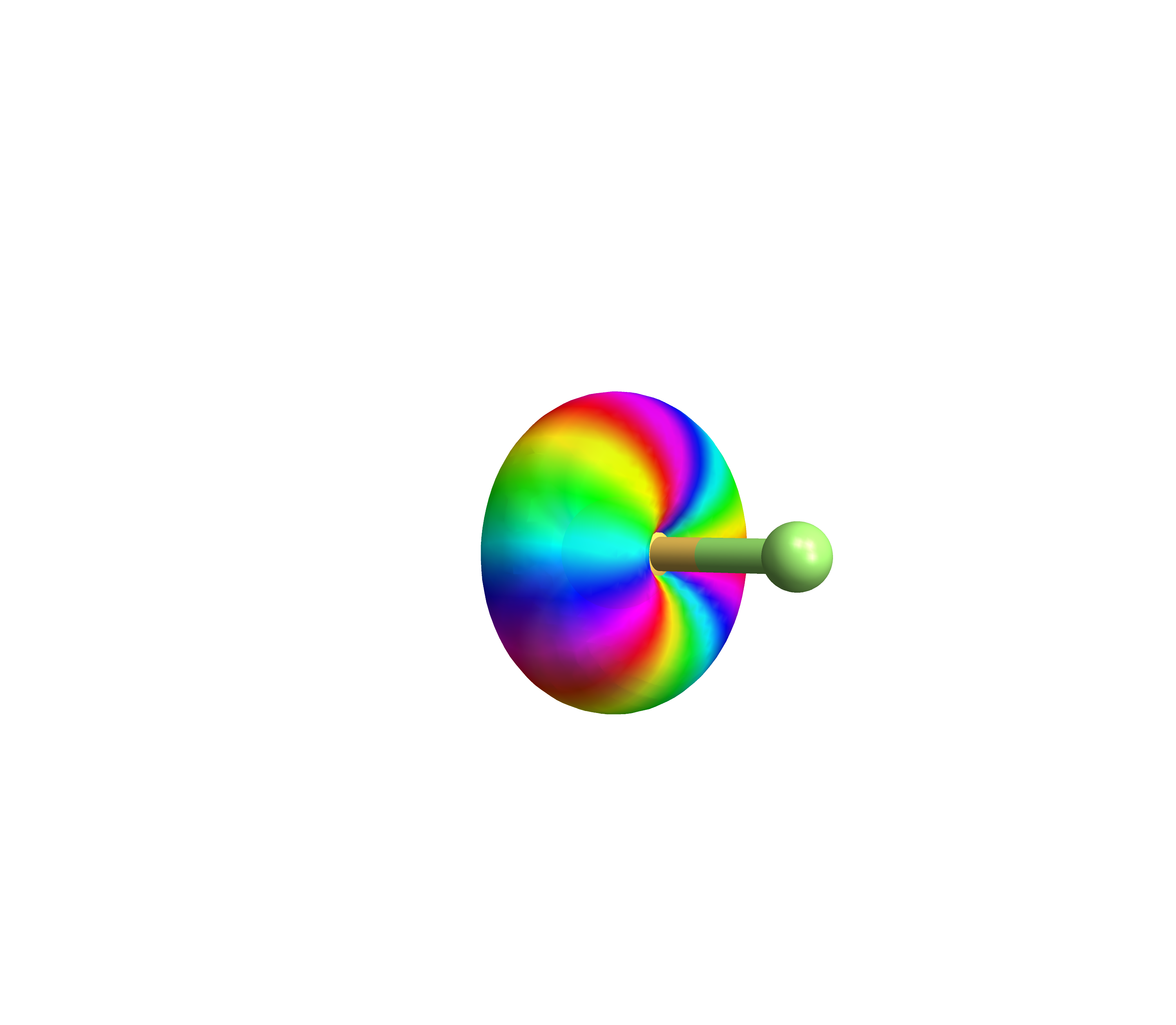}} \\
      \multirow{1}{*}{$(1){3/2}$}& $^2\Delta_{3/2}$ & 1.782 & 1.9068 & 0.27714 & 718 & 1050          & 1.46573 & 32p(F),64f    & \raisebox{-.5\height}{\includegraphics[trim={4.0cm 2.3cm 2.2cm 3.3cm},clip,height=1.4cm,angle=90]{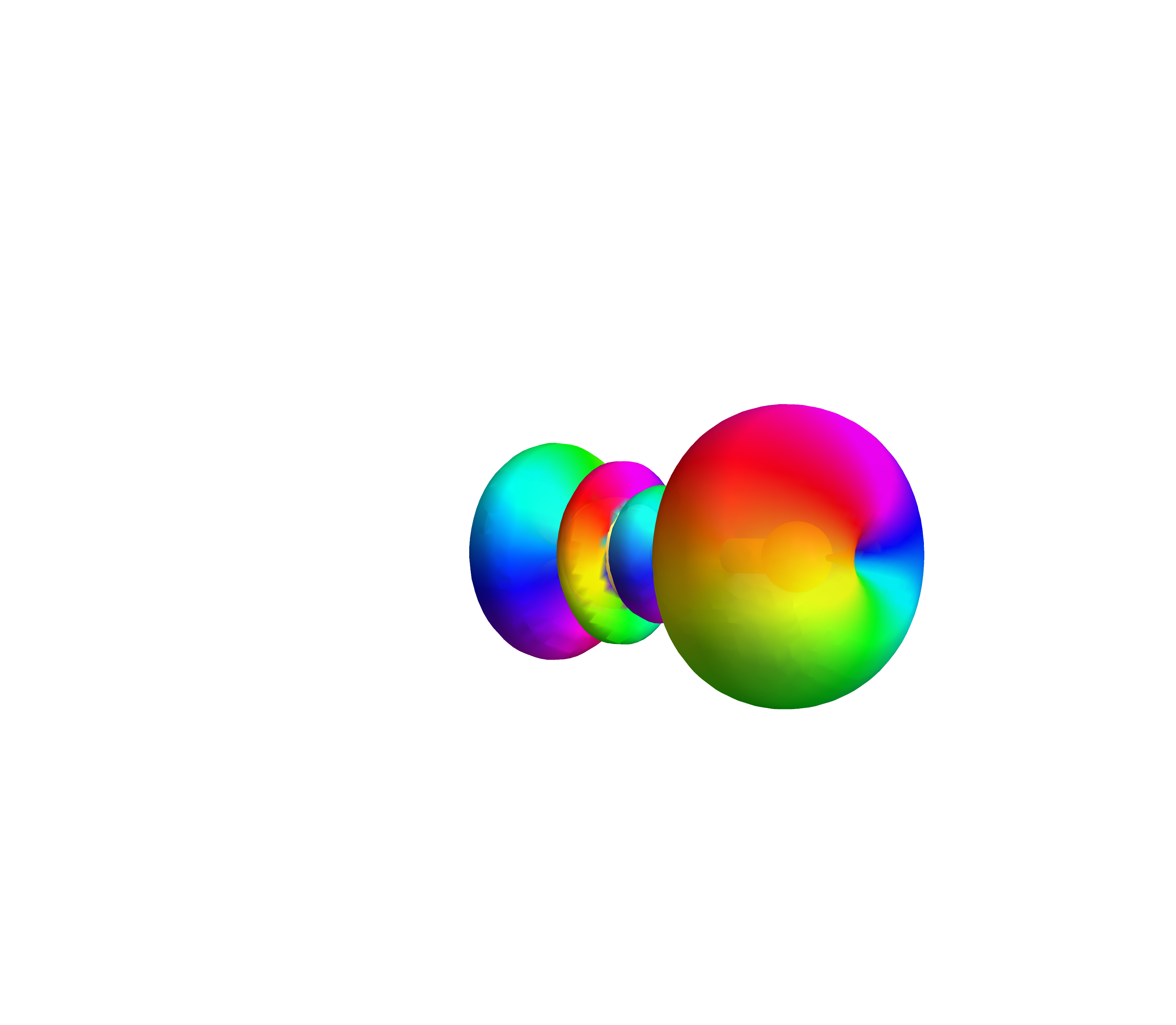}} & \raisebox{-.5\height}{\includegraphics[trim={4.0cm 2.3cm 2.2cm 3.3cm},clip,height=1.4cm,angle=90]{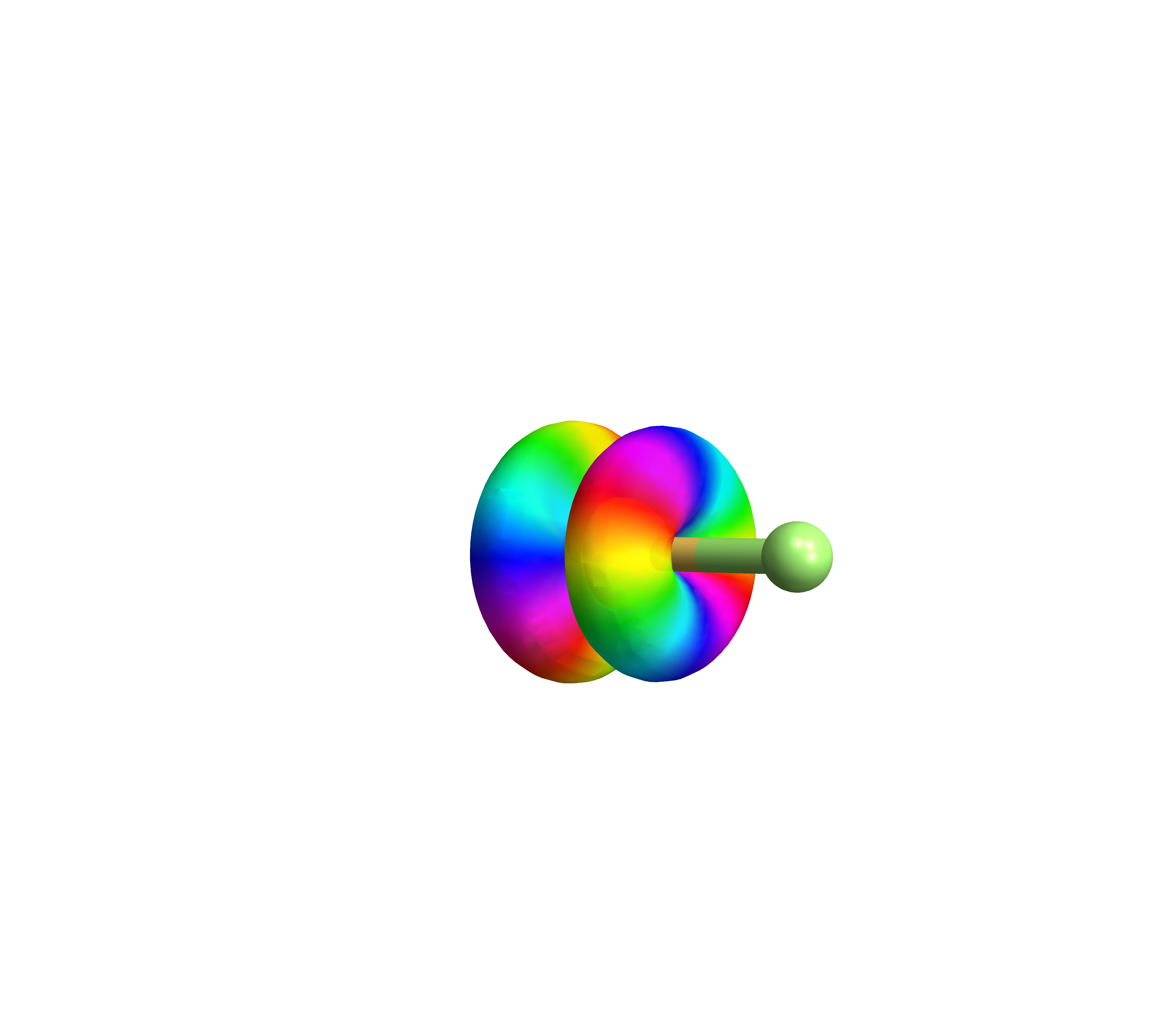}} \\
      \multirow{1}{*}{$(1)1/2$}  & $^2\Pi   _{1/2}$ & 0.587 & 1.9140 & 0.27506 & 716 & 1462          & 1.44524 & 47p(F),11d,37f& \raisebox{-.5\height}{\includegraphics[trim={4.0cm 2.3cm 2.2cm 3.3cm},clip,height=1.4cm,angle=90]{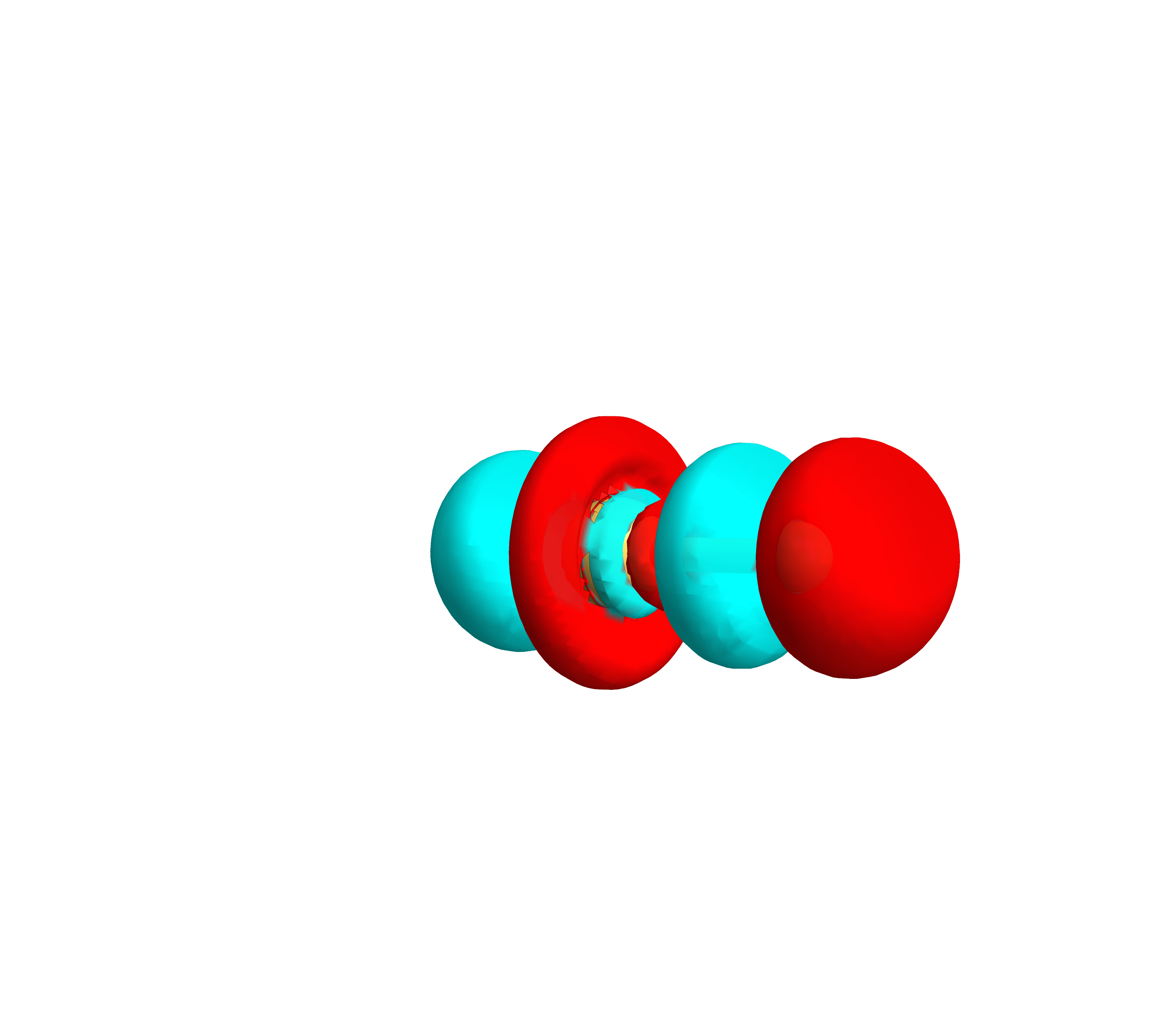}} & \raisebox{-.5\height}{\includegraphics[trim={4.0cm 2.3cm 2.2cm 3.3cm},clip,height=1.4cm,angle=90]{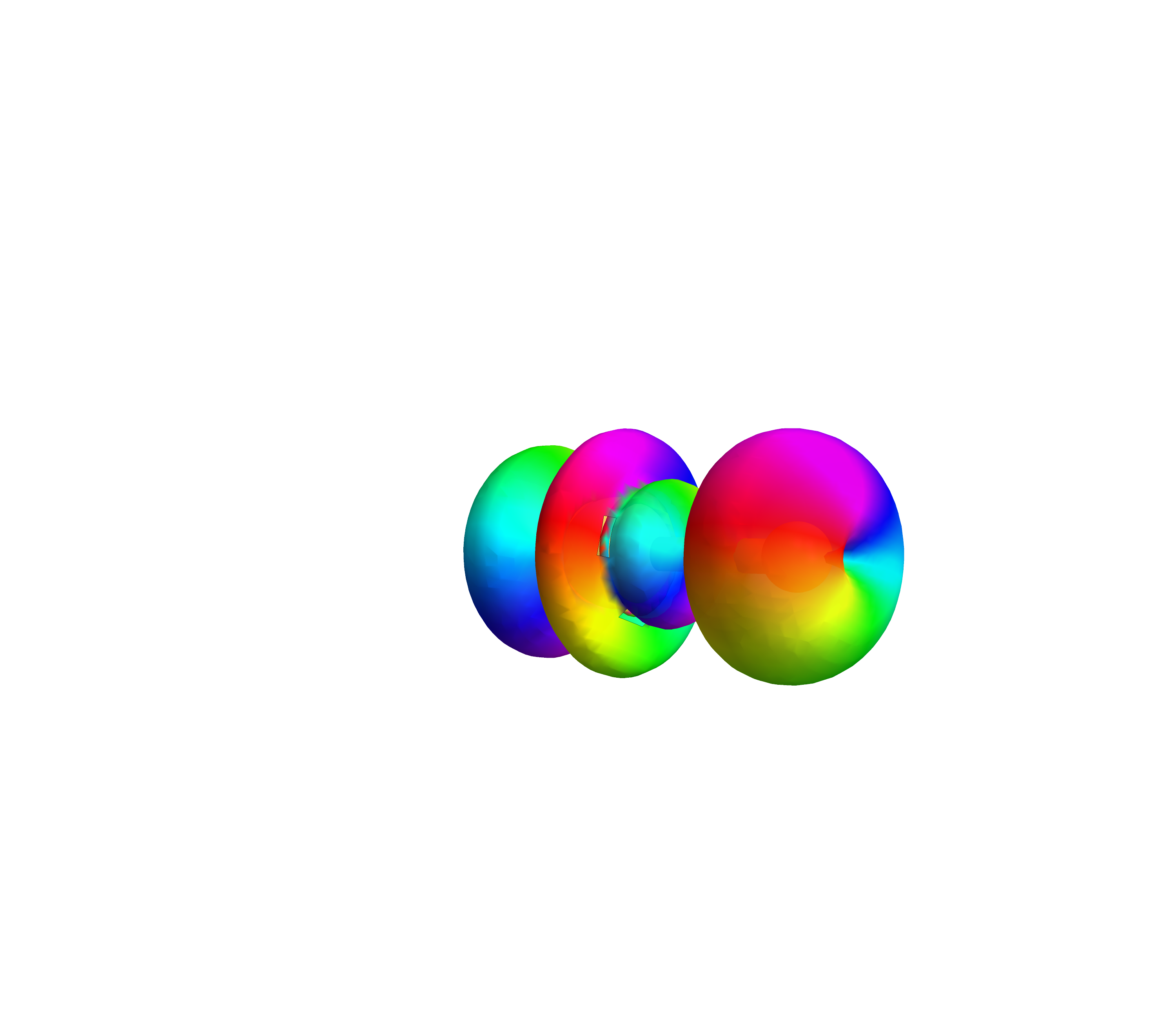}} \\
      \multirow{1}{*}{$(1)7/2$}  & $^2\Phi  _{7/2}$ & 3.000 & 1.9123 & 0.27555 & 723 & 2304          & 1.49529 & 99f           & \raisebox{-.5\height}{\includegraphics[trim={4.0cm 2.3cm 2.2cm 3.3cm},clip,height=1.4cm,angle=90]{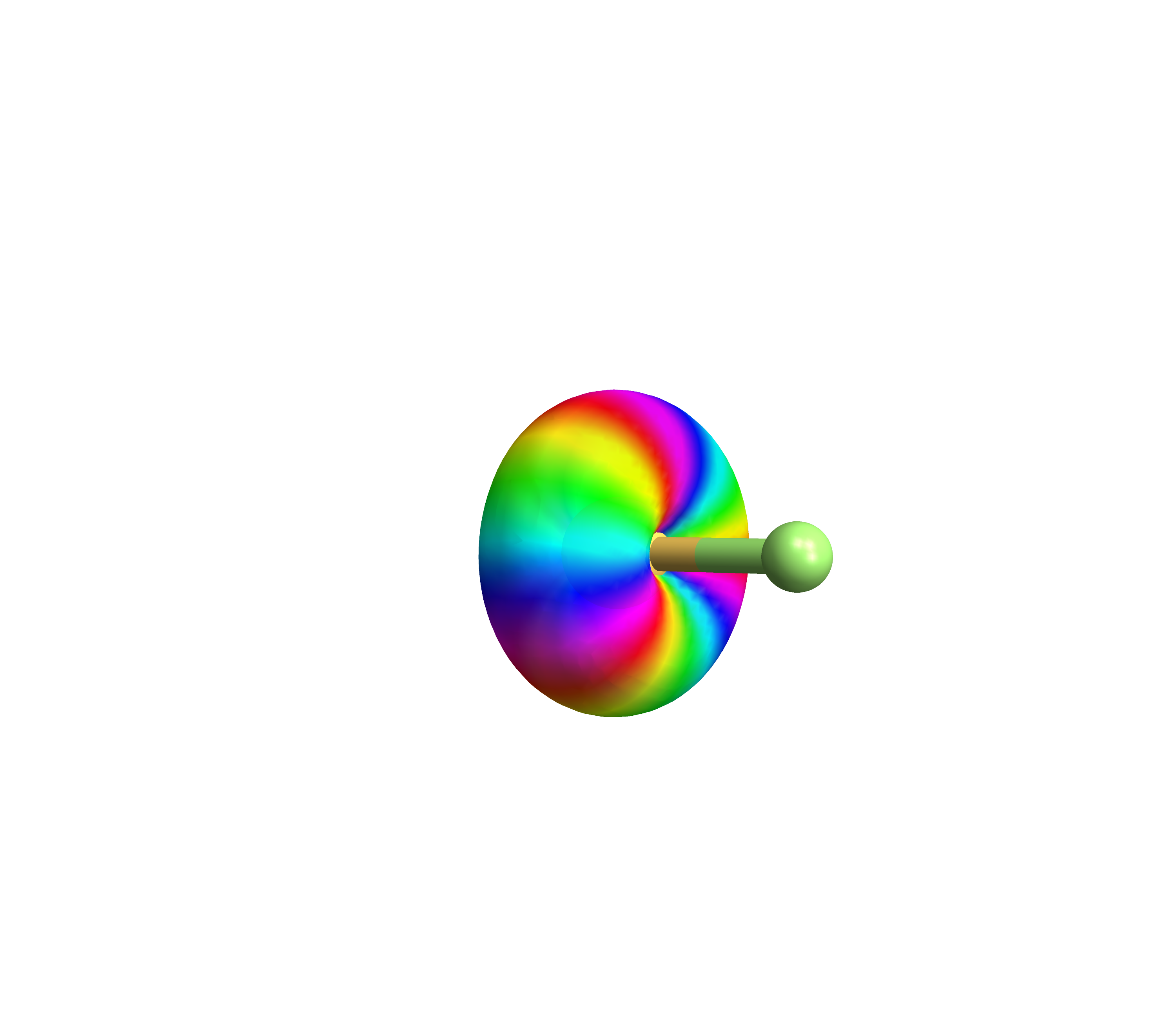}} & \raisebox{-.5\height}{\includegraphics[trim={4.0cm 2.3cm 2.2cm 3.3cm},clip,height=1.4cm,angle=90]{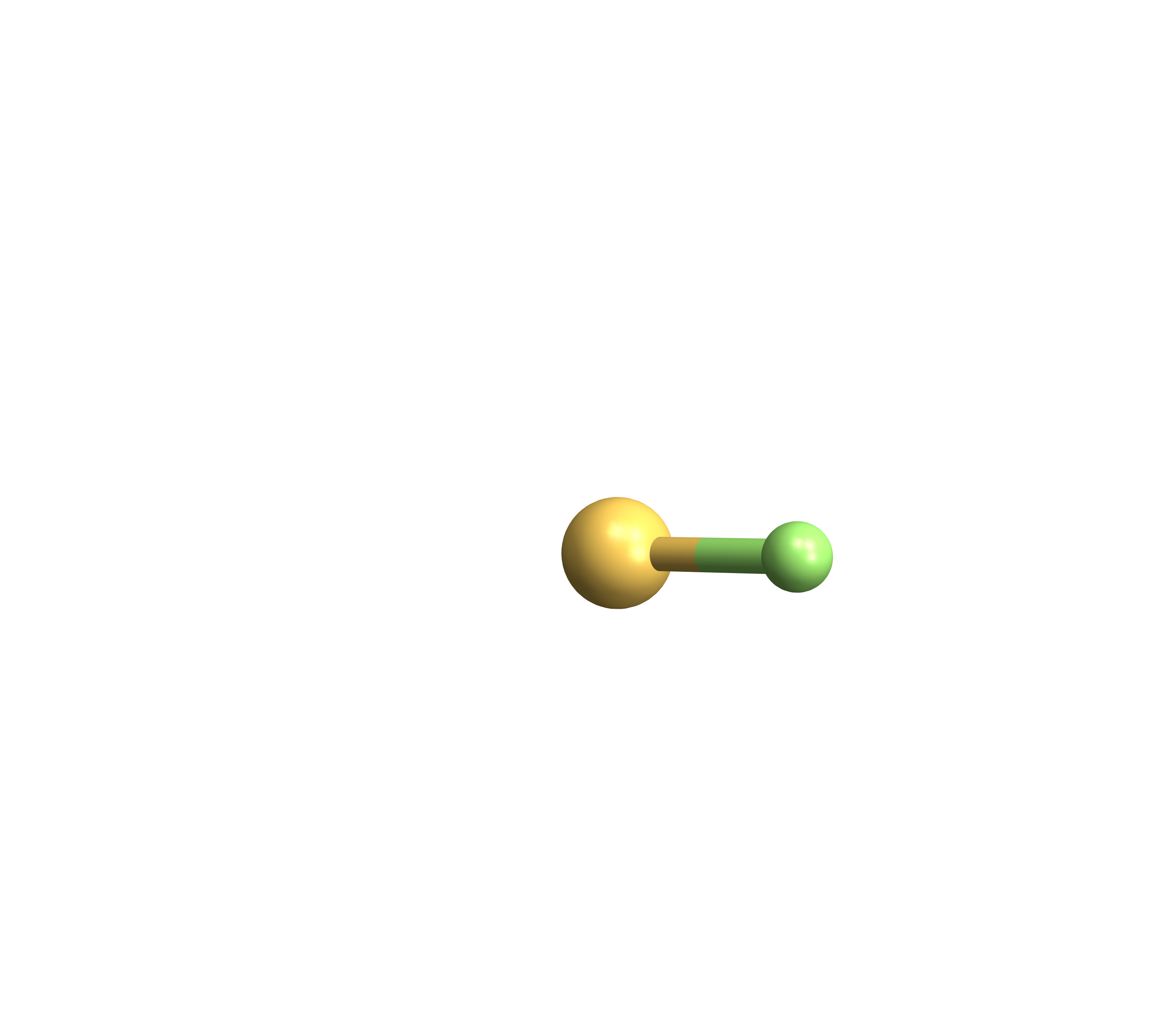}} \\
      \multirow{1}{*}{$(1)5/2$}  & $^2\Delta_{5/2}$ & 2.090 & 1.9058 & 0.27743 & 720 & 3306          & 1.46858 & 100f          & \raisebox{-.5\height}{\includegraphics[trim={4.0cm 2.3cm 2.2cm 3.3cm},clip,height=1.4cm,angle=90]{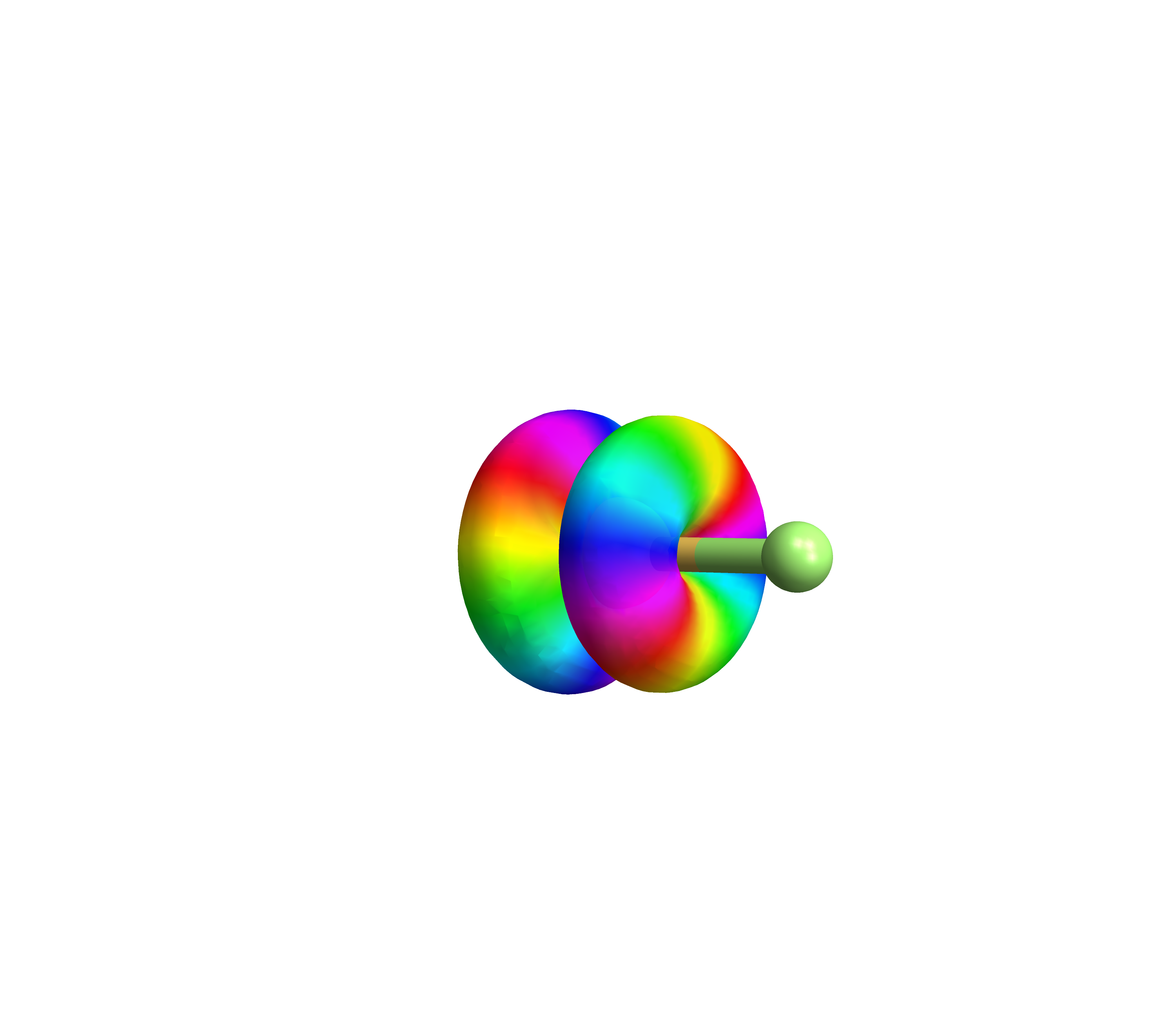}} & \raisebox{-.5\height}{\includegraphics[trim={4.0cm 2.3cm 2.2cm 3.3cm},clip,height=1.4cm,angle=90]{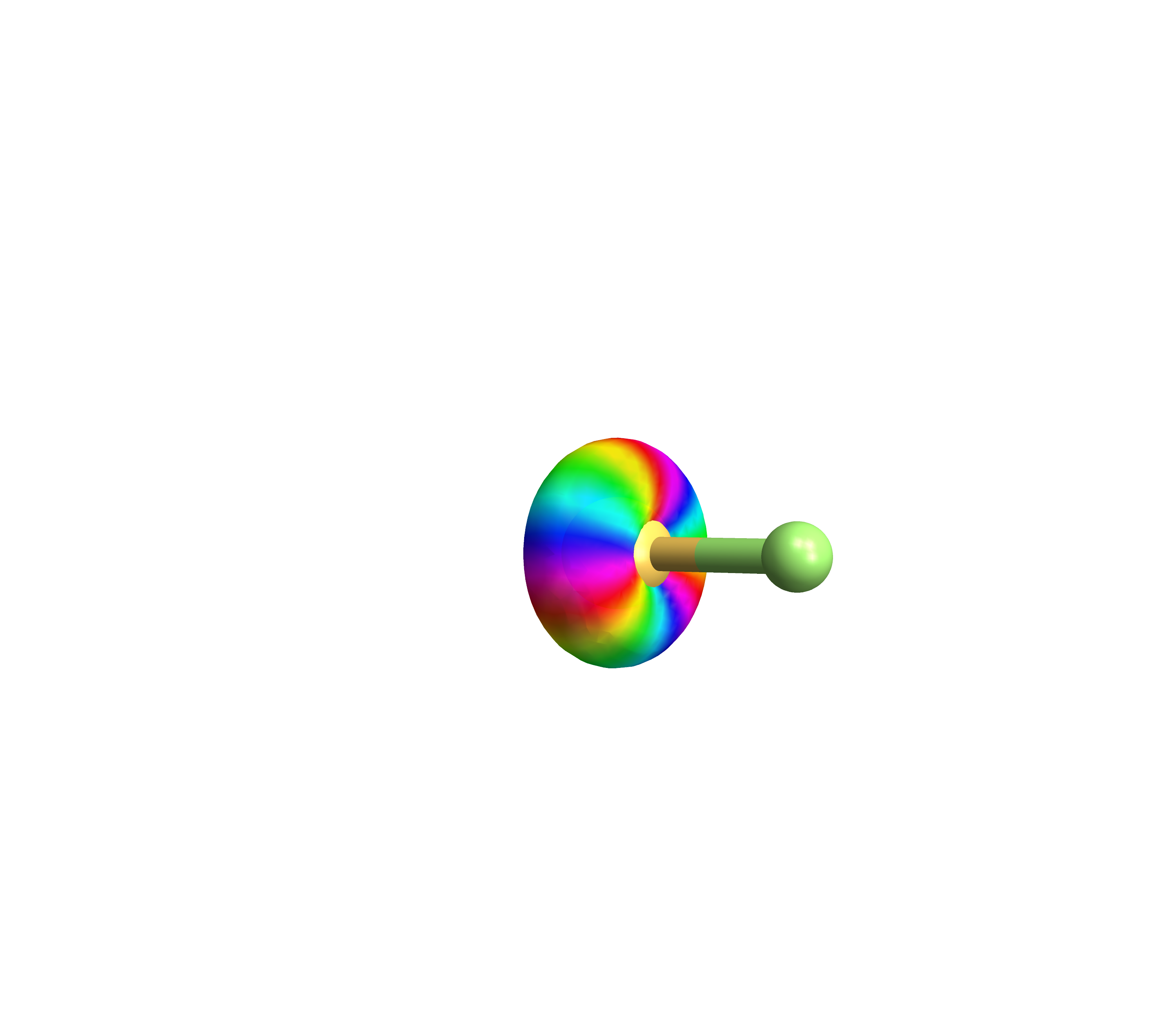}} \\
      \multirow{1}{*}{$(2)3/2$}  & $^2\Pi   _{3/2}$ & 1.203 & 1.9105 & 0.27607 & 718 & 3766          & 1.45758 & 34p(F),61f    & \raisebox{-.5\height}{\includegraphics[trim={4.0cm 2.3cm 2.2cm 3.3cm},clip,height=1.4cm,angle=90]{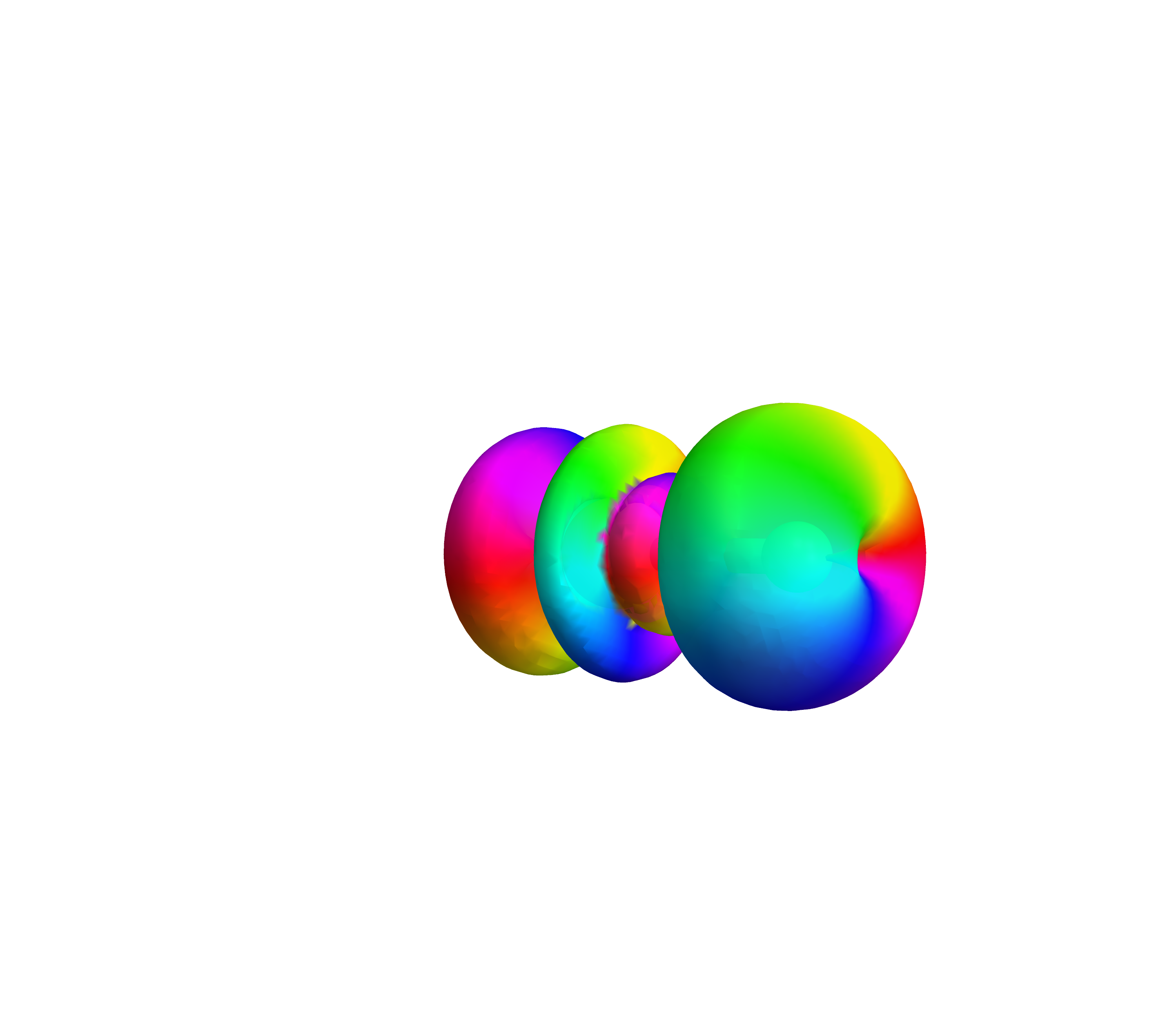}} & \raisebox{-.5\height}{\includegraphics[trim={4.0cm 2.3cm 2.2cm 3.3cm},clip,height=1.4cm,angle=90]{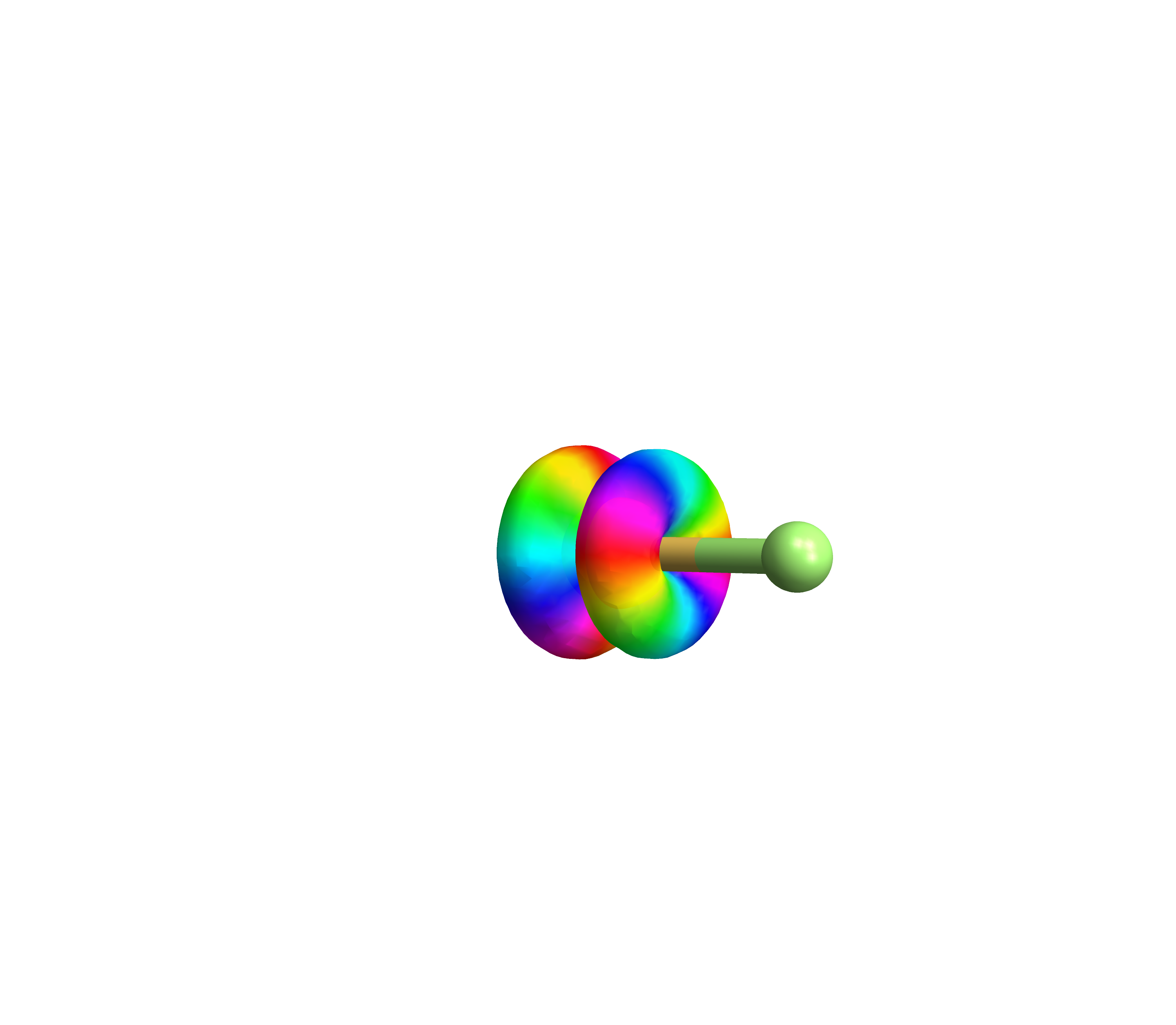}} \\
      \multirow{1}{*}{$(2)1/2$}  & $^2\Sigma_{1/2}$ & 0.400 & 1.9143 & 0.27497 & 716 & 3929          & 1.43199 & 40p(F),12d,44f& \raisebox{-.5\height}{\includegraphics[trim={4.0cm 2.3cm 2.2cm 3.3cm},clip,height=1.4cm,angle=90]{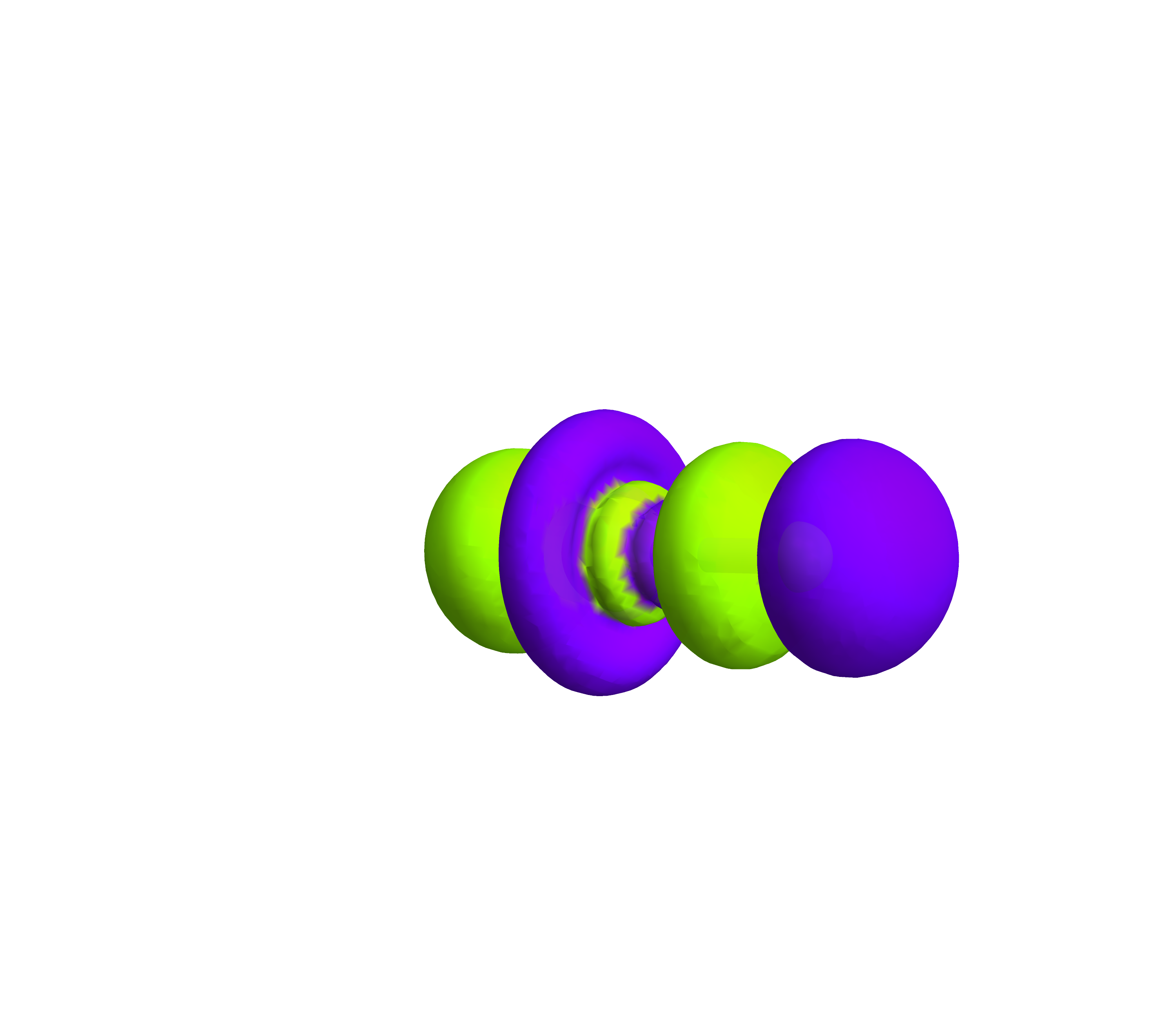}} & \raisebox{-.5\height}{\includegraphics[trim={4.0cm 2.3cm 2.2cm 3.3cm},clip,height=1.4cm,angle=90]{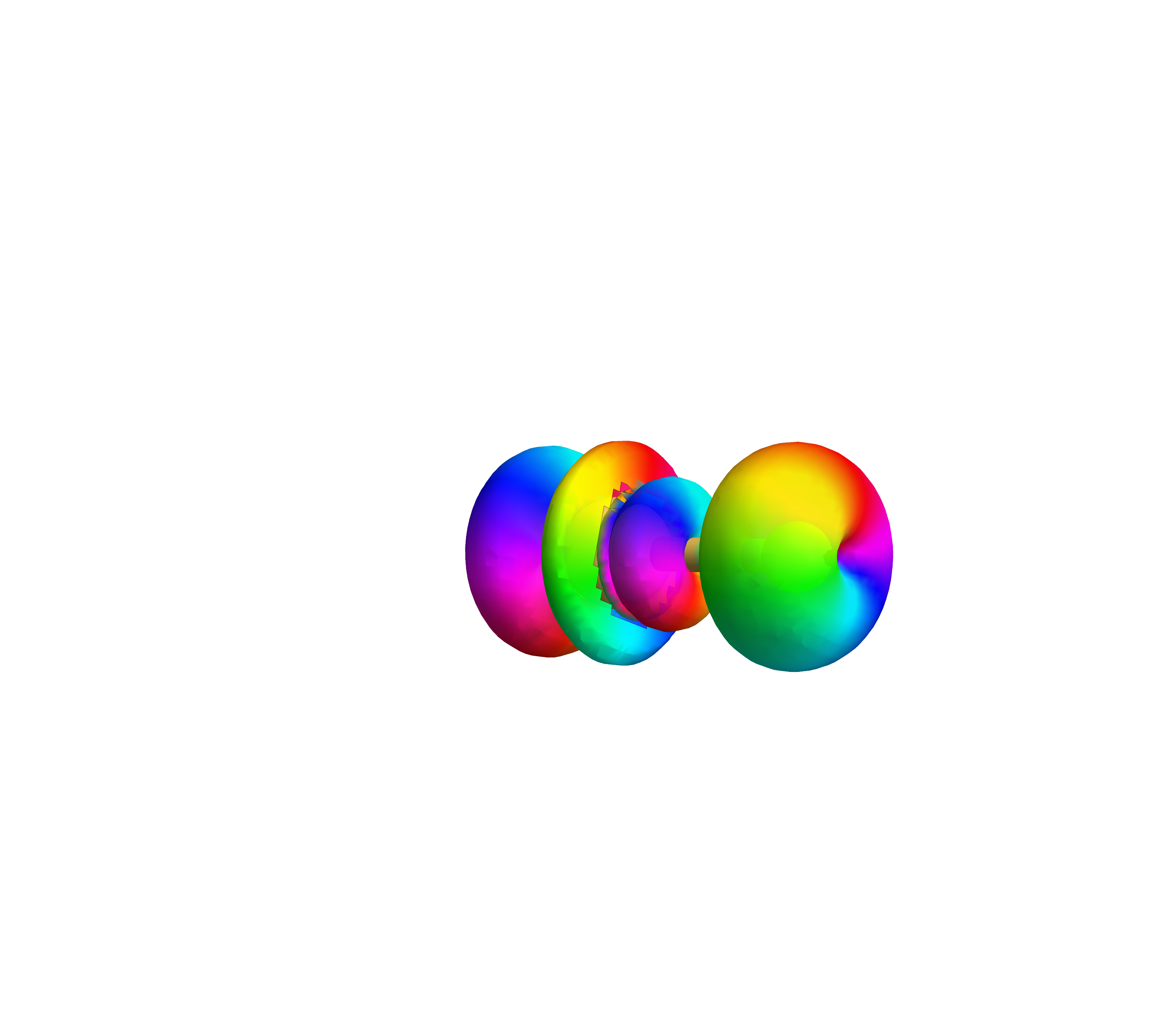}} \\
      \multirow{1}{*}{$(3)3/2$}  & $^2\Delta_{3/2}$ & 1.996 & 1.8475 & 0.29522 & 774 & 31890         & 1.17870 & 99d           & \raisebox{-.5\height}{\includegraphics[trim={4.0cm 2.3cm 2.2cm 3.3cm},clip,height=1.4cm,angle=90]{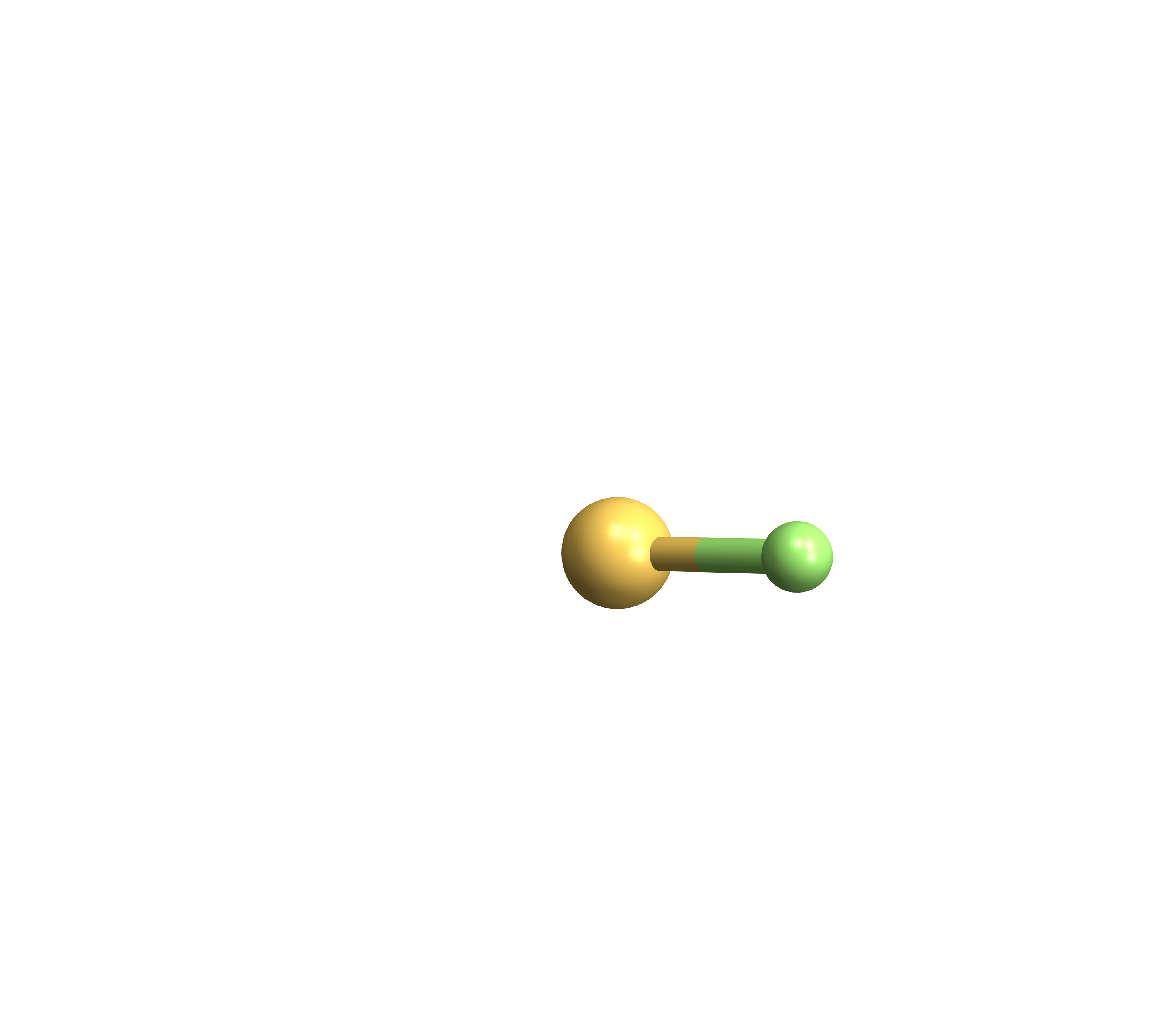}} & \raisebox{-.5\height}{\includegraphics[trim={4.0cm 2.3cm 2.2cm 3.3cm},clip,height=1.4cm,angle=90]{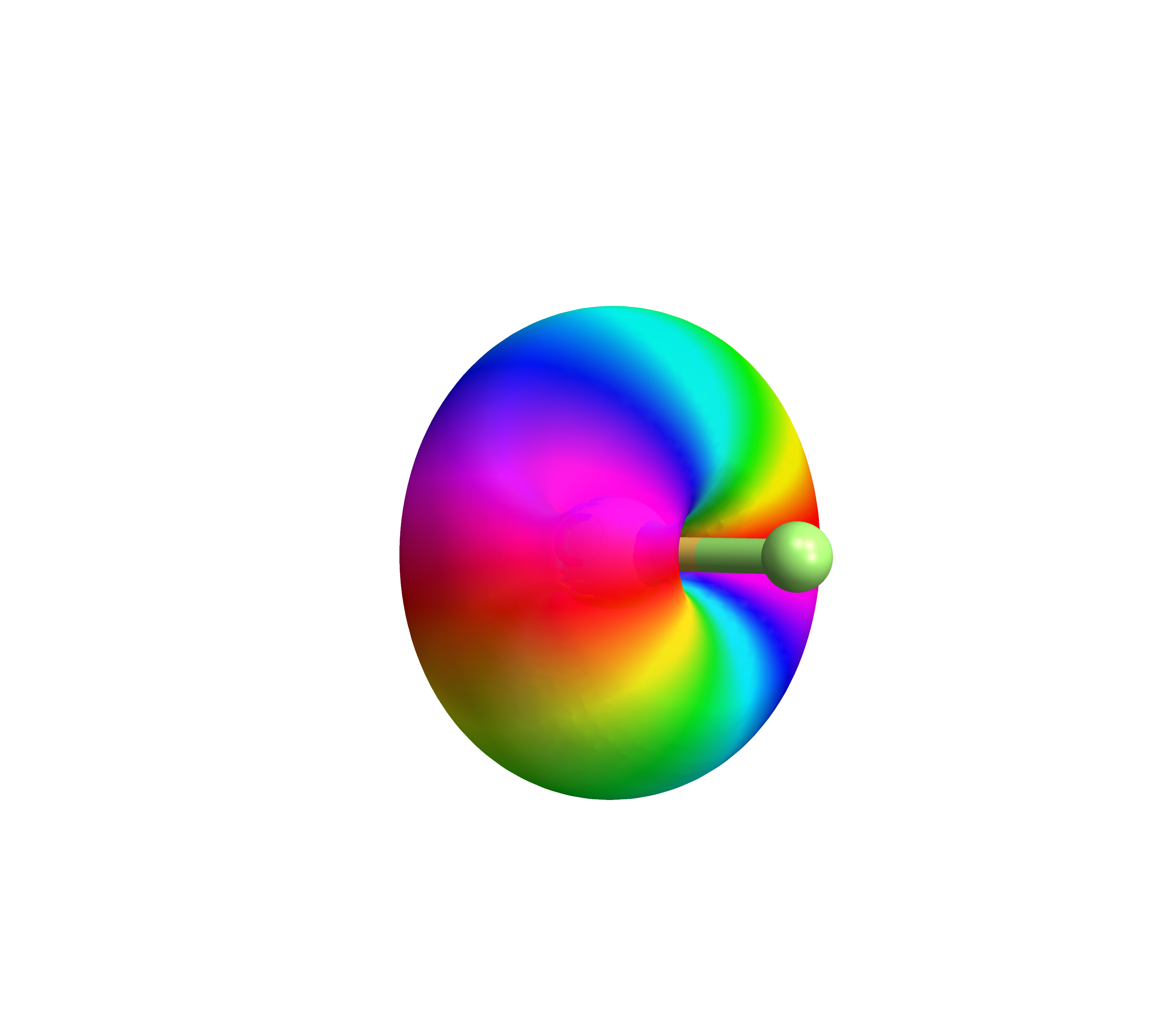}} \\
      \multirow{1}{*}{$(3)1/2$}  & $^2\Sigma_{1/2}$ & 0.006 & 1.8673 & 0.28900 & 687 & 49913         & 1.15844 & 64s,5p(F),28d & \raisebox{-.5\height}{\includegraphics[trim={3.0cm 1.8cm 2.2cm 2.8cm},clip,height=1.4cm,angle=90]{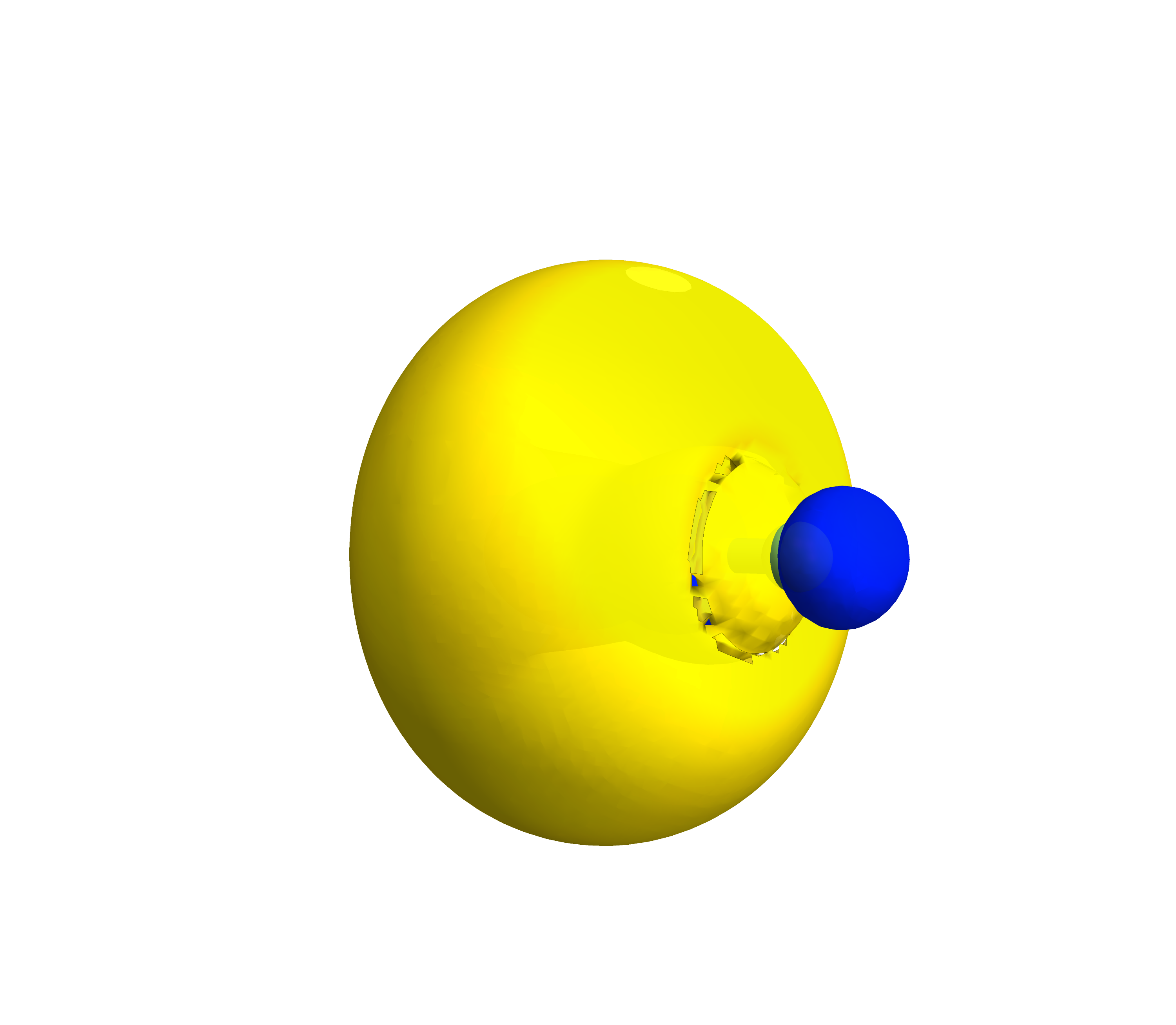}} & \raisebox{-.5\height}{\includegraphics[trim={3.0cm 2.3cm 2.2cm 3.3cm},clip,height=1.4cm,angle=90]{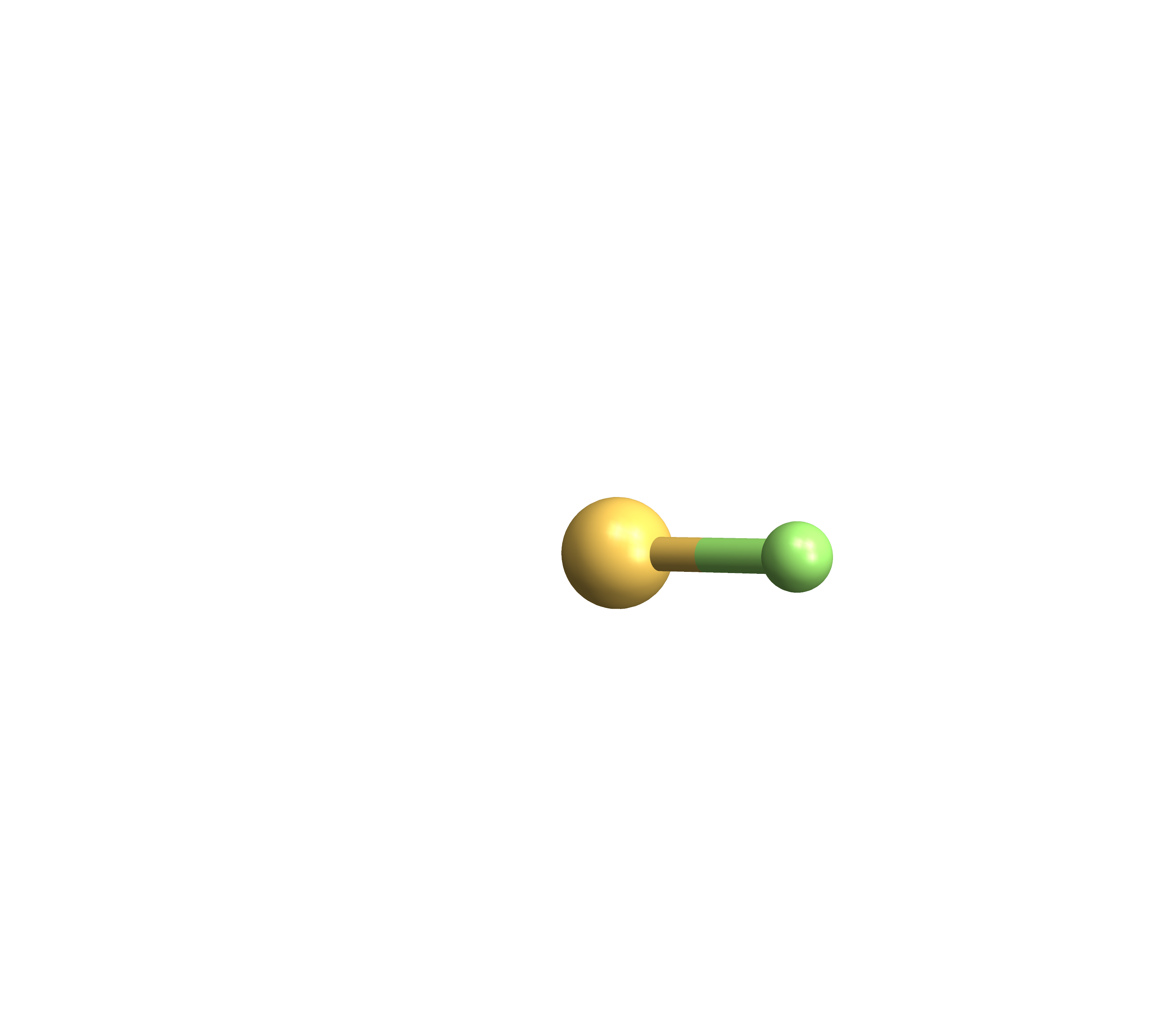}} \\
      \bottomrule
    \end{tabular}
    }%
\end{table*}

For the seven lowest electronic states found on the level of 2c-ZORA-cGHF, a Mulliken population analysis finds the singly occupied molecular orbital (SOMO) to be mostly  ormed by the atomic orbitals of f electrons on cerium. Here, the SOMO of the states $(\mathrm{X})5/2$, $(1)7/2$ and $(1)5/2$ arise completely from said atomic orbitals, whereas the rest has admixture of p atomic orbitals located at the fluorine center. This is also reflected in the visualization of the alpha and beta (or spin-up and spin-down) components of the SOMO. When focusing only on the angular part of the wavefunction localized at cerium, the hydrogenic wavefunctions for f-electrons are recognized in an ordered fashion. Thus, there are two blocks for the first seven electronic states: the SOMOs of the first three electronic states result from the f spinors with a total electronic angular momentum quantum number of $5/2$ and the next four with a total electronic angular momentum quantum number of $7/2$. Each block
amounts to the different possible axis projections of the total electronic angular momentum quantum number of the f spinor which form the ordered electronic states.  Thus, only states with SOMOs arising from f spinors with a magnetic quantum number of $3/2$ or $1/2$ can have contributions from the p spinor located at the fluorine center. In total these ultimately compose the states with the specific projection of the electronic angular momentum $\Omega$ found in the state specification and the term symbols. This admixture is further affecting the expectation values of the projection of the electronic orbital angular momentum  $\Lambda$ such that non-integer values are obtained. For \ce{PaF^3+} essentially the same structure and analysis is found.

In addition to the block of seven electronic states with considerable f spinor contributions to the SOMO, we obtained two of the expected five electronic states characterized by a SOMO with mainly d spinor contributions. Three further electronic states from this manifold can, thus, be expected with quantum numbers $\Omega=$ 5/2, 3/2 and 1/2. The contributions from the different spinors on cerium are reflected in the equilibrium bond lengths, that are non-deviating within the `f' block but differ compared to the states $(3)3/2$ and $(3)1/2$. A similar observation is made for the harmonic vibrational wavenumbers which are roughly $\SI{720}{cm^{-1}}$ for the first seven electronic states indicating the strong bond formed in \ce{CeF^2+}. Anharmonicities do not have large contributions (qualitatively seen in Figure~\ref{fig:pes_cef}) resulting in parallel potential energy surfaces for the lower states, which is a qualitative indicator for diagonal FC factors and, thus, for suitable laser-cooling transitions. Transitions with non-diagonal FC factors would require additional re-pump laser if used in a laser-cooling scheme. The number of additional lasers is estimated by the additional FC factors needed to obtain an accumulated FC factor close to unity. The closer the total FC factor is to unity, the more photons can be scattered with a single laser in an optical cycling loop \cite{isaev2020}.

However, due to this quasi-parallel, dense structure of the states built mostly from f spinors, transition dipole moments are predicted to be rather low for a few transitions, caused by a change in $\Omega$ greater than $1$, which is beyond the dipole approximation (Table~\ref{tab:tdm}). Although unfavorable for e.g. laser-cooling schemes , some of these transitions could be considered for probing the spatio-temporal variation of fundamental constants as they exhibit lifetimes in the seconds regions (e.g. transition from the electronic ground state to the state $(1)1/2$, Table~\ref{tab:fc}). Meanwhile a multitude of strong transitions exists from
the 'f' block to the upper two electronic states. 

In order to assess more on the suitability of spectroscopical techniques on 
specific transitions, various parameters influencing the fine structure are
estimated. This includes $\Omega$-doubling and the hyperfine components for 
(two isotopes of) cerium and fluorine (Table~\ref{tab:spec_props}). 

\begin{table*}[!htb]
  \caption{\justifying Crude estimations for the $\Omega$-doubling parameter $\Omega_\mathrm{p}$ together with computed values of the parallel component of the hyperfine structure constant. The isotopes $^{139}$Ce ($I=3/2$,
  $\mu=1.06~\mu_\mathrm{N}$), $^{141}$Ce ($I=7/2$, $\mu=1.09~\mu_\mathrm{N}$) and $^{19}$F ($I=1/2$, $\mu=2.628868~\mu_\mathrm{N}$) \cite{stone:2015} are considered. For discussion of the approximations used in estimating $\Omega_\mathrm{p}$ see Section~\ref{sec:comps}.}
  \label{tab:spec_props}
  %\resizebox{\textwidth}{!}{%
    \begin{tabular}{
      c
      S[table-format=-2.5,round-mode=places,round-precision=2]
      S[table-format=-2.5,round-mode=figures,round-precision=2]
      S[table-format=-2.5,round-mode=figures,round-precision=2]
      S[table-format=-2.5,round-mode=figures,round-precision=2]
      }
      \toprule
      State &
      {$\Omega_\mathrm{p}/2B$} &
      {$A_\parallel(^{139}\mathrm{Ce})\ /\ \mathrm{MHz}$} &
      {$A_\parallel(^{141}\mathrm{Ce})\ /\ \mathrm{MHz}$} &
      {$A_\parallel(^{19}\mathrm{F})\ /\ \mathrm{MHz}$} \\
      \midrule
      \multirow{1}{*}{$(X)5/2$} & 0.00016 & -393.734  &  -173.421 & -14.4237 \\
      \multirow{1}{*}{$(1)3/2$} & 0.00945 & -383.865  &  -169.074 & -12.7603 \\
      \multirow{1}{*}{$(1)1/2$} & 1.36859 & -379.086  &  -166.969 & -39.6627 \\
      \multirow{1}{*}{$(1)7/2$} & 0.00288 & -196.963  &  -86.753  & -1.55506 \\
      \multirow{1}{*}{$(1)5/2$} & 0.00383 & -203.391  &  -89.5841 & -2.47366 \\
      \multirow{1}{*}{$(2)3/2$} & 0.00034 & -196.236  &  -86.4325 & 0.017041 \\
      \multirow{1}{*}{$(2)1/2$} & 1.94654 & -191.673  &  -84.4229 & -9.25067 \\
      \multirow{1}{*}{$(3)3/2$} & 0.00011 & -627.168  &  -276.238 & -33.1699 \\
      \multirow{1}{*}{$(3)1/2$} & 0.58077 & -6446.51  &  -2839.38 & -296.108 \\
      \bottomrule
    \end{tabular}
  %}%
\end{table*}%

\section{\texorpdfstring{Sensitivity Towards $\mathcal{P,T}$-odd Properties}{}}\label{sec:4}

For the analysis of enhancement factors associated with signatures of new physics, we focus on the first-order terms in the spin–rotational Hamiltonian defined in Ref.~\cite{Gaul:2020}. Specifically, we consider the $\mathcal{P,T}$-odd contributions arising from a nuclear Schiff moment, characterized by the enhancement factor $W_{\mathcal S}$; from an electron electric dipole moment (eEDM), $W_{\mathrm d}$; from scalar-pseudoscalar, pseudoscalar-scalar, and tensor$\text{-}$pseudotensor nucleon$\text{-}$electron current interactions, described by $W_{\mathrm s}$, $W_{\mathrm p}$, and $W_{\mathrm T}$, respectively; as well as from the nuclear magnetic quadrupole moment, $W_{\mathcal M}$.

\setcounter{page}{12}

\begin{table*}[!htb]
\centering
\caption{\justifying Selected $\mathcal{P,T}$-odd properties obtained on the level of
2c-ZORA-cGHF. Here the steep s and p functions employed are crucial 
to describe the behavior of the wavefunction in the core region. 
}
\label{tab:pt}
\begin{tabular}{
c
c
S[table-format=-5.0]
S[table-format=-2.4,round-mode=figures,round-precision=2]
S[table-format=-3.3,round-mode=figures,round-precision=2]
S[table-format=-6.1,round-mode=figures,round-precision=2]
S[table-format=-2.6,round-mode=figures,round-precision=2]
S[table-format=-2.2,round-mode=figures,round-precision=2]
}
\toprule
State
& Term Symbol
& {$W_\mathcal{S} / \frac{e}{4\pi\epsilon_0a_0^4}$}
& {$W_\mathrm{d}/ \frac{10^{24}\mathrm{Hz}}{e\mathrm{cm}}$}
& {$W_\mathrm{s}/h\mathrm{kHz}$}
& {$W_\mathrm{T} / h\si{Hz}$}
& {$W_\mathcal{M}/ \frac{10^{33}h\si{Hz}}{c\ e\si{cm^2}}$}
& {$W_\mathrm{p} / \si{Hz}$}\\
\midrule
%%%%%%%%%%%%%%%%%%%%%%%%%%%%%%%%%%%%%%%%%%%%%%%%%%%%%%%%%%%%%%%%%%%%%%%%%%%%%%%%
HfF$^+$    & $^3\Delta_1$     &-15000 & 6.4$^{\dagger\dagger}$ & 23.4$^{\dagger\dagger}$  & -1146   & 0.56   & -4.1 \\ 
ThO        & $^3\Delta_1$     &-35000 & 24$^{\dagger\dagger}$  & 141$^{\dagger\dagger}$ & -3523   & 1.39    & -13.9\\ 
PaF$^{3+}$ \cite{zulch2022cool}& $^2\Phi  _{5/2}$ &-72000 & 0.66     & 4.16      & -6700   & 0.037   & -26.9$^\dagger$ \\ 
\hline 
(X)5/2    & $^2\Phi  _{5/2}$ &-7300  &   0.044  &  0.1218  &  -529.3 & 0.0047  & -1.6542\\ 
(1)3/2    & $^2\Delta_{3/2}$ &-6900  &   0.013  &  0.0782  &  -499.3 & -0.0052 & -1.5604\\ 
(1)1/2    & $^2\Pi   _{1/2}$ &-6600  &   0.040  &  0.2138  &  -474.2 & -0.0183 & -1.4820\\ 
(1)7/2    & $^2\Phi  _{7/2}$ &-7400  &   -0.028 &  -0.0783 &  -531.1 & -0.0036 & -1.6598\\ 
(1)5/2    & $^2\Delta_{5/2}$ &-7000  &   -0.009 &  -0.0293 &  -507.7 & 0.0014  & -1.5864\\ 
(2)3/2    & $^2\Pi   _{3/2}$ &-6700  &   -0.033 &  -0.1057 &  -485.5 & -0.0006 & -1.5170\\ 
(2)1/2    & $^2\Sigma_{1/2}$ &-6500  &   -0.098 &  -0.2974 &  -472.6 & -0.0045 & -1.4765\\ 
(3)3/2    & $^2\Delta_{3/2}$ &-7100  &   0.379  &  1.0214  &  -521.5 & 0.0388  & -1.6289\\ 
(3)1/2    & $^2\Sigma_{1/2}$ &-9400  &   3.276  &  8.1313  &  -661.3 & 0.5146  & -2.0703\\ 
\bottomrule                    
\end{tabular}
\begin{tablenotes}
\item $^\dagger$Computed here with the wavefunction from Ref.~\cite{zulch2022cool}.
\item $^{\dagger\dagger}$ Values computed herein with basis sets comparable to the ones used in the present work for CeF$^{2+}$. Deviations from previous results in Ref.~\cite{Gaul2024}, obtained on the same level of theory but with a larger basis set, are smaller than figures reported here.
\end{tablenotes}
\end{table*}

To place our newly calculated enhancement factors of $\ce{CeF^2+}$ in context, see Table~\ref{tab:pt}. We compare them to those of the science states of $\ce{HfF^+}$ and $\ce{ThO}$, which have played a central role in previous eEDM experiments, as well as to the ground state of $\ce{PaF^3+}$, proposed for measurements of the nuclear Schiff moment of \mbox{$\ce{^{229}Pa}$}~\cite{zulch2022cool}. The $\ce{PaF^3+}$ system benefits from a combination of strong nuclear enhancement, arising from the nuclear octupole deformation and a proposed low-lying parity doublet in $\ce{^{229}Pa}$, as well as strong molecular enhancement, reflected in the large magnitude of $W_{\mathcal S}$. The impact of these combined amplification mechanisms, along with the sensitivity of the system to specific enhancement factors, on reducing the hypervolume of allowed $\mathcal{P}$,$\mathcal{T}$-odd parameter space has been analyzed in detail in Ref.~\cite{Gaul2024}. The global analysis model applied therein suggests that future $\ce{{}^{229}PaF^3+}$ measurements could serve to reduce the coverage hypervolume by three to six orders of magnitude, compared to today's constraints imposed by all existing atomic and molecular experiments.

This potential of PaF$^{3+}$ to reduce the parameter space is because molecules such as $\ce{HfF^+}$ and $\ce{ThO}$ primarily provide sensitivity to electron spin-dependent interactions, including the eEDM and the scalar-pseudoscalar nucleon-electron coupling, due to substantial s- and p-character of the singly occupied molecular orbital at the heavy atom. The four-dimensional (or five-dimensional, depending on the model) subspace of the remaining $\mathcal{P},\mathcal{T}$-odd parameters is predominantly constrained by the $\ce{{}^{199}Hg}$ experiment \cite{Hg1992016, Hg1992017} and the neutron EDM experiment \cite{AbelNEDM}. Yet, relying primarily on only two sensitive measurements induces strong correlations within the remaining four (or five) dimensional parameter space~\cite{Degenkolb2024}, and is therefore intrinsically inadequate to robustly constrain a parameter subspace of this dimensionality. This stresses the demand for powerful additional experiments that can break these correlations and confine the otherwise comparatively loosely constrained model parameters, namely those from the purely hadronic sector together with the pseudoscalar-scalar as well as the tensor-pseudotensor coupling constants for nucleon-electron current interactions.

In the quest to further constrain the $\mathcal{P},\mathcal{T}$-odd parameter space using atomic and molecular probes, the global analysis of Ref.~\cite{Gaul2024} highlighted in this context the general importance of medium-heavy systems. Generally, these species are less sensitive to a single source of $\mathcal{P},\mathcal{T}$ violation due to the expected sensitivity scaling with nuclear charge $Z$. Nevertheless, they were found to constitute quintessential complementary probes, because molecular enhancement factors exhibit distinct $Z$ dependencies for different underlying sources. Therefore, `repeating' experiments on new systems with large-$Z$ atomic nuclei and comparable (ratios of) enhancement factors may have a smaller impact on the overall coverage hypervolume of $\mathcal{P},\mathcal{T}$-odd physics than studies of medium-heavy species involving lower-$Z$ nuclides, even when the $\text{latter’s}$ nominal sensitivities to individual sources are lower.

Although Ce, with $Z=58$, is slightly outside the $Z\leq54$  range explicitly addressed in Ref.~\cite{Gaul2024}, it is not unreasonable to assume that a similar rationale may also apply to CeF$^{2+}$. To guide the investigation of this expectation in upcoming global analyses, we provide our calculated molecular enhancement factors for $\ce{CeF^2+}$ in Table~\ref{tab:pt}. Contrary to $\ce{HfF^+}$ and $\ce{ThO}$, $\ce{CeF^2+}$ exhibits, as a $\ce{PaF^3+}$ analogue, diminished values for the electron spin-dependent enhancement factors $W_{\mathrm d}$ and $W_{\mathrm s}$ for the first seven electronic states due to the SOMO contributions from the f spinor in these states. Relative to $W_{\mathrm d}$ and $W_{\mathrm s}$, CeF$^{2+}$ -- like $\ce{PaF^3+}$ -- displays markedly larger enhancement factors $W_\mathrm{T}$ and $W_\mathcal{S}$. These electron spin-independent factors remain roughly the same in the various energetically low-lying states of $\ce{CeF^2+}$ at about half the value of $\ce{HfF^+}$. In contrast, the higher-lying (3)1/2 state displays significantly enhanced values of $W_{\mathrm d}$ and $W_{\mathrm s}$, consistent with increased s- and p-character in the SOMO and the corresponding rise in electron density at the nucleus. The combination of this diverse electronic structure and generally strong, albeit somewhat lower than in heavier systems, electron spin-independent enhancements positions $\ce{CeF^2+}$ as a valid platform for probing multiple classes of $\mathcal{P},\mathcal{T}$-odd interactions and for providing an additional set of constraints to complement both current experiments and future studies in heavier systems, such as the proposed $\ce{PaF^3+}$ experiment with octupole-deformed $^{229}$Pa.

\section{Prospects for Quantum Control}\label{sec:5}

Control of individual quantum states of polar diatomic molecules for measurements of $\mathcal{P,T}$-odd effects have been demonstrated to various extents in many molecular systems \cite{Roussy2023an,acme2018improved,hudson2011improved,Cho1991search,demille2001search}. The respective quantum control schemes highly depend on the details of the molecular parameters. Fortunately, the molecular parameters of $\mathrm{CeF}^{2+}$ as obtained in the present work are largely similar to those of HfF$^+$ and ThF$^+$, i.e., the molecules of choice exploited by the JILA group to measure the electron's electric dipole moment \cite{Roussy2023an,caldwell2023,ng2022spectroscopy, KB2025ThF}. Hence, established techniques can be largely ported over to analogous studies with $\mathrm{CeF}^{2+}$. For details, the interested reader is encouraged to refer to the JILA protocols laid out in Section II E of Ref.~\cite{caldwell2023} and Section II A of Ref.~\cite{ng2022spectroscopy}. Despite the similarities, some differences persist between $\mathrm{CeF}^{2+}$ and the aforementioned molecules, the implications of which are explored in subsequent sections. 

The following discussions do not aim to provide a definite quantum control scheme for CeF$^{2+}$, but solely serve to explore the feasibility of quantum control with known techniques. The discussions will loosely follow various aspects critical to the JILA protocol, but are generally applicable to any experimental scheme involving a rotating electric field used to polarize the molecules. We begin with an overview of the energy landscape in Section~\ref{sec:EnergyLandscape}, which serves as a point of reference for subsequent sections. Section~\ref{sec:PolarizingElectricField} examines the requirements of the rotating electric field needed to polarize the molecule. Section~\ref{sec:non-inertial-frame} discusses the implications of working with a rotating electric field with CeF$^{2+}$. Section~\ref{sec:StatePreparationAndDetection} explores the feasibility of performing (i) state preparation with an optical pumping scheme using the calculated electronic levels in this work and (ii) state detection with known techniques. Finally, Section~\ref{sec:HyperfineStructure} explains the additional complexity introduced when selecting a cerium isotope with non-zero nuclear spin. Unless stated otherwise, we will focus the discussions on the ground vibrational and rotational manifold of the $(X)5/2$ state of $^{140}\mathrm{Ce}^{19}\mathrm{F}^{2+}$, motivated by its prospect as a steppingstone to quantum control of the ground state of $^{229}\mathrm{Pa}^{19}\mathrm{F}^{3+}$ \cite{zulch2022cool}.

\subsection{Energy Landscape}\label{sec:EnergyLandscape}

\begin{figure*}[htb]
    \centering
    \includegraphics[width=\linewidth]{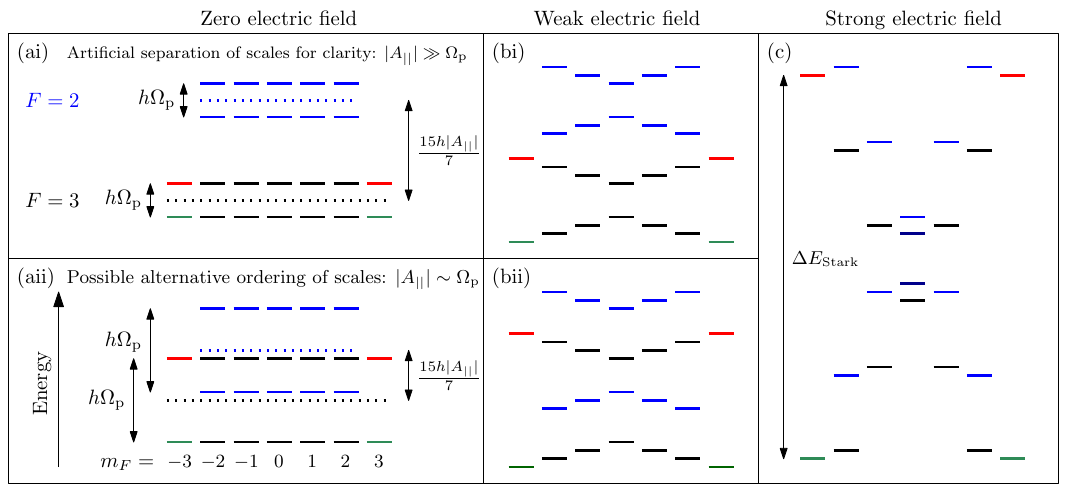}
    \caption{\justifying Energy landscape (not to scale) of the $(X)5/2~(v=0, J=5/2)$ manifold of $^{140}\mathrm{CeF}^{2+}$. The parameters $h\Omega_\mathrm{p}$ and $15h|A_{||}|/7$ correspond to the energy difference between the parity states and the hyperfine manifolds, respectively, where $h$ is the Planck constant. $\Delta E_\mathrm{Stark}$ corresponds to the Stark shift. The (a), (b), and (c) columns correspond to the cases with zero, weak ($\Delta E_\mathrm{Stark} \ll h \Omega_\mathrm{p}$), and strong ($\Delta E_\mathrm{Stark} \gg h \Omega_\mathrm{p}, h|A_{||}|$) external electric fields. In all panes, the vertical direction corresponds to an increase in energy, while the horizontal coordinate reflects the Zeeman sub-levels $m_F$. The dotted lines in the (a) column are to guide the eye to the hyperfine splitting. The (i) row shows the $\Omega$-doubling splitting $h\Omega_\mathrm{p}$ and hyperfine splitting $15h|A_{||}|/7$ in the context of the energy landscape. The (ii) row shows the same, but with a possible alternative ordering of the energy levels according to our estimated uncertainties of the parameters calculated in this paper. The blue lines indicate states corresponding to the $F=2$ hyperfine manifold, while those in black to the $F=3$ manifold, where $\mathbf{F} = \mathbf{J} + \mathbf{I}$, and $I=1/2$ for the nuclear spin of $^{19}$F; $^{140}$Ce has no nuclear spin. Two pairs of doublets corresponding to the two pairs of oppositely fully stretched Zeeman states are colored red and green. These shall be referred to as the upper and lower doublets, respectively. The hyperfine quantum number $F$ is strictly not a good quantum number in the presence of external electric fields, but we keep the color scheme in all columns to suggest asymptotic correspondence to the states in the case of diminishing electric field strength.}
    \label{fig:CeF_energy}
\end{figure*}

There are 24 energy eigenstates in the ground electronic, vibrational, and rotational manifold of $^{140}\mathrm{CeF}^{2+}$. These eigenstates and their responses to external electric fields are shown in Figure~\ref{fig:CeF_energy}. States that have desirable properties for the measurement of $\mathcal{P,T}$-odd effects are the fully stretched Zeeman states ($|F=3, m_F=\pm 3, \Omega=\pm5/2\rangle$, where $F$, $m_F$, and $\Omega$ are the hyperfine, Zeeman sub-level, and $\Omega$-doubling quantum numbers, respectively). These states, highlighted in red and green in Figure~\ref{fig:CeF_energy}, are of well defined molecular and spin orientations in the presence of polarizing electric fields.

Details of the shown energy landscape depend on molecular parameters of the $(X)5/2$ state in $^{140}\mathrm{CeF}^{2+}$ listed in Tables \ref{tab:elec_states} and \ref{tab:spec_props} of which the ones relevant for the present discussion are summarized in Table \ref{tab:molpara}. 
\begin{table*}[tb]
    \centering
    \caption{\justifying Calculated molecular parameters of the $(X)5/2$ state in $^{140}\mathrm{CeF}^{2+}$, reproduced from Tables \ref{tab:elec_states} and \ref{tab:spec_props} for convenience.}
    \label{tab:molpara}
    \begin{tabular}{l l l}
        \toprule
        Parameter & Value & Note \\
        \midrule
        Molecular dipole moment, $|\vec\mu_e|$ & $1.49~e a_0$ & Equivalent to $\approx 2~\mathrm{MHz}/(\mathrm{V} \hspace{1pt} \mathrm{cm}^{-1})$ \\
        Energy splitting between parity states, $\Omega_\mathrm{p}$ & $<0.01 \times 2B$ & Estimated to be $< 100~\mathrm{MHz}$. See discussion in main text. \\
        Hyperfine splitting constant, $A_{||}(^{19}\mathrm{F})$ & $-14$~MHz & - \\
        \bottomrule
    \end{tabular}
\end{table*}

While we have calculated the parity-doublet splitting arising from the direct coupling between Kramers partners in Table~\ref{tab:spec_props}, other factors may also contribute to the lifting of the degeneracy of the parity states in the $(X)5/2$ state. Accurately evaluating energy splittings of this magnitude is challenging. Due to the lack of experimental data on the parity-doublet splitting for the ground state of CeF$^{2+}$, but guided by measured parity-doublet splittings for $|\Omega|=5/2$ states in other molecules (e.g., Ref.~\cite{abbasi:2018}), we will assume $\Omega_\mathrm{p} \sim 10~\mathrm{MHz}$ for the purposes of subsequent discussions. It is worth noting that another relevant energy scale for our discussion is the $^{19}$F hyperfine splitting, shown in Table~\ref{tab:spec_props}, which is also on the order of 10~MHz. A parity-doublet splitting of order 10~MHz or smaller will not affect the qualitative conclusions of the following sections.

\subsection{Polarizing Electric Field}\label{sec:PolarizingElectricField}

In the absence of external electric fields, the eigenstates of the molecule are states of good parity, i.e., with no well defined molecular orientation. However, measurements of $\mathcal{P,T}$-odd effects often require the molecules to have well defined orientation. Applied external electric fields mix the parity states to result in eigenstates of good orientation with respect to the direction of the applied field. The field strength required to fully mix these states depends on the energy splitting between the parity states ($h \Omega_\mathrm{p}$, where $h$ is the Planck constant) and the molecular electric dipole moment ($\vec\mu_e$) in the body fixed frame. The numbers presented in Table \ref{tab:molpara} imply that the molecule is already sufficiently polarized --- well oriented in space --- at an electric field strength of only 25~$\mathrm{V}~\mathrm{cm}^{-1}$.
Working with ionic species introduces complications on how an external electric field can be applied to polarize the molecules while keeping them confined. Several implementations involving rotating electric fields \cite{leanhardt2011high,ng2022spectroscopy,zhou2024quantum} have been proposed and demonstrated to various extents. A rotating electric field on the order of 25~$\mathrm{V}~\mathrm{cm}^{-1}$ can be readily realized, as shown in the JILA experiments.

Ions in a rotating field exhibit rotational micromotion. The radius of the micromotion scales linearly with the field strength and charge state of the ions, and inversely proportional to the mass and square of the rotating frequency. It is desirable to keep the amplitude of the micromotion small such that the ions sample a smaller volume in space, allowing for less stringent design requirements on the spatial homogeneity of the applied fields. Doubly charged CeF$^{2+}$ implies a larger rotational micromotion than monocations like those used in the JILA experiments, but this can be compensated with a faster rotating frequency. In particular, the rotational micromotion can be kept to a radius of about 1~mm with typical field strengths of 60~$\mathrm{V}~\mathrm{cm}^{-1}$ and a rotating frequency of about 400~kHz, still within reasonably attainable limits for a typical setup.

\subsection{Non-inertial-frame Coupling}\label{sec:non-inertial-frame}

\begin{figure}[htb]
    \centering
    \includegraphics[width=1\columnwidth]{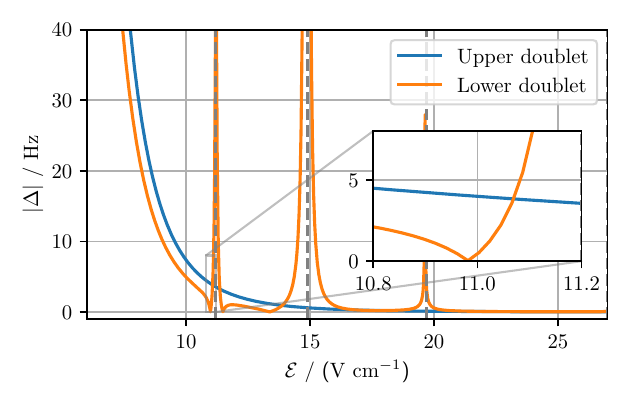}
    \caption{\justifying Absolute sizes of the effective coupling ($|\Delta|$) between oppositely stretched Zeeman states as a function of applied electric field strength ($\mathcal{E}$) with a rotating frequency of $f_\mathrm{rot} = 400~\mathrm{kHz}$. The upper and lower doublets are defined in the caption of Figure~\ref{fig:CeF_energy}. The vertical dashed lines indicate hyperfine crossings, where the upper doublet crosses a $F=2$ state in energy; the energy degeneracy results in a much stronger effective coupling. For the molecular parameters calculated in this paper, we show the first three hyperfine crossings as the energy levels go from those depicted in (ai) to (bi) to (c) in Figure~\ref{fig:CeF_energy}. The first corresponds to when the upper doublet crosses the $| F=2, m_F=\pm 2 \rangle$ states, while the second to the $| F=2, m_F= \pm 1 \rangle$ states and so on. Once again, $F$ is not a good quantum number in the presence of an external electric field, but we have labeled the states based on asymptotic correspondence. $\Delta$ values for the upper and lower doublets are not the same in general, however they can cross near hyperfine crossings, which can be useful in certain protocols like that in Ref.~\cite{zhou2024quantum}. The inset shows one such occurrence. $|\Delta|$ for both doublets take on vanishing values much lower than $\sim 1~\mathrm{Hz}$ beyond 20~$\mathrm{V}~\mathrm{cm}^{-1}$ except at hyperfine crossings.}
    \label{fig:Deltas}
\end{figure}

Measurements of $\mathcal{P,T}$-odd effects in molecular ions are typically performed in the rotating frame of the molecules, leading to a non-inertial-frame coupling that connects neighboring Zeeman states. This results in an effective coupling ($h \Delta$) between the oppositely stretched Zeeman states. We refer the interested reader to Section 4.4 in Ref.~\cite{leanhardt2011high} for a more thorough discussion of the physics mechanism that gives rise to $h \Delta$ and its implications on the measurement scheme. A plot of the effective coupling for both the upper and lower doublets (defined in the caption of Figure~\ref{fig:CeF_energy}) is shown in Figure~\ref{fig:Deltas}.

Each of the upper and lower doublets can be represented by an effective two-level Hamiltonian of the form $\mathcal{H}_\mathrm{eff.} = \left( \begin{smallmatrix} E_{m_F=3} & h\Delta \\ h\Delta & E_{m_F=-3} \end{smallmatrix} \right)$, where $E_{m_F=\pm 3}$ correspond to the energy levels of the $|F=3, m_F=\pm 3 \rangle$ states. There are two extreme regimes at which one can select to operate in: (i) $|h\Delta| \ll |E_{m_F=3} - E_{m_F=-3}|$ and (ii) $|h\Delta| \gg |E_{m_F=3} - E_{m_F=-3}|$. One might choose to work in the former regime where the eigenstates asymptotically approach $|m_F = \pm3\rangle$ when, for example, one is interested in measuring the energy difference between the oppositely stretched Zeeman states during the Ramsey free evolution with minimal interference from unwanted effects like $h\Delta$. Conversely, one might choose to operate in the latter regime where there is strong off-diagonal coupling between the $|m_F = \pm3\rangle$ state to, for example, effect a $\pi/2$ rotation of the Bloch vector on the Bloch sphere for a Ramsey interferometry experiment.

The size of the effective coupling typically scales with $\Delta \sim \Omega_\mathrm{p} \left( \frac{h f_\mathrm{rot}}{|\vec\mu_e| \mathcal{E}} \right)^{\delta m_F}$ \cite{leanhardt2011high}, where $h$ is the Planck constant, $\Omega_\mathrm{p}$ is the $\Omega$-doubling splitting constant, $f_\mathrm{rot}$ is the rotating frequency, $\mathcal{E}$ is the electric field strength, and $\delta m_F$ is the difference between the Zeeman sub-level quantum number between the two oppositely stretched Zeeman states, in this case $\delta m_F = 6$. The electric field strength can be used as an experimental knob to tune between the two extreme regimes related to the relative sizes of $|h\Delta|$ and $|E_{m_F=3} - E_{m_F=-3}|$. For example, to suppress the effect of $h\Delta$ during the measurements, one can operate at large electric field strengths. Of note, the hyperfine crossings all occur at low electric field strengths, far from typical operation parameters. As another example, one can ramp down to low electric field strengths to temporarily increase the coupling between the $|m_F = \pm3\rangle$ state to effect a $\pi/2$ rotation of the Bloch vector on the Bloch sphere before ramping back up again to high field strengths for the Ramsey free evolution part of the Ramsey interferometry experiment.

Given that the molecular parameters calculated in this work imply less clear separation of certain critical energy scales, higher order effects not included in our calculations could result in an actual ordering of energy scales that is different from those suggested by this work. Furthermore, the electric field strengths at which the hyperfine crossings occur can change drastically with slightly different values for the molecular parameters. A consequence, for example, is that the calculated ordering where the hyperfine splitting is larger than the $\Omega$-doubling splitting could actually be inverted, resulting in fewer hyperfine crossings (when evolving from (aii) to (bii) to (c) in Figure~\ref{fig:CeF_energy}). Measurements of the relevant molecular parameters are required to resolve this. Hence, Figure~\ref{fig:Deltas} is intended solely as a rough guide for the prospects of molecular control.

\subsection{State Preparation and Detection}\label{sec:StatePreparationAndDetection}

The CeF$^{2+}$ molecules produced with the techniques demonstrated in this work would result in a spread of population over many electronic and vibrational states in the molecule. A prerequisite to quantum control is to prepare the molecules in a well defined quantum state. As a first step to achieve this, the entropy of the internal states could be reduced by cooling the molecular ions with a cryogenic buffer gas leaked into a cryogenic Paul trap \cite{gerlich1995ion,chakrabarty2013novel,hansen2014efficient,gunther2017cryogenic,LECHNER2024169471}. Alternatively, one could obtain a smaller spread of internal state population through a careful ionization scheme that selectively populates the ions of the desired charge state in only the lowest energy states through energy considerations like those pursued in the JILA experiments. A combination of both approaches may also be considered. Regardless, for the present discussion, we assume that the molecules are found in the ground electronic and vibrational manifold, with its population spread over a few of the lowest rotational states.

Optical pumping is a standard technique to usher population around in molecular systems. This involves using laser and microwave systems to address certain transitions in the molecule, exploiting selection rules in the process. It is then crucial to find suitable intermediate states for this purpose. The suitability depends on a few considerations, including (i) the quantum numbers of the intermediate state, (ii) accessibility of the laser wavelengths required to address the transition, (iii) natural decay lifetime of the intermediate state, and (iv) branching ratio of spontaneous decay from the intermediate state to various lower lying states. 

\begin{figure*}[htb]
    \centering
    \includegraphics[width=2\columnwidth]{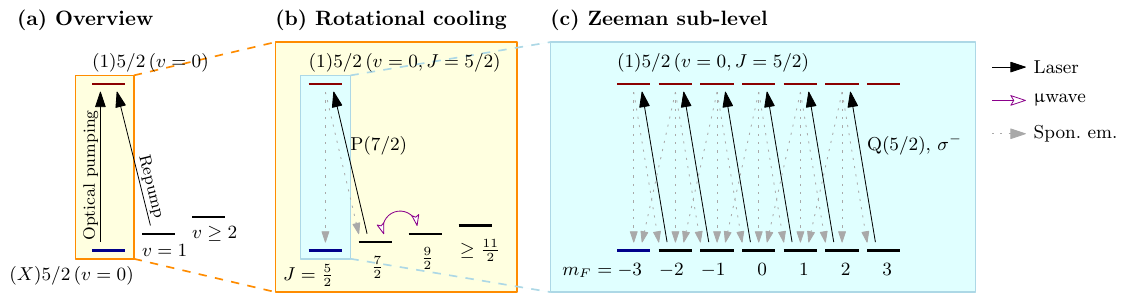}
    \caption{\justifying An example of an optical pumping scheme in CeF$^{2+}$ (not to scale). Figure design and optical pumping scheme are adapted from Ref.\ \cite{ng2022spectroscopy}. The $(1)5/2$ electronic state is used in this example as the excited electronic state for the optical pumping scheme. (a) An overview of the optical pumping scheme. Most of the lasers will be connecting the $(X)5/2\,(v=0)$ ground vibronic manifold to the $(1)5/2\,(v=0)$ vibronic manifold. Vibrational repump lasers can be used to bring back population lost to the excited vibrational states of the $(X)5/2$. (b) Rotational cooling can be done within the $(X)5/2\,(v=0)$ manifold by pumping the $J=7/2$ state through the excited state. Microwaves can couple the higher rotational states to transfer population in those states to the $J=5/2$ state. (c) Circularly polarized light can be used to pump all the population into a stretched Zeeman sub-level. The Q line connecting the ground and the excited states allows for a dark Zeeman sub-level ($m_F=-3$ in this example.), where population will accumulate through spontaneous decay from the excited state. Of note, the ground energy levels in panel (c) correspond to those in Figure~\ref{fig:CeF_energy} with the hyperfine manifolds and parity doublets unresolved.}
    \label{fig:CeF_opticalpumping}
\end{figure*}

As shown in Table~\ref{tab:fc}, most transitions between the nine lowest lying electronic states involve either UV or mid-IR transitions. However, the Franck--Condon factors for transitions at the UV wavelengths are rather unfavorable. Should one use one of those UV transitions for optical pumping, numerous vibrational repump lasers would have to be introduced. The increased experimental complexity would thus reduce practical feasibility. On the other hand, while commercial mid-IR lasers are readily available, their diminishing Einstein $A$ coefficients make many of the low lying states unappealing. Although work with trapped ions provides an opportunity to employ optical pumping schemes that require longer time scale as compared to beam experiments, various destabilizing mechanisms from experimental imperfections could result in an overall decoherence or population loss with time. This could limit the practical duration of pumping periods. To strike a balance between efficiency and feasibility, the $(1)5/2$ state could be used for optical pumping from the $(X)5/2$ state. A schematic diagram of an example of how optical pumping can be used to usher population around in CeF$^{2+}$ is shown in Figure~\ref{fig:CeF_opticalpumping}.

Despite having many transitions exhibiting diagonal Franck-Condon factors, their low Einstein $A$ coefficients imply low photon detection rates, making optical cycling a challenging option for state detection. As a remedy, action spectroscopy for readout through state-selective multiresonance photodissociation could be employed, similar to the one used in the JILA experiments. Alternatively, one could leverage on the ionic nature of the molecules to perform non-destructive state readout with quantum logic spectroscopy \cite{maison2022,fan2021,zhou2024quantum}, but technical difficulties still remain in combining quantum logic spectroscopy together with a rotating polarizing electric field.

Of note, the numbers shown in Table \ref{tab:fc} are similar in magnitude to those of $\mathrm{PaF}^{3+}$ \cite{zulch2022cool}. Hence, this offers an opportunity use the stable and more abundant $^{140}\mathrm{CeF}^{2+}$ molecule as a surrogate also to master quantum control of $\mathrm{PaF}^{3+}$.

\subsection{Hyperfine Structure from Cerium Isotopes with Nuclear Spin}\label{sec:HyperfineStructure}

A non-zero cerium nuclear spin is required for measurements of some $\mathcal{P,T}$-odd effects like the nuclear Schiff moment. All stable cerium isotopes incorporate an even number of protons and neutrons and, thus, exhibit a nuclear spin $I=0$. Radioactive cerium isotopes with an odd number of neutrons, such as $^{141}$Ce with a half-life of 32 days, have $I\neq0$ and would thus be sensitive to these $\mathcal{P,T}$-odd effects, too. However, the non-vanishing nuclear spin has consequences for the energy landscape. We shall discuss two of them below: (i) larger $\delta m_F$ and (ii) additional hyperfine structures. 

We briefly alluded to how $\Delta$ can be a resource in the context of a Ramsey interferometry experiment in Section~\ref{sec:non-inertial-frame}. An increased $\delta m_F$, however, would result in smaller $\Delta \sim \Omega_\mathrm{p} \left( \frac{h f_\mathrm{rot}}{|\vec\mu_e| \mathcal{E}} \right)^{\delta m_F}$ given that the term in parenthesis is usually less than unity under typical experimental parameters (variables defined in Section~\ref{sec:non-inertial-frame}). This necessitates a reduction in the electric field strength or an increase in the frequency of the rotating electric field to access a suitable value of $\Delta$ for the $\pi/2$ rotation on the Bloch sphere. The former could result in higher quantum decohering rates from effects of spurious electric fields due to the smaller magnitude of the polarizing electric field, while the latter could complicate trap electronics design. It is worth pointing out that sweeping $\Delta$ to low values is just one of the many ways to effect a $\pi/2$ rotation on the Bloch sphere. One could, for instance, use microwaves to couple the stretched Zeeman states through the unstretched states in the hyperfine manifold.

\begin{figure}[htb]
    \centering
    \includegraphics[width=1\columnwidth]{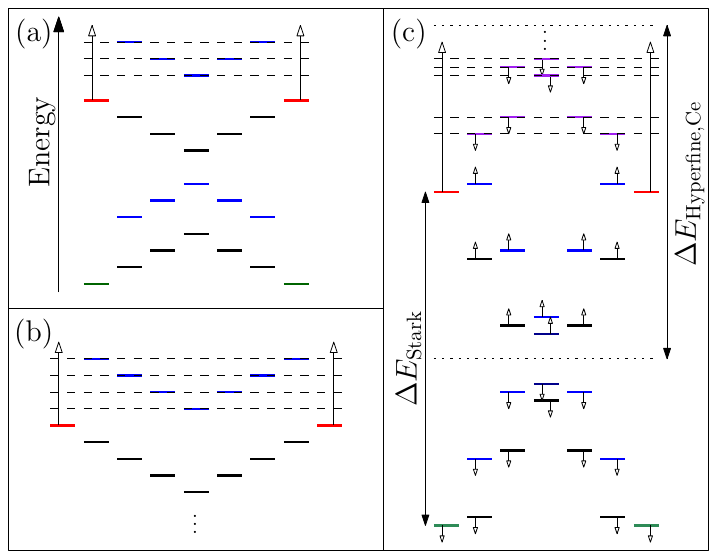}
    \caption{\justifying Additional hyperfine crossings due to a larger hyperfine structure from non-zero cerium nuclear spin (not to scale). Similar to Figure~\ref{fig:CeF_energy}, the vertical axis indicates increasing energy, while the horizontal coordinate represents the Zeeman sub-levels. In all panels, the vertical arrows with unfilled arrow head indicate the direction of the Stark shift with increasing electric field strength. The dashed horizontal lines provide a visual suggestion of the number of hyperfine crossings that the upper doublet (colored in red) will undergo with increasing electric field strength. (a) Adapted from panel (bii) of Figure~\ref{fig:CeF_energy}. (b) Same as (a), but with larger hyperfine quantum numbers $F$ due to the non-zero cerium nuclear spin. Only the top half of the energy levels are shown for brevity. (c) Adapted from panel (c) of Figure~\ref{fig:CeF_energy}. The dotted lines are to guide the eye to the hyperfine splitting due to the non-zero cerium nuclear spin. The purple lines correspond to those from a different $F_1$ hyperfine manifold (defined in main text); only the lower half of the energy levels are shown for brevity. Once the Stark splittings are comparable to the hyperfine splitting from the non-zero cerium nuclear spin, the upper doublet starts to cross the high-field-seeking states from the neighboring $F_1$ hyperfine manifold.}
    \label{fig:HyperfineCrossing}
\end{figure}

Similarly, the introduction of additional hyperfine structures could also increase the number of hyperfine crossings, where the upper doublet crosses the energy levels of the unstretched Zeeman states from the other hyperfine manifolds with increasing electric field strength. The increased number of hyperfine crossings with non-zero cerium nuclear spin arises for two reasons. First, the $F$ hyperfine quantum number for the most stretched Zeeman state would be bigger with a non-zero cerium nuclear spin ($\mathbf{F} = \mathbf{F_1} + \mathbf{I_\mathrm{^{19}F}}$, $\mathbf{F_1} = \mathbf{J} + \mathbf{I_\mathrm{Ce}}$, where $I_\mathrm{^{19}F,Ce}$ are the nuclear spins of $^{19}$F and cerium, respectively). This would result in more hyperfine crossings with the neighboring $F$ hyperfine manifold, as depicted in panel (b) in contrast with panel (a) of Figure~\ref{fig:HyperfineCrossing}.
Second, we see that our calculated hyperfine splitting constants from cerium with non-zero nuclear spin are on the order of $\Delta E_\mathrm{Hyperfine,Ce} \sim 100~\mathrm{MHz}$, similar to our expected Stark splittings. This implies that states from neighboring $F_1$ will start crossing in energy when Stark splittings are on the order of 100~MHz, as depicted in panel (c) of Figure~\ref{fig:HyperfineCrossing}. As we have seen from Section~\ref{sec:non-inertial-frame}, $\Delta$ saturates to a high value at hyperfine crossings, thus one typically wants to operate far away from the hyperfine crossings. More careful treatment with experimentally measured molecular parameters will be required to identify suitable experimental parameters.

Interestingly, hyperfine crossings can be used to one's advantage for state transfer. It is usually the case where state preparation is more efficiently done on the largest $F$, which necessarily implies the largest $F_1$. However, the $F_1$ manifold that has the largest sensitivity to the new physics of interest may be the smaller one, e.g., in the case of Schiff moments where the sensitivity goes as $\mathbf{I_\mathrm{Ce}}\cdot\hat{n}$, where $\hat{n}$ is the internuclear axis. In situations like this, one can envision first preparing the molecules in the stretched states of the largest $F$ (and hence largest $F_1$), sweep the electric field strength for a hyperfine crossing of the stretched states with the neighboring $F_1$ manifold, and perform an adiabatic rapid passage transfer of population between the $F_1$ manifolds.

Evidently, we note that the larger space that comes with additional hyperfine structures presents itself as an untapped resource. With the rapid advancement of the field of quantum state manipulation in molecules, many groups have successfully taken advantage of the rich internal structures of molecules for (i) manipulation of internal quantum states, (ii) manipulation of external degrees of freedom, and (iii) suppression of systematic effects through co-magnetometry, among many others. Moreover, various work also exploit energy level crossings in atoms to probe fundamental physics \cite{kozlov2018highly}. Abundant energy level crossings that come with additional hyperfine structures may open doors to such tools in molecules. One can thus remain optimistic that the challenges which arise from larger internal quantum space also bring new opportunities.

\section{Conclusion}\label{sec:6}

The non-radioactive analogue of tricationic protactinium monofluoride (PaF$^{3+}$), namely dicationic cerium monofluoride (CeF$^{2+}$), was produced in the gas phase. This was achieved by adapting TRIUMF's nuclear-physics infrastructure to develop a molecular formation method compatible with minute quantities of accelerator-produced, short-lived radionuclides and with standard instrumentation at radioactive ion beam facilities. Thus, this experiment employed  methodology which could potentially also be applied to obtain radioactive PaF$^{3+}$ incorporating octupole-deformed $^{229}$Pa. For our experimental studies, triply charged Ce$^{3+}$ cations were produced at TRIUMF's off-line ion source and delivered to the TITAN ion-trap facility. There, CeF$^{2+}$ molecules were synthesized in an RFQ cooler-buncher in which sulfur hexafluoride (SF$_6$) was added to the helium buffer gas as the reactant for the incoming Ce beam. Downstream of the cooler-buncher, TITAN's Multiple-Reflection Time-Of-Flight mass spectrometer was utilized for its high mass resolving power exceeding $R=m/\Delta m>10^5$ to confirm the successful formation of CeF$^{2+}$. 

As predicted by our \textit{ab initio} quantum chemistry calculations, electronic structure and properties of CeF$^{2+}$ exhibit large similarities (PaF$^{3+}$), namely a compressed, quasi-parallel structure was found for the seven lowest electronic states below $\SI{6000}{cm^{-1}}$. \mbox{$^{229}$PaF$^{3+}$} had recently been proposed as an intriguing molecular system for new-physics searches \cite{zulch2022cool} but is presently challenging to address directly in experiments. The found similarity between PaF$^{3+}$ and CeF$^{2+}$ thus represents an opportunity to master the techniques of molecular formation and quantum control with stable and now readily available CeF$^{2+}$ in preparations of future precision experiments with PaF$^{3+}$.

Moreover, our calculations of CeF$^{2+}$ indicate sizable molecular enhancement factors to $\mathcal{P,T}$-odd new-physics phenomena, albeit at a somewhat reduced magnitude compared to valence-isoelectronic PaF$^{3+}$. In particular, the singly occupied molecular orbital (SOMO) in the lowest seven electronic states is formed from the f spinors on the metal atoms, combined with the p spinors of the fluorine where formation is allowed due to similar projection of electronic angular momenta on the internuclear axis. This suppresses enhancement factors for electron spin-dependent $\mathcal{P,T}$-odd properties where electron density is needed within the nucleus, and, therefore, pronounces in contrast quark-sector $\mathcal{P,T}$-odd properties like the nuclear Schiff moment. The next two predicted excited states lie beyond $\SI{30000}{cm^{-1}}$ with SOMOs formed from d and s spinors at the metal center and, thus, exhibit larger enhancement factors for e.g., the electric dipole moment of the electron. Precision experiments with CeF$^{2+}$ hence promise to impose additional constraints on new physics parameters through a comprehensive analysis of measurement results from various atomic and molecular systems. Transitions between the seven lowest states in CeF$^{2+}$ have almost diagonal Franck--Condon factors but low transition dipole moments suitable for e.g. potential experiments catching the variation of fundamental constants \cite{zuelch2025}. 
In contrast high transition dipole moments are found for transitions to the upper two states, with less favorable Franck--Condon factors. 

The molecular parameters for CeF$^{2+}$ were found to be similar to those used in the JILA eEDM experiments utilizing trapped molecular ions. Hence, established techniques can readily be used with minimal modifications for quantum state manipulation of CeF$^{2+}$.
We expect the same to apply to PaF$^{3+}$ through the valence-isoelectronic relation with CeF$^{2+}$. 
We note the following properties of CeF$^{2+}$ provide untapped opportunities for new ways of quantum state manipulation: (i) the possibility of performing optical cycling transitions, and (ii) richer hyperfine structures due to additional nuclear spins. In the present work, we evaluated the additional complexity of performing quantum state preparation and detection with denser hyperfine structures to be within reasonably manageable regime. However, we call for experimental measurements of the molecular parameters before any concrete quantum state manipulation scheme can be conceived. To this end, the successful formation of CeF$^{2+}$ marks the foundational perquisite for future work on its molecular spectroscopy. In support of these spectroscopic studies, employing calculations of higher level of sophistication --- for example on the computation of $\Omega$-doubling parameter --- could on the theoretical side further constrain the parameters obtained by our present calculations. 

We note that EDM experiments in diatomic molecules have so far relied primarily on $^1\Sigma$, $^2\Sigma$, $^2\Pi$, and $^3\Delta_1$ electronic states. These states have well-understood advantages (and limitations) that have been developed over years of work in molecular EDM searches. However, not all diatomic molecules possess these states as low-lying levels suitable for experiments; our work on CeF$^{2+}$ shows one such example. In particular, as one moves into the heavier actinides, low-lying $\Phi$ states become increasingly common. Given the strong scientific potential of heavy actinide systems, a systematic investigation of the structure and EDM-relevant properties of $\Phi$ states with CeF$^{2+}$ would both complement and broaden the reach of ongoing molecular EDM searches.

In addition to its intrinsic physics merits, our work establishes (stable) CeF$^{2+}$ as a vital platform to develop experimental and theoretical frameworks that are largely applicable to (radioactive) PaF$^{3+}$ and, thus, opens the door to measure $\mathcal{P,T}$-odd effects in  PaF$^{3+}$. The success of the molecular formation of gaseous CeF$^{2+}$ using infrastructure and methods readily available at radioactive ion beam facilities provides valuable insight into the viability of transferring our methodology to the synthesis of $^{229}$PaF$^{3+}$. Based on the thermodynamic considerations of the reaction pathways involving the incoming Pa cations and the neutral SF$_6$ gas, it is proposed that the development of a dedicated gas exchange cell along with the production and injection of Pa$^{4+}$ ions will result in the formation of the desired molecule and charge state. Similarly to the experimental protocol which was followed to obtain UF$^{3+}$ \cite{Schroder2004}, PaF$^{3+}$ may also be produced from a singly charged molecule by electron impact ionisation via low energy electrons provided, for example, by TITAN's EBIT. All molecular formation techniques demonstrated with CeF molecules at TRIUMF in this work can be readily applied once protactinium beams become available on site. This paves the way for a new class of high-precision studies, combining the unique advantages of control techniques with molecules and exotic radionuclides in the search for new physics beyond the Standard Model of particle physics.

\section*{Acknowledgement}
We would like to express our gratitude to the technical teams at TRIUMF. We are grateful to E. A. Cornell for fruitful discussions as well as to Jan L{\"{o}}rin{\v{c}}{\`{i}}k for sharing insights into ion-beam formation from argon-ion bombardment of a macroscopic CeF3 crystal.

For the computational work, we acknowledge computing time provided at the NHR Center NHR@SW at Goethe-University Frankfurt. This was funded by the Federal Ministry of Education and Research and the state governments participating on the basis of the resolutions of the GWK for national high performance computing at universities (\url{http://www.nhr-verein.de/unsere-partner}). Financial support by the Deutsche Forschungsgemeinschaft (DFG, German Research Foundation) with the project number 445296313 is gratefully acknowledged.

The experimental work is supported  by the Natural Sciences and Engineering Research Council (NSERC) of Canada and through TRIUMF by the National Research Council (NRC) of Canada. OLIS beam development is funded by an NSERC Discovery Grant (C. Charles, \#SAPIN-2022-00026). 

\section{Computational Details}\label{sec:comps}

The ionization energy of CeF$^{2+}$ is computed on the unrestricted coupled cluster level with singles and doubles amplitudes included iteratively and the triples amplitudes perturbatively [UCCSD(T)] and UCCSD(T)-F12b as implemented within the quantum chemistry program package \textsc{Molpro} \cite{molpro2012a,molpro2019a} with a restricted open-shell Hartree--Fock (ROHF) reference wavefunction. Scalar relativistic effects are
included with an effective core potential on the Ce center (small core, ECP28 \cite{cao2002}) and on the Pa center (small core, ECP60 \cite{cao2003theoretical}) complemented by an atomic natural orbital valence basis set. At the F-atom an augmented correlation-consistent polarized basis set with quadruple-$\zeta$ quality (aug-cc-pVQZ) \cite{dunning:1989} was employed with the specific aug-cc-pVQZ-F12 basis set for the explicitly correlated coupled cluster calculations. For the S-atom an aug-cc-pVDZ basis set was used for the UCCSD(T) computations and an aug-cc-pVTZ-F12 basis set in the F12b computations.  Core electrons were frozen in the UCCSD(T) computations, whereas
all electrons were included in the correlation treatment on the F12 level. The ROHF reference wavefunction was optimized self-consistently up to a
change in the gradient of the orbital rotation of $10^{-13}$. The cluster amplitudes were optimized up to a change in energy of
$10^{-9}~\si{\hartree}$, while the bond lengths were computed on the level of UCCSD(T) up to a change in energy of $10^{-6}~\si{\hartree}$ between two  consecutive steps of a rational function optimization procedure. Symmetry constraints were imposed in the optimization of polyatomic molecules, which prevented complete dissociation of the SF$_6^+$ molecule and prohibited Jahn-Teller distortions in SF$_5$. Due to these constraints harmonic vibrational frequencies were not computed for the final structures of polyatomic molecules.

Quasi-relativistic calculations were performed on the level of complex generalized Hartree--Fock (cGHF) with two-component (2c) wavefunctions within the framework of the zeroth order regular approximation (ZORA) employing a model potential to alleviate the gauge dependence as proposed by van W\"ullen \cite{wullen:1998} and additional damping \cite{liu:2002} as implemented in a modified version \cite{wullen:2010} of the program package Turbomole \cite{ahlrichs:1989}. At the $^{140}$Ce, $^{178}$Hf, $^{232}$Th, $^{19}$F, $^{16}$O-atom Dyall's core-valence triple-$\zeta$ basis set was used \cite{dyall:2004,dyall:2007,gomes:2010}. For the metal centers Ce, Hf, Th mono-augmentation in an even-tempered manner was selected including additional steep s and p functions in order to describe the core-near region better (Table~\ref{tab:exponents}).
This is especially necessary for properties computed in the vicinity of the nucleus like $W_\mathcal{S}$. The wavefunction was optimized up to a change  in energy of $10^{-9}~\si{\hartree}$ and a change in the spin-orbit energy lower than $10^{-9}~\si{\%}$ between two cycles within the self-consistent-field (SCF) calculation. Bond lengths were optimized via the energies up to a change of the cartesian gradient of $10^{-5}~\si{\hartree}/\si{a_0}$. In all (quasi-)relativistic calculations a normalized spherical Gaussian nuclear density distribution as suggested by Visscher and Dyall  \cite{visscher:1997} was employed having the form  $\rho_A \left( \vec{r} \right) = \frac{\zeta_A^{3/2}}{\pi ^{3/2}} \text{e}^{-\zeta_A \left| \vec{r} - \vec{r}_A \right| ^2}$ with $\zeta_A = \frac{3}{2 r^2_{\text{nuc},A}}$ and the root-mean-square radius $r_{\text{nuc},A}$. Corresponding parameters were taken from Ref.~\cite{visscher:1997}, which uses an empirical formula for $r_{\text{nuc},A}$ that depends only on the atomic mass number \cite{johnson:1985}. The same assumption is made for the finite nucleus magnetization density. The isotopes $^{140}$Ce and $^{19}$F were employed unless stated otherwise. At this stage no effects emerging from higher electric nuclear moments are considered.

With the maximum overlap method (MOM) electronic excited states were computed by choosing occupation numbers according to the maximum overlap with the determinant of the initial guess (IMOM \cite{barca:2018}) or the determinant of the previous SCF cycle (standard MOM \cite{gilbert:2008}). If in respect to the previous SCF cycle the change in the differential density was above $5\times10^{-5}$ the IMOM approach was used and the standard MOM otherwise. On this level of theory atomic orbital contributions to the molecular spinors were obtained with a Mulliken population analysis.

$\mathcal{P,T}$-odd properties were computed with the toolbox approach outlined in Ref.~\cite{Gaul:2020} except for $W_\mathcal{S}$
where a finite nucleus model was used as described in Ref.~\cite{flambaum:2002} and an additional factor of 6 was introduced in order to align with the definition of current literature. The $\Omega$-doubling parameter $\Omega_\mathrm{p}$ is estimated for molecules aligned along the principal axis, which for convenience is defined as the $z$-axis, by computing the off-diagonal element of the electron angular momentum operator $\hat{\mathbf{J}}^\mathrm{e}$ \cite{kozlov1987} between Kramers partners
\begin{equation}
     \Omega_\mathrm{p} = 2 B \left<\Phi \middle| \hat{\mathbf{J}}^\mathrm{e} \middle| \tilde{\Phi} \right>_{x,y}
     \label{eq:omegadoubling}
\end{equation}
with the rotational constant $B$ as a pre-factor. Here, $x$, $y$ indicate the projection along the corresponding axis. The Kramers partner $\tilde{\Phi}$ is directly constructed from $\Phi$ and L\"owdin rules for nonorthogonal Slater determinants are used to compute the matrix element.
This approach is suitable for systems with a projection of the electron angular momentum on the internuclear axis of $\left| \Omega \right| = 1/2$, but misses additional coupling terms for $\left| \Omega \right| > 1/2$ because only direct couplings between Kramers partners are computed. 
Results for $\left| \Omega \right| = 1/2$ within the present scheme could also be improved upon by e.g. computing the matrix element in equation~\ref{eq:omegadoubling} with the spin-quantization axis chosen perpendicular to the internuclear axis \cite{verma2013}.

Applying L\"owdin rules for transition matrix elements of single particle operators computed between nonorthogonal single-determinant wave functions, we obtained transition electric dipole moments $\vec{\mu}$ for two self-consistently optimized cGHF determinants as outlined in Ref.~\cite{klues:2016}.

Franck--Condon factors, vibrational wavenumbers and, further, Einstein coefficients and radiative life times for the vibronic transition analysis are computed using the harmonic force constants in the case of the electronic structure of CeF$^{2+}$ and by solving the vibrational Schr\"odinger equation in a one-dimensional discrete variable representation (DVR, \cite{meyer:1970}) implemented in a development version of the program package \textsc{hotFCHT} \cite{berger:1998} using 1000 evenly spaced grid points for the vibronic transitions between the ground states of CeF$^{2+}$ and CeF$^{3+}$ with the grid starting at \SI{1.807}{\bohr} and ending at \SI{7.226}{\bohr}. For the excited state in CeF$^{3+}$ the same number of grid points and start of the grid were used but the grid ended at \SI{4.710}{\bohr}.
  
%bst derived from apsrev4-2.bst (revtex)
%Control: key (0)
%Control: author (72) initials jnrlst
%Control: editor formatted (1) identically to author
%Control: production of article title (1) required
%Control: page (0) single
%Control: year (1) truncated
%Control: production of eprint (0) enabled
%

\begin{table*}[!htb]
  \centering
  \caption{Vertical excitation energies $\Delta\tilde{\nu}_\mathrm{e}$, the sum
  of the Franck-Condon factors $f^{(i)}=\sum_if_{i-0}$ and Einstein coefficients $A$ of the
  vertical transitions from the equilibrium ground state of state (X)5/2.
  Spontaneous decay life-times $\tau_\mathrm{e}$ are estimated as $\tau_\mathrm{e}=\left( \sum_a^i
  A_i^a\right)^{-1}$ including all possible spontaneous emission path ways, where the individual transitions occur from state $i$ to state $a$.}
  \label{tab:fc}
  \resizebox{\textwidth}{!}{%
    \begin{tabular}{
      l l || 
      S[group-digits=false,table-text-alignment=right,input-symbols= (),table-format=5.0e+1]|
      S[group-digits=false,table-text-alignment=right,input-symbols= (),table-format=5.0e+1]|
      S[group-digits=false,table-text-alignment=right,input-symbols= (),table-format=5.0e+1]|
      S[group-digits=false,table-text-alignment=right,input-symbols= (),table-format=5.0e+1]|
      S[group-digits=false,table-text-alignment=right,input-symbols= (),table-format=5.0e+1]|
      S[group-digits=false,table-text-alignment=right,input-symbols= (),table-format=5.0e+1]|
      S[group-digits=false,table-text-alignment=right,input-symbols= (),table-format=5.0e+1]|
      S[group-digits=false,table-text-alignment=right,input-symbols= (),table-format=5.0e+1]
      }
      \toprule
                              &                               &       {$(1)3/2$}      &       {$(1)1/2$}      &       {$(1)7/2$}      &       {$(1)5/2$}      &       {$(2)3/2$}      &       {$(2)1/2$}      &       {$(3)3/2$}      &       {$(3)1/2$}      \\
      \midrule
                              & $\tau/s$                      &        4e-01  &        1e+00  &        1e+00  &        2e-03  &        2e-01  &        8e-02  &        7e-08  &        2e-07  \\
      \midrule
      \multirow{5}{*}{($\mathrm{X}$)5/2} & $\Delta\tilde{\nu}_\mathrm{e}/\si{cm^{-1}}$   &   1050     &   1462                &   2304                &   3306                &   3766                &   3929                &  31890                &  49913                \\
                              & $\Delta\tilde{\nu}_\mathrm{e}/\si{nm}$        &   9524                &   6840                &   4340                &   3025                &   2655                &   2545                &    314                &    200                \\
                              & $A/\si{Hz}$                   & 2e+00                 & 4e-01                 & 9e-01                 & 6e+02                 & 2e+00                 & 2e-02                 & 8e+06                 & 1e-01                 \\
                              & $f^{(0)}$                     & 0.99527               & 0.99928               & 0.99999               & 0.99312               & 0.99961               & 0.99900               & 0.46209               & 0.70543               \\
                              & $f^{(1)}$                     & 1.00000               & 0.99999               & 1.00000               & 0.99999               & 1.00000               & 0.99999               & 0.80624               & 0.95758               \\
      \hline
      
      \multirow{5}{*}{(1)3/2} & $\Delta\tilde{\nu}_\mathrm{e}/\si{cm^{-1}}$   &                       &    412                &   1254                &   2256                &   2716                &   2879                &  30840                &  48863                \\
                              & $\Delta\tilde{\nu}_\mathrm{e}/\si{nm}$        &                       &  24272                &   7974                &   4433                &   3682                &   3473                &    324                &    205                \\
                              & $A/\si{Hz}$                   &                       & 3e-01                 & 2e-04                 & 5e+00                 & 1e+00                 & 8e+00                 & 4e+06                 & 7e+02                 \\
                              & $f^{(0)}$                     &                       & 0.99095               & 0.99472               & 0.99980               & 0.99760               & 0.99003               & 0.52103               & 0.76156               \\
                              & $f^{(1)}$                     &                       & 0.99997               & 0.99997               & 1.00000               & 1.00000               & 0.99997               & 0.84769               & 0.97347               \\
      \hline
      
      \multirow{5}{*}{(1)1/2} & $\Delta\tilde{\nu}_\mathrm{e}/\si{cm^{-1}}$   &                       &                       &    842                &   1844                &   2304                &   2467                &  30428                &  48451                \\
                              & $\Delta\tilde{\nu}_\mathrm{e}/\si{nm}$        &                       &                       &  11876                &   5423                &   4340                &   4054                &    329                &    206                \\
                              & $A/\si{Hz}$                   &                       &                       & 4e-04                 & 1e-01                 & 5e-01                 & 5e+00                 & 3e+05                 & 3e+06                 \\
                              & $f^{(0)}$                     &                       &                       & 0.99947               & 0.98807               & 0.99785               & 0.99998               & 0.44187               & 0.68456               \\
                              & $f^{(1)}$                     &                       &                       & 0.99999               & 0.99989               & 0.99999               & 1.00000               & 0.78831               & 0.94921               \\
      \hline
      
      \multirow{5}{*}{(1)7/2} & $\Delta\tilde{\nu}_\mathrm{e}/\si{cm^{-1}}$   &                       &                       &                       &   1002                &   1462                &   1625                &  29586                &  47609                \\
                              & $\Delta\tilde{\nu}_\mathrm{e}/\si{nm}$        &                       &                       &                       &   9980                &   6840                &   6154                &    338                &    210                \\
                              & $A/\si{Hz}$                   &                       &                       &                       & 6e-01                 & 7e-05                 & 4e-05                 & 2e+00                 & 5e-02                 \\
                              & $f^{(0)}$                     &                       &                       &                       & 0.99247               & 0.99944               & 0.99923               & 0.45881               & 0.70218               \\
                              & $f^{(1)}$                     &                       &                       &                       & 0.99998               & 1.00000               & 0.99999               & 0.80378               & 0.95656               \\
      \hline
      
      \multirow{5}{*}{(1)5/2} & $\Delta\tilde{\nu}_\mathrm{e}/\si{cm^{-1}}$   &                       &                       &                       &                       &    460                &    623                &  28584                &  46607                \\
                              & $\Delta\tilde{\nu}_\mathrm{e}/\si{nm}$        &                       &                       &                       &                       &  21739                &  16051                &    350                &    215                \\
                              & $A/\si{Hz}$                   &                       &                       &                       &                       & 3e-01                 & 2e-04                 & 5e+05                 & 1e+01                 \\
                              & $f^{(0)}$                     &                       &                       &                       &                       & 0.99601               & 0.98702               & 0.53286               & 0.77239               \\
                              & $f^{(1)}$                     &                       &                       &                       &                       & 1.00000               & 0.99995               & 0.85578               & 0.97635               \\
      \hline
      
      \multirow{5}{*}{(2)3/2} & $\Delta\tilde{\nu}_\mathrm{e}/\si{cm^{-1}}$   &                       &                       &                       &                       &                       &    163                &  28124                &  46147                \\
                              & $\Delta\tilde{\nu}_\mathrm{e}/\si{nm}$        &                       &                       &                       &                       &                       &  61350                &    356                &    217                \\
                              & $A/\si{Hz}$                   &                       &                       &                       &                       &                       & 1e-02                 & 4e+05                 & 1e+01                 \\
                              & $f^{(0)}$                     &                       &                       &                       &                       &                       & 0.99739               & 0.47947               & 0.72228               \\
                              & $f^{(1)}$                     &                       &                       &                       &                       &                       & 1.00000               & 0.81845               & 0.96238               \\
      \hline
      
      \multirow{5}{*}{(2)1/2} & $\Delta\tilde{\nu}_\mathrm{e}/\si{cm^{-1}}$   &                       &                       &                       &                       &                       &                       &  27961                &  45984                \\
                              & $\Delta\tilde{\nu}_\mathrm{e}/\si{nm}$        &                       &                       &                       &                       &                       &                       &    358                &    217                \\
                              & $A/\si{Hz}$                   &                       &                       &                       &                       &                       &                       & 1e+05                 & 2e+06                 \\
                              & $f^{(0)}$                     &                       &                       &                       &                       &                       &                       & 0.43799               & 0.68057               \\
                              & $f^{(1)}$                     &                       &                       &                       &                       &                       &                       & 0.78513               & 0.94776               \\
      \hline
      
      \multirow{5}{*}{(3)3/2} & $\Delta\tilde{\nu}_\mathrm{e}/\si{cm^{-1}}$   &                       &                       &                       &                       &                       &                       &                       &  18023                \\
                              & $\Delta\tilde{\nu}_\mathrm{e}/\si{nm}$        &                       &                       &                       &                       &                       &                       &                       &    555                \\
                              & $A/\si{Hz}$                   &                       &                       &                       &                       &                       &                       &                       & 7e-03                 \\
                              & $f^{(0)}$                     &                       &                       &                       &                       &                       &                       &                       & 0.92981               \\
                              & $f^{(1)}$                     &                       &                       &                       &                       &                       &                       &                       & 0.99976               \\
      \bottomrule
    \end{tabular}
  }%
\end{table*}

\begin{table*}[!htb]
\centering
\caption{Electric transition dipole moments $\left| \vec{\mu}(i \leftarrow j)\right|^2$
of the vertical transitions from the equilibrium structure of the electronic ground state
(X)$5/2$}
\label{tab:tdm}
\resizebox{\textwidth}{!}{%
\begin{tabular}{
 c | 
S[table-format=1.1e-2,round-mode=figures,round-precision=2]     
S[table-format=1.1e-2,round-mode=figures,round-precision=2]   
S[table-format=1.1e-2,round-mode=figures,round-precision=2]   
S[table-format=1.1e-2,round-mode=figures,round-precision=2]   
S[table-format=1.1e-2,round-mode=figures,round-precision=2]   
S[table-format=1.1e-2,round-mode=figures,round-precision=2]   
S[table-format=1.1e-2,round-mode=figures,round-precision=2]   
S[table-format=1.1e-2,round-mode=figures,round-precision=2] 
}
\toprule
$\left| \vec{\mu}(i \leftarrow j)\right|^2 / (\mathrm{e}^2\mathrm{a}_0^2)$
                & {(1)3/2}      & {(1)1/2}      & {(1)7/2}      & {(1)5/2}      & {(2)3/2}      & {(2)1/2}      & {(3)3/2}      & {(3)1/2}      \\
\midrule
($\mathrm{X}$)5/2 & 1.0e-03     & 6.1e-05       & 3.5e-05       & 8.0e-03       & 2.1e-05       & 1.3e-07       & 1.2e-01       & 4.6e-10       \\
(1)3/2            &             & 2.2e-03       & 3.9e-08       & 2.3e-04       & 2.8e-05       & 1.6e-04       & 6.7e-02       & 2.8e-06       \\
(1)1/2            &             &               & 3.6e-07       & 1.0e-05       & 2.1e-05       & 1.8e-04       & 5.7e-03       & 1.3e-02       \\
(1)7/2            &             &               &               & 3.0e-04       & 1.0e-08       & 4.5e-09       & 3.5e-08       & 2.5e-10       \\
(1)5/2            &             &               &               &               & 1.8e-03       & 3.2e-07       & 9.7e-03       & 4.7e-08       \\
(2)3/2            &             &               &               &               &               & 1.4e-03       & 8.4e-03       & 5.2e-08       \\
(2)1/2            &             &               &               &               &               &               & 2.3e-03       & 1.1e-02       \\
(3)3/2            &             &               &               &               &               &               &               & 5.8e-10       \\
\bottomrule

\end{tabular}
}
\end{table*}

\begin{table*}[!htb]
\caption{\justifying Adiabatic electronic ionization energies (IE) and dissociation energies (DE) for relevant
dissociation channels computed on the level of ROHF-UCCSD(T) are given together
with literature values where possible. In order to further gauge the validity
of the results ROHF-UCCSD(T)-F12b calculations are shown for the
S$_{0-1}$F$_n^{m+}$ species.}
\label{tab:dissociations}
\begin{threeparttable}
\centering
\begin{tabular}{
l
S[table-format=3.5,round-mode=figures,round-precision=3]
S[table-format=1.3,round-mode=figures,round-precision=3]
S[table-format=1.3,round-mode=figures,round-precision=3]
}
\toprule
 & {Lit.}  & {ROHF-UCCSD(T)} & {ROHF-UCCSD(T)-F12b}       \\
\midrule
IE(SF$_6$)                                & 15.116*~\tnote{a)}       & 16.01**$^\dagger$  & 15.87**    \\
IE(SF$_5$)                                & 9.60*~\tnote{b)}         &  9.86***$^\dagger$   & 9.45$^{\dagger}$    \\
DE(SF$_6\rightarrow$SF$_5^+$+F$^-$)       &                          & 10.97***$^\dagger$  & 10.93    \\
DE(SF$_6^+\rightarrow$SF$_5^+$+F)         & -0.99*~\tnote{c)}        & -1.65***$^\dagger$**  & -1.57**    \\
\hline   
IE(F)                                     & 17.423~\tnote{d)}        & 17.35 & 17.40    \\
EA(F)                                     & 3.40119~\tnote{e)}       & 3.38  & 3.38     \\
\hline   
IE(Ce$^{+}$)                              & 10.956~\tnote{f)}       & 9.60*$^\ddagger$             \\
IE(Ce$^{2+}$)                             & 20.198~\tnote{g)}       & 19.34*$^\ddagger$        \\
IE(CeF$^{+}$)                             &                          & 11.2*            \\
IE(CeF$^{2+}$)                            &                          & 24.13*           \\
DE(CeF$^{2+}\rightarrow$ Ce$^{2+}+$F)     &                          & 6.45*$^\ddagger$            \\
DE(CeF$^{2+}\rightarrow$ Ce$^{+}+$F$^+$ ) &                          & 14.20            \\
DE(CeF$^{2+}\rightarrow$ Ce$^{3+}+$F$^-$) &                          & 22.42            \\
DE(CeF$^+   \rightarrow$ Ce$^{+}+$F )     &                          & 8.08*            \\
DE(CeF$^{3+}\rightarrow$ Ce$^{3+}+$F      &                          & 1.66*             \\
DE(CeF$^{3+}\rightarrow$ Ce$^{2+}+$F$^+$) &                          & -0.33$^\ddagger$            \\

\hline   
IE(Pa$^{2+}$)                             & 17.6~\tnote{h)}          & 18.04*~\tnote{i)}\\
IE(Pa$^{3+}$)                             & 30.9~\tnote{h)}          & 29.59*~\tnote{i)}\\
IE(PaF$^{2+}$)                            &                          & 19.93*           \\
IE(PaF$^{3+}$)                            &                          & 31.71*           \\
DE(PaF$^{3+}\rightarrow$ Pa$^{3+}+$F)     & 4.9~\tnote{j)}          & 5.80*~\tnote{i)}  \\
DE(PaF$^{3+}\rightarrow$ Pa$^{2+}+$F$^+$) & 4.2~\tnote{j)}           & 5.11~\tnote{i)}  \\
DE(PaF$^{3+}\rightarrow$ Pa$^{4+}+$F$^-$) & 31.8~\tnote{j)}          & 32.01~\tnote{i)} \\
DE(PaF$^{2+}\rightarrow$ Pa$^{2+}+$F)     &                          & 7.69*            \\
DE(PaF$^{4+}\rightarrow$ Pa$^{4+}+$F)     & 3.7~\tnote{j)}           & 3.68*~\tnote{i)}   \\
DE(PaF$^{4+}\rightarrow$ Pa$^{3+}+$F$^+$) & -9.7~\tnote{j)}          & -8.57~\tnote{i)}  \\
\bottomrule
\end{tabular}
\begin{tablenotes}
  \item[a)] Threshold photoelectron spectroscopy \cite{yencha:1997}; 
  \item[b)] Collision-induced dissociation/charge-transfer reactions \cite{fisher:1992};
  \item[c)] Threshold photoelectron-photoion coincidence spectroscopy \cite{creasey:1993}; 
  \item[d)] Absorption spectra; \cite{huffman:1967};
  \item[e)] Photodetachment microscopy \cite{blondel:2001}; 
  \item[f)] Interpolation \cite{johnson:2017}; 
  \item[g)] Interpolation \cite{sugar1973};
  \item[h)] Complete Active Space Self-Consistent Field computations employing scalar-relativistic effective core potentials \cite{cao2003theoretical};
  \item[i)] ROHF-UCCSD(T) \cite{zulch2022cool}; 
  \item[j)] ROHF-UCCSD(T) including predictions of spin-orbit coupling \cite{zulch2022cool};
  \item[*] Values used to generate Figure~\ref{fig:reactionenergy};
  \item[**] Structure of SF$_6^+$ optimized with $O_\mathrm{h}$ [UCCSD(T)] or $D_{4\mathrm{h}}$ [UCCSD(T)-F12b] symmetry constraints;
  \item[***] SF$_5^+$ structure optimized on ROHF level of theory with $C_{3\mathrm{v}}$ symmetry constraints;
  \item[$\dagger$] Structures of SF$_6$, SF$_6^+$, SF$_5$ optimized on ROHF-UCCSD(T) level of theory with basis set of double-$\zeta$ quality;
  \item[$\ddagger$] Lowest energy state of Ce$^{2+}$ was obtained for df configuration instead of the experimentally reported f$^2$ configuration.
\end{tablenotes}
\end{threeparttable}
\end{table*}

\begin{figure*}[!htb]
\centering
\begin{tabular}{cc}
  ai) \vspace{.3cm}\hspace{-1cm}\raisebox{-\height-0.2cm}{\includegraphics[width=0.4\textwidth]{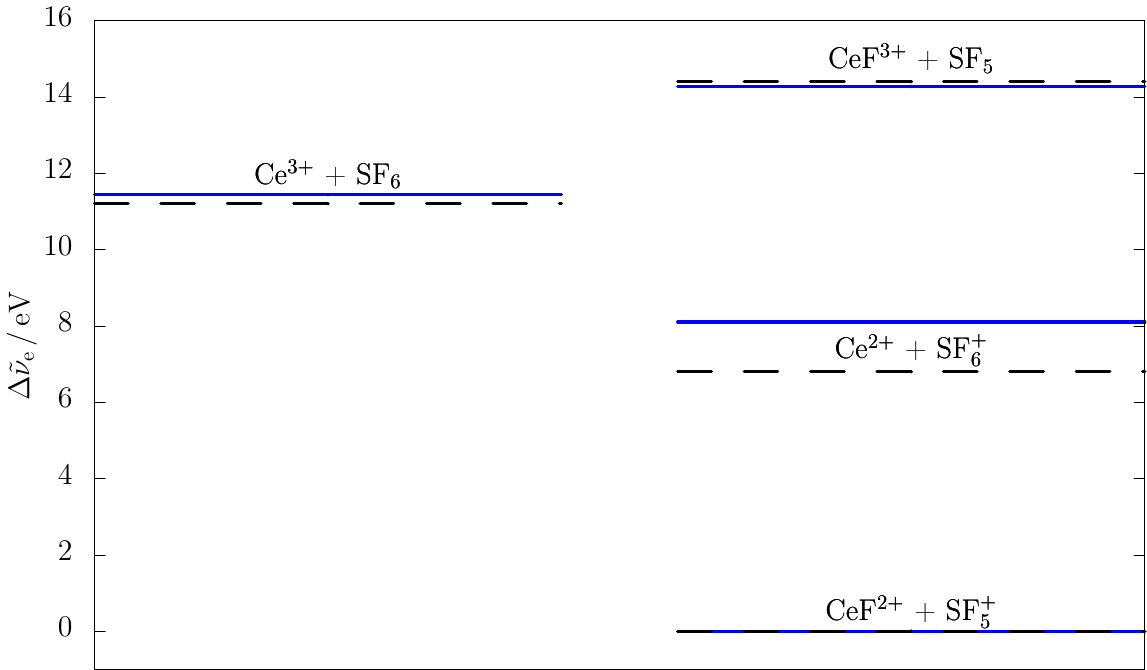}} &\hspace{0.7cm} aii) \vspace{.3cm}\hspace{-1cm}\raisebox{-\height-0.2cm}{\includegraphics[width=0.4\textwidth]{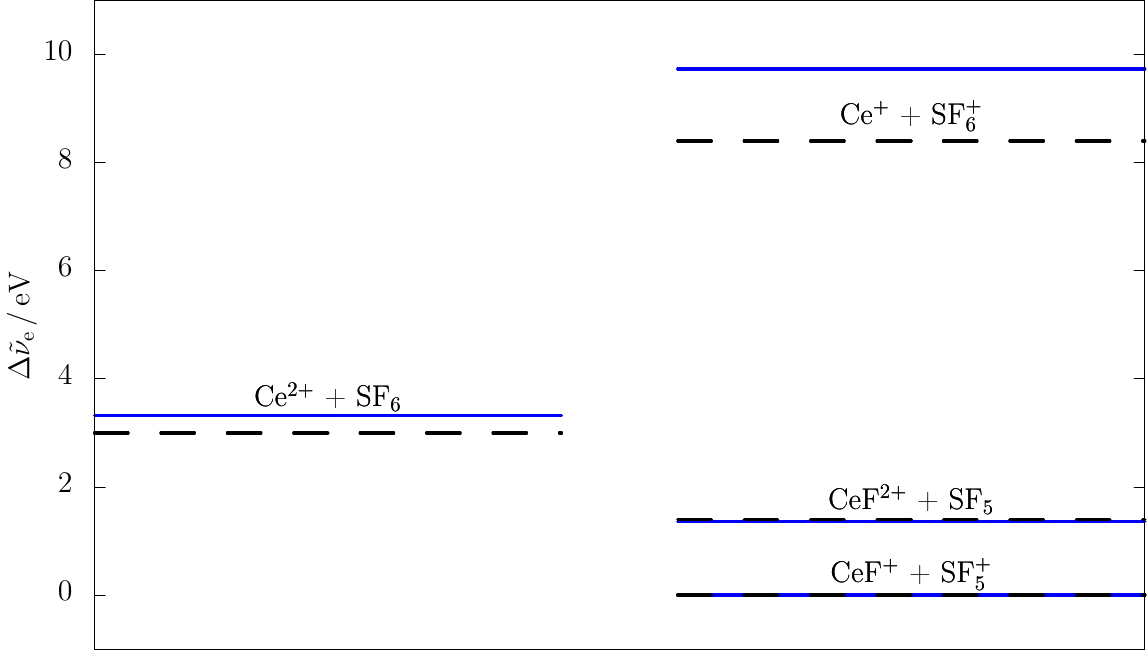} }\\
  bi) \vspace{.3cm}\hspace{-1cm}\raisebox{-\height-0.2cm}{\includegraphics[width=0.4\textwidth]{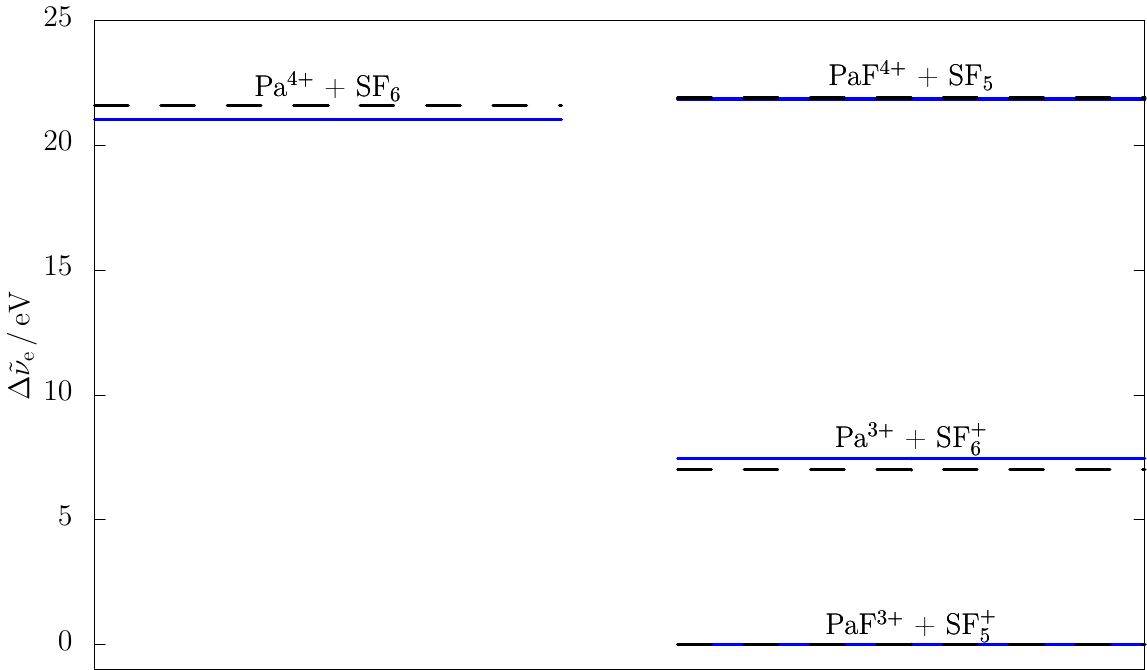}} &\hspace{0.7cm} bii) \vspace{.3cm}\hspace{-1cm}\raisebox{-\height-0.2cm}{\includegraphics[width=0.4\textwidth]{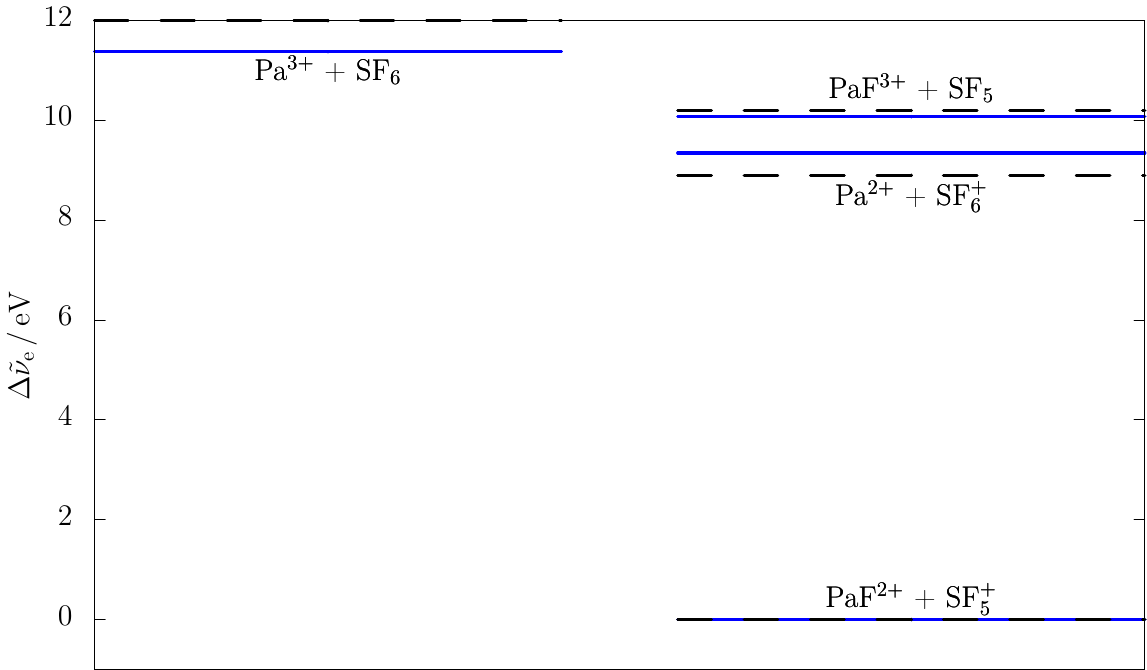}}\\
\end{tabular}
\caption{\justifying Energy level diagrams for selected reactions of cerium [row (a)] and protactinium [row (b)] in various charge states with neutral sulfur hexafluoride. The dashed black lines indicate the values used in the main text, while blue lines represent a pure theoretical scheme employing the UCCSD(T) values from Table~\ref{tab:dissociations}. The lowest product, which is the desired product in all cases, is set to \SI{0}{eV}.}
\label{fig13}
\end{figure*}

\begin{table}
\caption{Additional exponents used in Dyall's core valence basis set of
triple-$\zeta$ quality for Ce, Hf, Th.}
\label{tab:exponents}
\begin{tabular}{
c
S[table-format=1.10e-2]
S[table-format=1.10e-2]
S[table-format=1.10e-2]
}
\toprule
    & {Ce}  & {Hf} & {Th} \\
\midrule
\multirow{8}{*}{s}
& 7.2668207200e+09 & 6.4738719000e+09 & 5.7354665400e+09 \\
& 2.4222735733e+09 & 2.1579573000e+09 & 1.9118221800e+09 \\
& 8.0742452444e+08 & 7.1931910000e+08 & 6.3727406000e+08 \\
& 2.6914150815e+08 & 2.3977303333e+08 & 2.1242468667e+08 \\
& 8.9713836049e+07 & 7.9924344444e+07 & 7.0808228889e+07 \\
& 8.1191021625e+07 & 7.2331531722e+07 & 6.4081447144e+07 \\
& 4.5846867850e+07 & 4.0977274500e+07 & 3.6310500850e+07 \\
& 8.6786343088e-03 & 1.2157877205e-02 & 9.8318371004e-03 \\
\\
\multirow{8}{*}{p}
& 1.1511538800e+09 & 2.0743327100e+09 & 5.1735851300e+09 \\
& 3.8371796000e+08 & 6.9144423667e+08 & 1.7245283767e+09 \\
& 1.2790598667e+08 & 2.3048141222e+08 & 5.7484279222e+08 \\
& 4.2635328889e+07 & 7.6827137407e+07 & 1.9161426407e+08 \\
& 1.4211776296e+07 & 2.5609045802e+07 & 6.3871421358e+07 \\
& 1.2861657548e+07 & 2.3176186451e+07 & 5.7803636329e+07 \\
& 6.6950757550e+06 & 1.2321569025e+07 & 3.3346063950e+07 \\
& 6.9333511916e-03 & 8.7031078719e-03 & 6.7864632963e-03 \\
\\
\multirow{1}{*}{d}
& 2.0719550217e-02 & 2.0493874947e-02 & 1.7849913907e-02 \\
\\
\multirow{1}{*}{f}
& 2.6580533524e-02 & 1.5045628570e-02 & 5.7689905741e-02 \\
\\
\multirow{1}{*}{g}
& 5.1197513509e-02 & 2.6382053555e-02 & 2.3779934821e-02 \\
\\
\multirow{1}{*}{h}
& 9.9539783312e-02 & 4.1977418000e+00 & 5.0971980000e-01 \\
\bottomrule
\end{tabular}
\end{table}

\end{document}